\documentclass[useAMS,usenatbib]{mn2e}
\pdfoutput=1

\usepackage{rotating}
\usepackage{aas_macros}

\def\ergcm2s{\ifmmode {\rm\,erg\,cm^{-2}\,s^{-1}}\else
                ${\rm\,ergs\,cm^{-2}\,s^{-1}}$\fi}

\newcommand{\lya}{\ifmmode {\rm\,Ly\alpha}\else
                Ly$\alpha$\fi}
\newcommand{\msol}{M$_{\sun}$}
\newcommand{\oiii}{[O\,{\sc iii}]}
\newcommand{\oii}{[O\,{\sc ii}]}
\newcommand{\zsol}{Z$_{\sun}$}
\newcommand{\lsol}{L$_{\sun}$}
\newcommand{\csq}{$\chi^2$}

\newcommand{\fluxl}{erg s$^{-1}$ cm$^{-2}$ \AA$^{-1}$}
\newcommand{\lf}{erg s$^{-1}$ cm$^{-2}$}
\newcommand{\kms}{km s$^{-1}$}
\newcommand{\gfil}{\textit{g$^{\prime}$}}
\newcommand{\ifil}{\textit{i$^{\prime}$}}
\newcommand{\zfil}{\textit{z$^{\prime}$}}
\newcommand{\rfil}{\textit{r$^{\prime}$}}
\newcommand{\ufil}{\textit{u$^{\ast}$}}
\newcommand{\kfil}{\textit{K$_s$}}
\newcommand{\um}{$\mu$m}
\newcommand{\ages}{age$_{\mathrm{SFR}}$}

\title{Galactic winds and stellar populations in Lyman-alpha emitting
  galaxies at z $\sim$ 3.1}
\author[E. McLinden et al.]{E. M. McLinden, $^{1}$\thanks{E-mail:
    mclinden@astro.as.utexas.edu}
J. E. Rhoads,$^{2}$ 
S. Malhotra,$^{2}$ 
S. L. Finkelstein,$^{3}$ 
M. L. A. Richardson,$^{2}$ 
\newauthor B. Smith,$^{2}$ 
V. S. Tilvi,$^{4}$ \\
$^{1}$ McDonald Observatory, The University of Texas at Austin, Austin, TX 78712\\
$^{2}$ School of Earth and Space Exploration,  Arizona  State University,  Tempe, AZ  85287\\
$^{3}$ Department of Astronomy, The University of Texas at Austin, Austin, TX 78712\\
$^{4}$ George P. and Cynthia Woods Mitchell Institute for Fundamental\\
Physics and Astronomy, and Department of Physics and Astronomy, Texas\\
A\&M University, College Station, TX, 77843
}
\begin{document}

\date{}

\pagerange{\pageref{firstpage}--\pageref{lastpage}} \pubyear{2013}

\maketitle

\label{firstpage}

\begin{abstract}
We present a sample of 33 spectroscopically confirmed z $\sim$ 3.1 \lya\ emitting galaxies (LAEs) in the COSMOS field.  This paper details the narrowband survey we conducted to detect the LAE sample, the optical spectroscopy we performed to confirm the nature of these LAEs, and a new near-infrared spectroscopic detection of the \oiii\ 5007 \AA\ line in one of these LAEs.  This detection is in addition to two \oiii\ detections in two z $\sim$ 3.1 LAEs we have reported on previously \citet{mcl11}.    The bulk of the paper  then presents detailed constraints on the physical characteristics of the entire LAE sample from spectral energy distribution (SED) fitting.  These characteristics include mass, age, star-formation history, dust content and metallicity.  We also detail an approach to account for nebular emission lines in the SED fitting process - wherein our models predict the strength of the \oiii\ line in an LAE spectrum.  We are able to study the success of this prediction because we can compare the model predictions to our actual near-infrared observations both in galaxies that have \oiii\ detections and those that yielded non-detections.  We  find a median stellar mass of 6.9 $\times$ 10$^{8}$ \msol\ and a median star formation rate weighted stellar population age of 4.5 $\times$ 10$^{6}$ years.  In addition to SED fitting, we quantify the velocity offset between the \oiii\ and \lya\ lines in the galaxy with the new \oiii\ detection, finding that the \lya\ line is shifted 52 \kms\ red-ward of the \oiii\ line, which defines the systemic velocity of the galaxy.  
\end{abstract}


\section{Introduction}
High redshift Lyman alpha emitting galaxies (LAEs) are now routinely
identified via narrowband detection methods (e.g. Cowie \& Hu, 1998,
Malhotra \& Rhoads, 2002, 2004, Ouchi et al. 2003, Gawiser et
al. 2006).  Now that samples of these galaxies can be more easily
compiled at a variety of redshifts, attention has turned to deriving
the physical characteristics of these galaxies by fitting spectral
energy distributions (SEDs) to their observed photometry of these
galaxies \citep{gaw06, gaw07, pirz07, nil07, nil11, fink07, fink08,
fink09, fink11b, lai07, lai08, ono10a, acq12}.

The majority of early work in SED fitting (e.g. Gawiser et al. 2006,
Finkelstein et al. 2007, Nilsson et al. 2007)  relied on deriving
average LAE characteristics from stacked LAE samples, but stacked
analyses may not reveal the full distribution of LAE characteristics.
Most efforts to date have found LAEs to be largely young or of
intermediate ages and having characteristically small masses
\citep{pirz07, gaw07, fink09, cow11}, but SED fitting procedures tend to vary from author to author,
making direct comparisons of derived characteristics difficult from
sample to sample.  In addition, SED fitting procedures for
starbursting galaxies have been evolving recently to account for
contamination of observed photometry from  nebular emission
lines. \citet{sd09} 
and others have demonstrated that failure to include these lines,
produced from hot gas in star forming regions, can drastically alter
the ages and masses derived from SED fitting.

In this paper we present a simple way to account for nebular emission
during SED fitting in our sample of 33 spectroscopically confirmed z
$\sim$ 3.1 LAEs.  The technique we outline in this paper allows us to
predict the strength of the \oiii\ nebular emission line, which we can
compare to the NIR detections and upper limits we have made of this
line in six z $\sim$ 3.1 LAEs.

In Section \ref{sec:obsdat} we present the extensive observations that form the foundation of this paper, including a narrowband survey to find LAE candidates, optical spectroscopy to confirm LAE candidates and NIR spectroscopy to look for rest-frame optical nebular emission lines in these LAEs.  We also present our data reduction techniques in this section.  In Section \ref{sec:res1} we present our results from optical and NIR spectroscopy, including a new \oiii\ measurement in one LAE and the subsequent velocity offset between \oiii\ and \lya\ that we measure in this object.  Section \ref{sec:sed} outlines our methods for SED fitting, including the introduction of a new method to account for nebular emission lines in the SED fitting process.  We present our results from SED fitting in Section \ref{sec:res2}.  Finally, in Section \ref{sec:disc} we discuss the ability of our SED fitting process to match our observations of the \oiii\ line in LAEs.  We also compare our SED results to those presented by other authors.

Where relevant, we adopt the standard cosmological parameters H$_{0} =$ 70 km s$^{-1}$ Mpc$^{-1}$, $\Omega_{m} = $ 0.3, and $\Omega_{\Lambda} =$ 0.7 \citep{sper}.  Also we use the following vacuum wavelengths, 1215.67  \AA\ for \lya,  3727.092/3729.875 \AA\ for [O\,{\sc ii}], 4862.683 \AA\ for H$\beta$  and 4960.295/5008.240 for [O\,{\sc iii}] from the Atomic Line List v2.04\footnote{http://www.pa.uky.edu/$\sim$peter/atomic/index.html}.  All quoted magnitudes are AB magnitudes.

\section{OBSERVATION AND DATA}\label{sec:obsdat}

\subsection{Narrowband Survey}
We collected data for our narrowband (NB) survey in 2007 (PI Finkelstein) and 2009 (PI McLinden) using the 90-inch Bok telescope with the 90Prime Camera \citep{wil04} at Steward Observatory.   The survey was completed in the COSMOS field centered at R.A. 10:00:28.6 and decl. +02:12:21.0 (J2000) \citep{cap07}. The NB data were collected on UT 2007 February 21 and 22.  The rest of the NB data, described below, were collected on UT 2009 February 27, 28 and 2009 March 1.  We used the KPNO \oiii\ filter, centered at 5025 \AA, with a narrow bandpass of 55\AA, to select \lya\ emission from z = 3.11 -- 3.16.  The 90Prime instrument was originally outfitted with a 1 deg$^2$ field of view from four 4096 pixel x 4096 pixel CCDs.  This was the instrument setup for our 2007 observations. Due to instrument failure however, our 2009 observations were made with only a single 4064 pixel x 4064 pixel CCD, providing less coverage and therefore less depth than we had initially anticipated.  The pixel scale for 90Prime is 0.45\arcsec\ pixel$^{-1}$. 

To reduce the narrowband data we used the MSCRED package in IRAF.   The data reduction process included bias subtraction, overscan subtraction, flat-fielding and cross talk correction using CCDPROC.  We applied astrometry corrections using the USNO B1.0 catalog with the IRAF tasks MSCTPEAK and MSCCMATCH.  Cosmic ray rejection proceeded using the JMCCREJ algorithm developed by Rhoads (2000).  Complete bad pixel masks, including manually added satellite trails, were created and applied to each frame before stacking.  MSCIMAGE was used to resample individual exposures onto a common pixel grid. Scaling was determined using MSCIMATCH.  Before stacking the images, we applied skyflats in CCDPROC and did a sky subtraction using MSCSKYSUB. Finally, we used MSCSTACK to stack each individual frame into a single final exposure.  A total of 50 frames, representing 16.67 hours of integration, were stacked to create this final 1.96 deg$^2$ image.   We find a 5$\sigma$ depth of 23.2 magnitudes in a 3\arcsec\ diameter aperture, which corresponds to a line flux lower limit of $\sim 1.2 \times 10^{-16} \ergcm2s$ for pure emission line sources.  The point spread function FWHM in our final stack is $\sim$ 3.62 pixels, corresponding to 1.63\arcsec.

\subsection{Broadband Data for Candidate Selection}
\begin{figure*}
\centering
\includegraphics[scale=0.5]{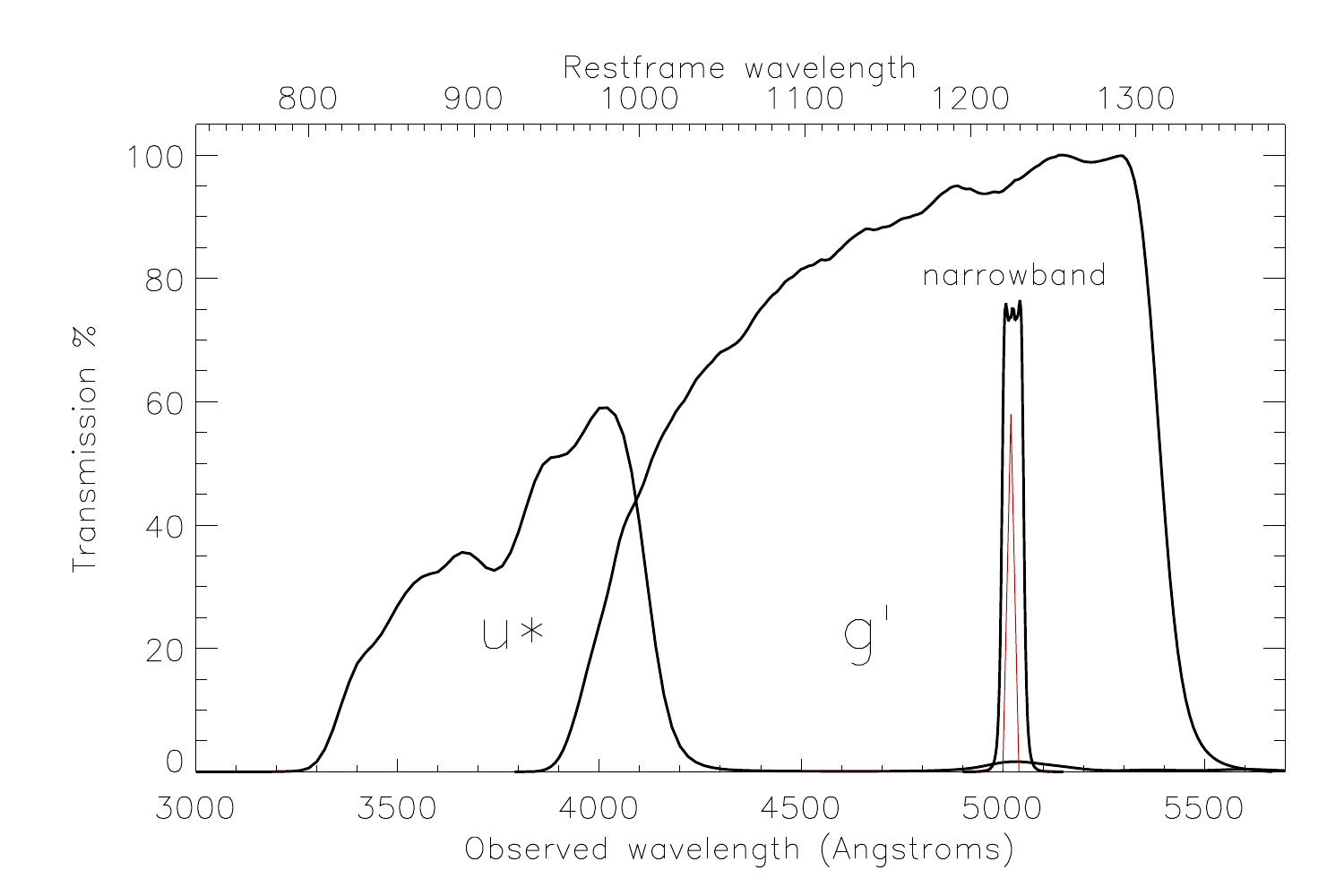}
\caption{Transmission curves for the \ufil, \gfil\ and narrowband filters.  The CFHT \ufil\ filter is centered at 3798 \AA\ (d$\lambda=$ 720\AA), the Subaru \gfil\ filter is centered at 4780 \AA (d$\lambda=$ 1265\AA), and  the KPNO \oiii\ narrowband filter ($\lambda=$ 5025 \AA, d$\lambda=$ 55\AA) used for our narrowband survey lies within the \gfil\ filter.  Also shown is an mock \lya\ line (not to scale) in red, inside the narrowband filter.}
\label{fig:filters}
\end{figure*}

Our narrowband survey is complemented by a plethora of publicly
available data in the COSMOS field. In particular, we used  \ufil\ and \gfil\ band images from the NASA/IPAC archive\footnote{http://irsa.ipac.caltech.edu/data/COSMOS/datasets.html} in concert with our narrowband survey to select LAEs as described in Section \ref{sec:select} below.  The \ufil\ band images come from the MegaPrime instrument \citep{bo03} on the 3.6 m Canada-France-Hawaii Telescope. The \ufil\ images have a 5$\sigma$ depth in a 3\arcsec\ aperture of 26.4 \citep{cap07}.  The \ufil\ filter is centered at 3798 \AA\ and has a bandpass of 720 \AA.  The \gfil\ images come from Suprime-Cam on the 8.3 m Subaru telescope.  The 5$\sigma$ depth in a 3\arcsec\ aperture for the \gfil\ images is 27.0 \citep{cap07}.  The \gfil\ filter is centered at 4780 \AA\ and has a bandpass of 1265 \AA.  The filter transmission curves for the \ufil, \gfil, and narrowband are shown in Figure \ref{fig:filters}.  Note that one of the wide broadband filters, the \gfil\ filter, encompasses the \oiii\ narrowband and the other broadband filter, the \ufil\ band,  is fully blue-ward of the narrowband filter and \lya\ line.  This filter setup is essential for our selection of LAEs via narrowband imaging because an LAE at 3.11 $\ge$ z $\ge$ 3.16 ought to have an excess of flux in the narrowband when compared to the \gfil\ band, due to the location of the \lya\ line.  The LAE SED should also be attenuated blue-ward of the \lya\ line due to \lya\ forest absorption.   Our use of the \ufil\ and \gfil\ filters with our narrowband data allows us to detect both this flux excess and attenuation as detailed in Section \ref{sec:select} below.

\subsection{LAE Candidate Selection via Source Extractor}\label{sec:select}
We selected LAE candidates based on a combination of their narrowband and broadband photometry. To do this we used Source Extractor (SExtractor, Bertin \& Arnouts 1996) to detect objects and extract their photometry.  We used aperture photometry measurements (FLUX\_APER) from SExtractor, in a 3\arcsec\ diameter aperture.  Objects were extracted from the central 1.44 deg$^2$ of our narrowband survey, avoiding some of the shallower edges of our survey.

We extracted fluxes for all objects detected in the narrowband image by running SExtractor in dual-image mode.  In dual-image mode, our narrowband image was the `detection' image and a second image, either the narrowband, \ufil\ or \gfil\ image, was `the measurement' image.  The detection image determines where objects are found, the measurement image is used to measure fluxes at those locations.  In order to run SExtractor in dual image mode, both images must have the same pixel scale.  To make this possible, we registered the \ufil\ and \gfil\ images to the narrowband image with the IRAF tasks WCSMAP and GEOTRAN, where WCSMAP computes a spatial transformation function from the WCS information of the images and GEOTRAN performs this geometric transformation.  This process changes the resolution of broadband images from their native resolution of 0.15\arcsec\ pixel$^{-1}$ to the 0.45\arcsec\ pixel$^{-1}$ resolution of the narrowband image.  Such a transformation means measurements can be made in the exact same pixels from image to image.  The NASA/IPAC COSMOS archive also includes maps of image RMS for all of our broadbands, so we used these as WEIGHT\_IMAGES in SExtractor with the SExtractor parameter WEIGHT\_TYPE set to MAP\_RMS.  We created RMS maps for our narrowband image using the CHECK\_IMAGE feature of SExtractor.

The final set of confirmed LAEs presented in this paper is a compilation of objects from multiple LAE selections.  Our earliest selection of LAEs was performed on an a preliminary reduction of our narrowband data that only included the 2007 data. Later selections were performed on reductions of the narrowband data that contained the full 16.67 hours of data.  Our selection criteria have also evolved since the preliminary selection, as we have honed in on criteria more likely to yield confirmations in optical spectroscopy given our specific combination of very deep broadband images (\ufil, \gfil) and our shallower narrowband image.  In addition, we re-reduced the narrowband datato try to improve the quality of the final product.  This improvement was achieved with additional flat fielding and chip-by-chip sky subtraction, as well as improved image weighting, and image scaling.  Our basic LAE selection criteria are: 

\begin{equation}
\frac{f_{NB}}{\delta f_{NB}} \ge 5 \textrm{ and } \frac{f_g}{\delta f_g} \ge 2
\end{equation}
\begin{equation}
\frac{f_{NB}}{f_g} \ge 2
\end{equation}
\begin{equation}
\frac{f_{NB} - f_g}{\sqrt{\delta f_{NB}^2 + \delta f_{g}^2}} \ge 4
\end{equation}
\begin{equation}
f_u \le 10^{-4/5}f_g + 2\delta f_u
\end{equation}

where $f_u$ is flux in the u$^*$ band, $f_g$ is flux in the \gfil\ band, $f_{NB}$ is flux in the narrowband band, $\delta f_u$ is flux error in the u$^*$ band, $\delta f_g$ is flux error in the \gfil\ band, and $\delta f_{NB}$ is flux error in the narrowband.  In other words, to be an LAE candidate, an object must (1) be detected at the 5$\sigma$ level in the narrowband and at the 2$\sigma$ level in the g band, (2) have an excess of flux density in the narrowband compared to the g band (corresponding to rest-frame equivalent width $\ge$ 14.7 \AA) (3) that flux excess must be significant at the 4$\sigma$ level, and (4) the flux blue-ward of the \lya\ line must be attenuated in a manner congruent with expected \lya\ forest absorption.  For z $\sim$ 3.1 this means that the \ufil\ band should be at least two magnitudes fainter than the \gfil\ band \citep{mad95}, but can amount to somewhat less than 2 magnitudes when you incorporate the \ufil\ error bars.  These criteria are based on those developed by Rhoads \& Malhotra (2001).  We note that the requirement of a detection in the \gfil\ band is not a requirement that there be continuum detection, as the presence of the \lya\ line can be sufficient to cause a detection in the \gfil\ band.  12 of the objects in sample presented in this paper were initially selected with these criteria (labeled as selection 1 in Table \ref{phottbl}).  Three  additional (unique) objects in the sample were  selected with a less stringent fourth criterion, i.e. $f_u \le 10^{-4/5}f_g + 3\delta f_u$ (selection 2). 14 more (unique) LAEs in the sample were  selected with an also less stringent fourth criterion, $m_u - m_g > 0.5$ (selection 3).  Because the u$^*$ band data are so much deeper than our narrowband data we found these less stringent requirements on the suppression of the u band flux to be useful.

In addition to the traditional narrowband selection criteria detailed above, we also experimented with finding LAEs using a broadband detection as the initial requirement.  This was possible again because the publicly available broadband data were so much deeper than our narrowband survey.  Three of the objects in our current sample were selected this way (selection 4).  The criteria in this case are as follows:

\begin{equation}
\frac{f_g}{\delta f_g} \ge 5
\end{equation}
\begin{equation}
\frac{f_{NB}}{f_g} \ge 1.445
\end{equation}
\begin{equation}
\frac{f_{NB} - f_g}{\sqrt{\delta f_{NB}^2 + \delta f_{g}^2}} \ge 2
\end{equation}
\begin{equation}
\frac{f_u}{f_g} < 10^{-2/5}
\end{equation}

In other words, the first requirement is a \gfil\ detection, not a narrowband detection as is the case for our narrowband detection criteria.  In addition, the \gfil\ detection is required at a higher significance (5$\sigma$) than the \gfil\ requirements in the narrowband criteria above.  Because we are requiring a \gfil\ detection as the preliminary criterion for these objects, we  re-ran SExtractor, still in dual-image mode, but now with the \gfil\ image as the `detection' image, and either the \gfil, \ufil, or narrowband image as the `measurement' image.  A strong emission line object in the narrowband  should be well detected in the \gfil\ that encompasses the narrowband. The second criterion still requires that an excess of flux be present in the narrowband compared to the g-band, but the minimum magnitude of this excess is lowered, and the significance of the excess is also lowered (from 4$\sigma$ to 2$\sigma$). Essentially, this only requires a rest-frame equivalent width of $\ge$ 6.4 \AA. Finally the \ufil\ flux must still be less than the \gfil\ flux, but the difference need not be as large, given the depth of the \ufil\ band.

While the sample of  LAEs discussed in this paper comes from a compilation of objects selected from multiple data reductions and different selection iterations, we emphasize that each LAE discussed here has been confirmed spectroscopically (as discussed in Section \ref{sec:optspec}).  The compilation of multiple extractions is simply a result of the long-term nature of this project and an interest in improving our selection process and results.  We can state, a posteriori, a broad set of selection criteria that each object in our sample is subject to, by comparing the four selections and combining the least stringent set of criteria from across the four selections.  This leaves us with the four criteria shown below, which all 33 of our confirmed LAEs satisfy:

\begin{equation}
\frac{f_g}{\delta f_g} \ge 2
\end{equation}
\begin{equation}
\frac{f_{NB}}{f_g} \ge 1.445
\end{equation}
\begin{equation}
\frac{f_{NB} - f_g}{\sqrt{\delta f_{NB}^2 + \delta f_{g}^2}} \ge 2
\end{equation}
\begin{equation}
\frac{f_u}{f_g} < 10^{-2/10}
\end{equation}


\begin{table*}
\begin{tabular}{|l|c|c|c|c|c|c|c|}
Object & Flux$_{NB}$ & Flux$_{g}$ & Flux$_{u}$ & EW$^{a}$  & Selection & LUCIFER$^{b}$ & NIRSPEC$^{c}$\\
 &  &  $\mu$Jy & & \AA\ & & seconds & seconds \\
\hline
LAE\_J100049.56+021647.1  &       1.41  $\pm$       0.28  &       0.62   $\pm$     0.052  &       0.16  $\pm$      0.028  &        19.  $\pm$         7. &4 &  & \\
LAE\_J095859.33+014522.0  &       0.71  $\pm$       0.15  &       0.37   $\pm$     0.049  &       0.13  $\pm$      0.047  &        14.  $\pm$         7.  &4&  & \\
LAE\_J100212.99+020137.7  &       1.58  $\pm$       0.23  &       0.29   $\pm$     0.050  &       0.14  $\pm$      0.037  &        78.  $\pm$        23.  &3&  & \\
LAE\_J095929.41+020323.5 (LAE6559)  &       1.53  $\pm$       0.23  &       0.21   $\pm$     0.043  &       0.07  $\pm$      0.043  &       121.  $\pm$        43. &1,2,3 & 1680$^{\ast}$ & 1800$^{\ast}$ \\
LAE\_J095944.02+015618.8  &       1.27  $\pm$       0.20  &       0.21   $\pm$     0.053  &       0.03  $\pm$      0.067  &        89.  $\pm$        35. &3 &  & \\
LAE\_J095930.52+015611.0 (LAE7745)  &       3.42  $\pm$       0.20  &       0.50   $\pm$     0.047  &       0.12  $\pm$      0.045  &       111.  $\pm$        17. &1,3  & 1680  & \\
LAE\_J100217.05+015531.7  &       1.35  $\pm$       0.22  &       0.40   $\pm$     0.062  &       0.21  $\pm$      0.039  &        37.  $\pm$        11.  &3&  & \\
LAE\_J100157.87+021450.0  &       1.34  $\pm$       0.23  &       0.18   $\pm$     0.059  &       0.02  $\pm$      0.079  &       131.  $\pm$        70.  &3&  & \\
LAE\_J100124.36+021920.8 (LAE40844) &       6.78  $\pm$       0.37  &       1.25   $\pm$     0.073  &       0.22  $\pm$      0.074  &        78.  $\pm$         8. &1,2  & 1200 & \\
LAE\_J095847.81+021218.2  &       1.73  $\pm$       0.26  &       0.40   $\pm$     0.019  &       0.09  $\pm$      0.054  &        55.  $\pm$        11.  &1&  & \\
LAE\_J095904.93+015355.4  &       0.99  $\pm$       0.14  &       0.14   $\pm$     0.011  &       0.03  $\pm$      0.069  &       121.  $\pm$        26.  &1&  & \\
LAE\_J095910.90+020631.6 (LAE14310) &       3.45  $\pm$       0.38  &       0.58   $\pm$     0.073  &       0.24  $\pm$      0.072  &        89.  $\pm$        20. &1,2  & 6000$^{\ast}$  & \\
LAE\_J095921.06+022143.4  &       1.13  $\pm$       0.16  &       0.25   $\pm$     0.015  &       0.07  $\pm$      0.052  &        57.  $\pm$        11.   &1&  & \\
LAE\_J095948.47+022420.8  &       1.11  $\pm$       0.15  &       0.16   $\pm$     0.011  &       0.00  $\pm$      0.001  &       114.  $\pm$        22.  &1,2&  & \\
LAE\_J100019.07+022523.9 (LAE27878)  &       1.68  $\pm$       0.25  &       0.24   $\pm$     0.046  &       0.09  $\pm$      0.051  &       118.  $\pm$        40. &1,2  & 4800  & \\
LAE\_J100100.35+022834.7  &       2.48  $\pm$       0.25  &       0.46   $\pm$     0.014  &       0.09  $\pm$      0.066  &        76.  $\pm$        10.  &1&  & \\
LAE\_J100146.04+022949.0  &       0.90  $\pm$       0.14  &       0.16   $\pm$     0.011  &       0.07  $\pm$      0.041  &        79.  $\pm$        17.  &1&  & \\
LAE\_J095843.11+020312.3  &       1.70  $\pm$       0.22  &       0.36   $\pm$     0.014  &       0.13  $\pm$      0.039  &        63.  $\pm$        11.  &1&  & \\
LAE\_J100128.11+015804.7  &       1.36  $\pm$       0.21  &       0.31   $\pm$     0.040  &       0.02  $\pm$      0.125  &        58.  $\pm$        15.  &2&  & \\
LAE\_J100017.84+022506.1 (LAE27910)  &       1.57  $\pm$       0.22  &       0.30   $\pm$     0.042  &       0.09  $\pm$      0.045  &        73.  $\pm$        19. &2  & 1800$^{\ast}$ & \\
LAE\_J095839.92+023531.3  &       1.53  $\pm$       0.25  &       0.43   $\pm$     0.017  &       0.16  $\pm$      0.053  &        40.  $\pm$         9.  &1&  & \\
LAE\_J095838.90+015858.2  &       1.08  $\pm$       0.18  &       0.08   $\pm$     0.009  &       0.03  $\pm$      0.048  &       452.  $\pm$       198.  &1,2&  & \\
LAE\_J100020.70+022927.0  &       1.13  $\pm$       0.22  &       0.18   $\pm$     0.041  &       0.14  $\pm$      0.040  &        98.  $\pm$        39.  &2&  & \\
LAE\_J095812.33+014737.6  &       1.10  $\pm$       0.18  &       0.59   $\pm$     0.048  &       0.22  $\pm$      0.057  &        13.  $\pm$         5.  &4&  & \\
LAE\_J095920.42+013917.1  &       1.10  $\pm$       0.16  &       0.58   $\pm$     0.054  &       0.20  $\pm$      0.056  &        13.  $\pm$         4.  &4&  & \\
LAE\_J095846.72+013706.1  &       1.18  $\pm$       0.18  &       0.24   $\pm$     0.072  &       0.04  $\pm$      0.106  &        66.  $\pm$        29.  &3&  & \\
LAE\_J095923.79+013045.6  &       1.37  $\pm$       0.20  &       0.16   $\pm$     0.061  &       0.02  $\pm$      0.133  &       154.  $\pm$        94.  &3&  & \\
LAE\_J100213.17+013226.8  &       1.19  $\pm$       0.23  &       0.18   $\pm$     0.062  &       0.04  $\pm$      0.087  &       105.  $\pm$        57.  &3&  & \\
LAE\_J095838.94+014107.9  &       1.03  $\pm$       0.17  &       0.21   $\pm$     0.045  &       0.07  $\pm$      0.075  &        69.  $\pm$        25.  &3&  & \\
LAE\_J095834.43+013845.6  &       2.01  $\pm$       0.19  &       0.22   $\pm$     0.046  &       0.09  $\pm$      0.063  &       182.  $\pm$        67.  &3&  & \\
LAE\_J100302.10+022406.7  &       3.87  $\pm$       0.39  &       0.39   $\pm$     0.046  &       0.20  $\pm$      0.052  &       206.  $\pm$        50.  &3&  & \\
LAE\_J100157.45+013556.2  &       2.09  $\pm$       0.19  &       0.39   $\pm$     0.078  &       0.17  $\pm$      0.068  &        75.  $\pm$        22.  &3&  & \\
LAE\_J100152.14+013533.2  &       1.36  $\pm$       0.19  &       0.41   $\pm$     0.077  &       0.14  $\pm$      0.068  &        36.  $\pm$        11.  &3&  & \\
\hline
\multicolumn{8}{l}{\textsuperscript{a}\footnotesize{rest-frame}}\\
\multicolumn{8}{l}{\textsuperscript{b}\footnotesize{Exposure time using LUCIFER on LBT}}\\
\multicolumn{8}{l}{\textsuperscript{c}\footnotesize{Exposure time using NIRSPEC on Keck}}\\
\multicolumn{8}{l}{\textsuperscript{$\ast$}\footnotesize{Non-detection}}\\
\end{tabular}
\caption{SExtractor photometry of confirmed LAEs\label{phottbl} and near-infrared followup details.}
\end{table*}

\subsection{Optical Spectroscopy}\label{sec:optspec}
We obtained optical spectroscopy of our LAE candidates using the Hectospec multi-fiber spectrograph \citep{fab05} at the 6.5m MMT Observatory (a joint facility of the Smithsonian Astrophysical Observatory and the University of Arizona)  in 2009 and 2011. Hectospec has 300 optical fibers, a 1 deg$^2$ field of view, and spectral coverage from 3650 - 9200 \AA.  We used the 270 lines per mm grating for our observations.  This setup has a blaze wavelength of $\sim$ 5200 \AA\ and dispersion of 1.21 \AA\ pixel$^{-1}$.  The resolution of the instrument is $\sim$ 6 \AA.  Optical spectroscopy allows us to the confirm the presence of the \lya\ line in the candidate's spectrum, thereby assuring us the object is indeed an LAE at z $\sim$ 3.1.  We rule out \oii\ emitters at z $\sim$ 0.34 and \oiii\ emitters at z $\sim$ 0 by looking for other optical lines that would be present in such cases.  Also, the presence of high ionization lines in addition to the \lya\ line, such as C\,\textsc{iv} redshifted to $\lambda$ $\sim$ 6350 \AA, also help us distinguish between starforming galaxies at z $\sim$ 3 and objects that are likely AGN. Our initial Hectospec data were obtained on UT 2009 February 16 and 21 and 2009 April 26 and 27 (PI Malhotra). Our reductions for the 2009 data combine 120 minutes of observations per object. Our newest LAE candidates were observed on UT 2011 March 25th and 26th (PI McLinden). Our reductions for the 2011 data combine either 150 or 330 minutes of observations for each object.

\subsubsection{Reduction of Optical Spectra}
We reduced the optical spectra of our LAE candidates observed in 2011 using HSRED, an IDL-based reduction package\footnote{http://astro.princton.edu/~rcool/hsred.}  HSRED is mostly based on SPECROAD, SAO's Hectospec reduction package.  The reduction process bias corrects and flatfields the fibers and removes cosmic rays.  Traces of the 300 fibers are made from the domeflats and a wavelength solution is derived from a HeNeAr arc lamp exposure using a 5th order Legendre polynomial.     Accurate  sky models are determined from dedicated sky fibers included in each observation.  Sky subtracted 1-D spectra are extracted.  The average residual from the wavelength calibration is $\sim$ 0.2 \AA.  Median combined spectra are created by combining multiple observations that have the same instrument/fiber setup.  See \citet{pap06} for more detail on each of these steps.

We chose to flux calibrate our optical spectra outside the reduction pipeline.  We scaled a  G8III spectral type Pickles model \citep{pic98} spectrum to match the \textit{V} $\sim$ 5.36 magnitude of our observed G8III spectral type standard star.  Before scaling, the Pickles model has zero magnitude in Vega magnitudes.  We divided the scaled down Pickles spectrum by the standard star's spectrum in counts to create a sensitivity curve.  We then multiplied each reduced, uncalibrated optical spectrum (in counts) by this sensitivity curve to get a flux calibrated spectrum in \fluxl.  The 33 LAEs discussed in this paper include 18 objects that were observed, reduced and confirmed in 2011.

The LAEs observed with Hectospec in 2009 were previously reduced with the External SPECROAD\footnote{http://iparrizar.mnstate.edu/$\sim$juan/research/ESPECROAD/index.php} pipeline developed by Juan Cabanela, as described previously in \citet{mcl11}.  ESPECROAD  applies bias, dark and flat field corrections and wavelength calibration (using He-Ne-Ar arc lamps).

Our 2009 data were not flux calibrated, but we were able to use LAEs that were observed both in 2011 and 2009 to go back and flux calibrate the 2009 data.  Four objects were observed in both years, and we chose to use the two brightest objects, with the highest signal to noise ratios, to derive a scale factor that would appropriately calibrate the 2009 data.  To derive this scale factor we compared the \lya\ line flux in these two bright LAEs, in the flux calibrated (2011) data and the non-flux calibrated (2009) data.  The line flux in the uncalibrated case is in units of counts $\cdot$ \AA.  The line flux in the calibrated spectra are in units \lf.  The scale factor is then this calibrated line flux divided by the uncalibrated line flux, yielding a constant with units \fluxl\ counts$^{-1}$.  Therefore, when this constant is multiplied by an uncalibrated spectrum with units counts, the result is an appropriately scaled spectrum with units \fluxl.  The constants from the brightest two LAEs, derived as described above, were averaged.  The averaged value was then used to flux calibrate the rest of the 2009 data.  This procedure was used to flux calibrate a total of 15 LAEs from 2009, among our larger sample of 33 confirmed LAEs.  The 1D optical spectra of all 33 confirmed LAEs are shown in Figures \ref{fig:opspec1} -- \ref{fig:opspec3}.

\begin{figure*}
\centering
\includegraphics[width=7in,trim=0cm 8cm 0cm 0cm,clip]{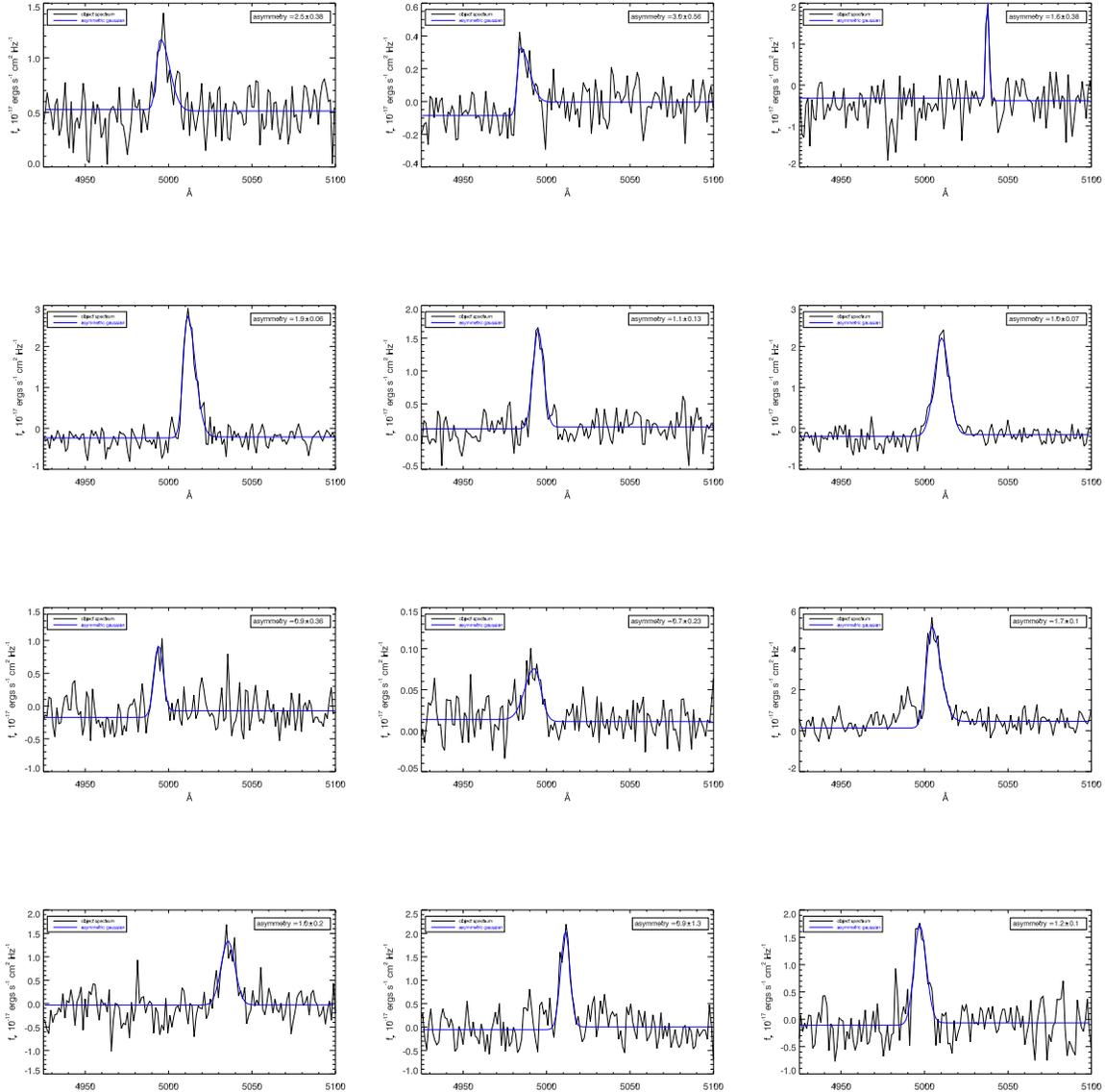}
\caption{1D optical spectra of first 12 confirmed LAEs.  Order of objects matches order of objects in Tables \ref{phottbl}, \ref{sedtbl1}, and \ref{sedtbl2} (reading spectra from top left to bottom right).  Asymmetry of \lya\ line is shown at top left of each panel, observed spectrum is in black, best-fit asymmetric Gaussian is overlaid in blue.}
\label{fig:opspec1}
\end{figure*}

\begin{figure*}
\centering
\includegraphics[width=7in,trim=0cm 8cm 0cm 0cm,clip]{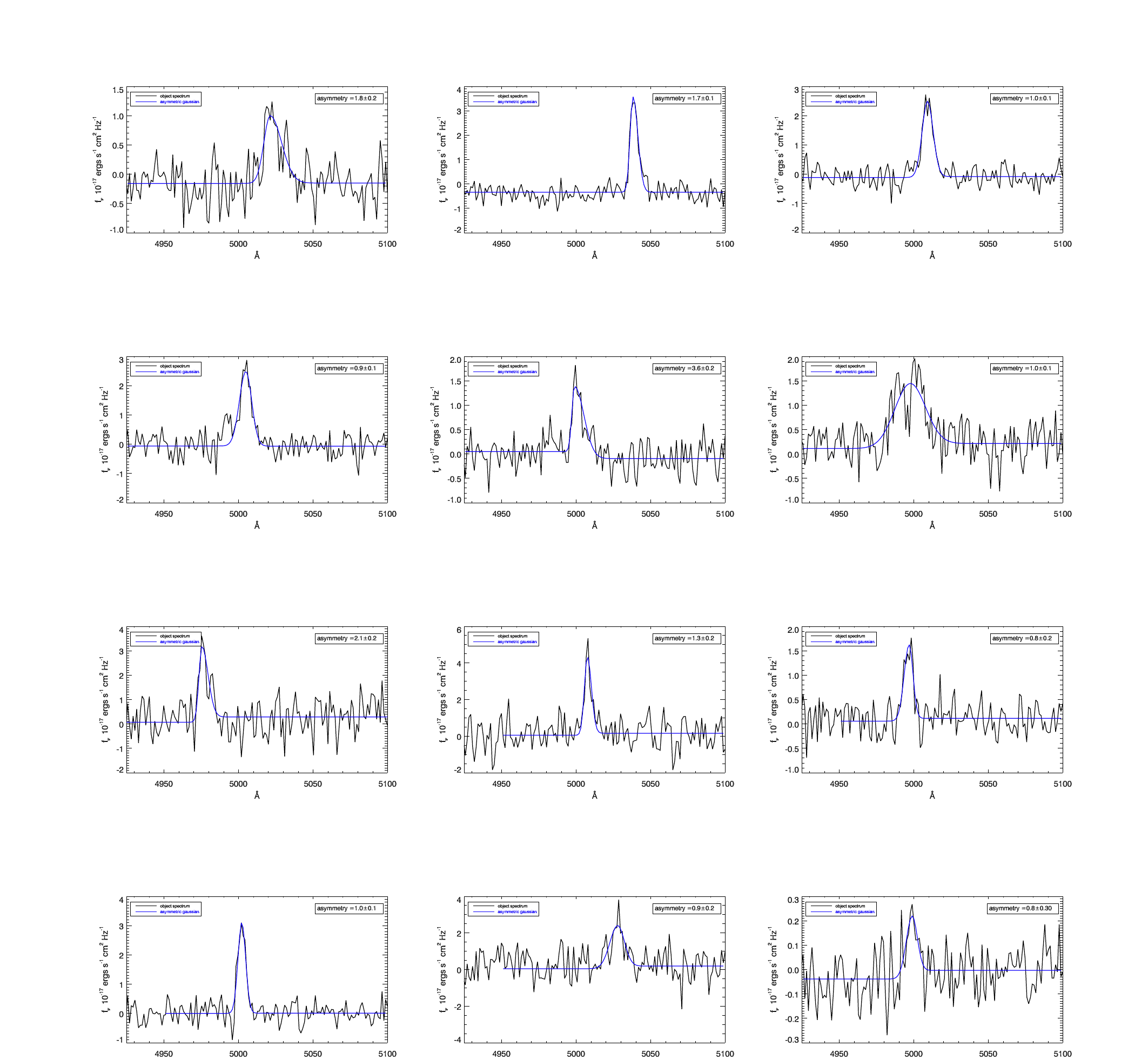}
\caption{Continuation of Figure \ref{fig:opspec1} - 1D optical spectra of next 12 confirmed LAEs.  Order of objects matches order of objects in Tables \ref{phottbl}, \ref{sedtbl1}, and \ref{sedtbl2} (reading spectra from top left to bottom right). Asymmetry of \lya\ line is shown at top left of each panel, observed spectrum is in black, best-fit asymmetric Gaussian is overlaid in blue.}
\label{fig:opspec2}
\end{figure*}

\begin{figure*}
\centering
\includegraphics[width=7in,trim=0cm 12cm 0cm 0cm,clip]{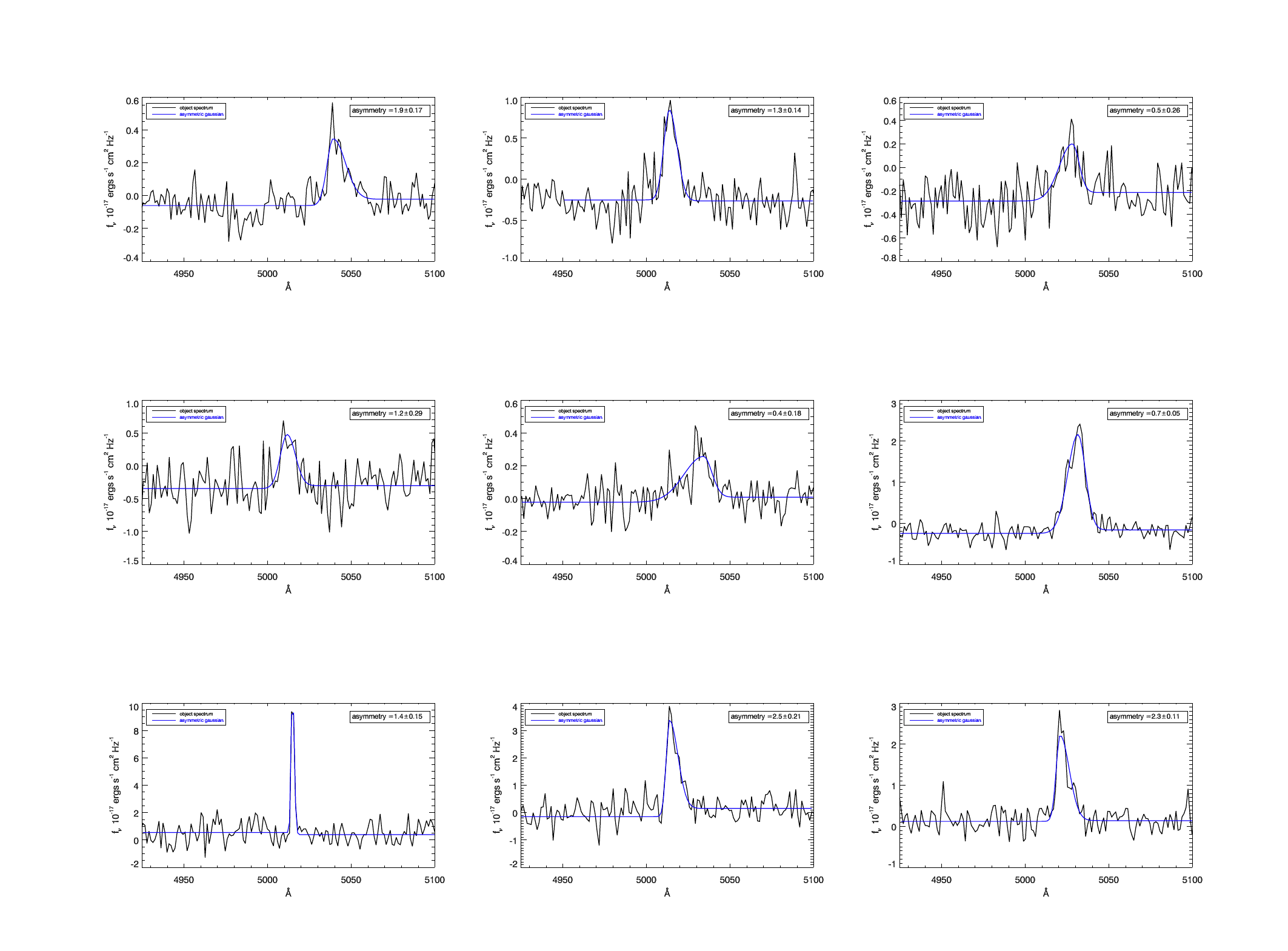}
\caption{Continuation of Figure \ref{fig:opspec1} and Figure \ref{fig:opspec2} - 1D optical spectra of final 9 confirmed LAEs.  Order of objects matches order of objects in Tables \ref{phottbl}, \ref{sedtbl1}, and \ref{sedtbl2} (reading spectra from top left to bottom right). Asymmetry of \lya\ line is shown at top left of each panel, observed spectrum is in black, best-fit asymmetric Gaussian is overlaid in blue.}
\label{fig:opspec3}
\end{figure*}

\subsection{Construction of the Final Sample}\label{sec:final}
Combining the object selection methods and  spectroscopic confirmations discussed above, we have a sample of 33 LAEs.  The photometry (from SExtractor) for these 33 confirmed LAEs is shown in Table \ref{phottbl}. This total does not include two \lya-emitting objects (LAE25972, LAE42795) that were removed because they are likely AGN (see Section\ref{sec:agn}). The AGN are excluded from discussion of our SED fitting results (Section \ref{sec:sed}) as their physical characteristics cannot be derived from comparison to star-forming SED models. We note that eight of our 33 LAEs have poor agreement between measured spectroscopic and photometric \lya\ line fluxes; 
they are not removed from the sample but are labeled as such later in this paper (Tables \ref{sedtbl1} and \ref{sedtbl2}).  An additional five of these 33 LAEs  have possible multiple components and/or morphology indicative of possible interacting sources \citep{mal12}.  This was determined by finding objects that had multiple matches within 2\arcsec\ in the COSMOS ACS Catalog \citep{leau07}.  We confirmed the multi-component morphology with visual inspection of the corresponding HST ACS F814W images \citep{koe07,mas10}.  Note that fitting SED models to photometry that may be from multiple sources can certainly affect what characteristics are derived from SED fitting results.  These five objects are also labeled in Tables \ref{sedtbl1} and \ref{sedtbl2}.

\subsection{New NIR Spectroscopy}
We observed five z $\sim$ 3.1 \lya\ emitting objects from our sample of 33 LAEs in the near-infrared (NIR).  These observations are in addition to the three LAEs previously observed in the NIR with LUCIFER, as detailed in McLinden et al (2011, henceforth Mc11).  We made our NIR observations using LUCIFER (LBT NIR Spectrograph Utility with Camera and Integral-Field Unit for Extragalactic Research) on the 8.4m LBT \citep{sei03,ag10} and using NIRSPEC on the 10m Keck~II telescope \citep{mc98}.     The previously observed LAEs in Mc11 were LAE40844, LAE27878, and LAE14310.  LAE40844 and LAE27878 yielded detections of the  \oiii\ line.  Of the five new observations, two yielded \oiii\ detections, but one of these \oiii-detected objects was among the objects removed as likely an AGN (see Section \ref{sec:agn}).  The other new detection, henceforth  LAE7745, appears to be a typical star-forming LAE, and will be discussed in more detail below.  No emission lines were detected in the other three observed objects, henceforth  LAE25972, LAE6559 and LAE27910.  

\subsection{New LUICFER Data}
We used the longslit mode of LUCIFER for two of our new NIR observations in the same manner as our previous LUCIFER observations \citep{mcl11} - with a 1$\arcsec$ slit utilizing the \textit{H}+\textit{K} grating with 200 lines/mm and the N1.8 camera.   The image scale of the N1.8 camera is 0.25$\arcsec$/pixel.   LAE25972 was observed over ten 120-second frames.  LAE7745 and LAE6559  were observed over seven 240-second frames. The longslit was oriented usch that each LAE shared the longslit with a bright (R $|sim$ 12 -- 18) continuum object.

\subsubsection{2D Reduction of NIR LUCIFER Spectra}
We reduced the 2-D LUCIFER spectra using  NIRSPEC\_REDUCE, a package of IDL scripts written by \citet{bec06}.   NIRSPEC\_REDUCE follows the methodology of \citet{kel03} for optimal sky subtraction.   In this
technique the sky subtraction is performed by sub-sampling the raw (distortion uncorrected) spectra thereby improving
the sky-subtraction significantly.  
We customized the scripts to accommodate LUCIFER data.  The first three scripts in the reduction process, NIRSPEC\_SLITGRID, NIRSPEC\_WAVEGRID and NIRSPEC\_FLATFIXER  were all modified to deal with LUCIFER's 2048 x 2048 pixel array as opposed to NIRSPEC's 1024 x 1024 pixel array.  NIRSPEC\_SLITGRID transforms x and y-coordinates to coordinates of slit position and NIRSPEC\_WAVEGRID transforms x and y-coordinates to coordinates of uncalibrated wavelength.  NIRSPEC\_FLATFIXER creates a median combined normalized flat and a separate file containing the variance in the median combined normalized flat.  We used three five-second Halo2 flats for each reduction.  

The final script in the process, the one that actually performs the sky subtraction, LONGSLIT\_REDUCE, was not directly modified.  Parameters for a specific instrument can be supplied to this script via an external LONGSLIT\_REDUCE.inc file.  Therefore, appropriate values for LUCIFER for information such as array size, gain, slit width, observatory location etc. can be easily supplied without modifying the actual script.  As noted in the README file supplied with the NIRSPEC\_REDUCE package, to subtract an accurate sky model this program processes a raw frame, locates and masks objects, iteratively fits the sky in a single frame to get sky levels and iteratively fits the sky in a differenced frame and then subtracts the fit.  The program can also provide wavelength calibration and extract a 1-D spectrum but we only used this package to produce reduced sky-subtracted 2-D frames.

Finally, individual frames for each object, output from NIRSPEC\_REDUCE,  were median combined with IRAF task IMCOMBINE.  Nods along the slit were removed by providing integer pixel offsets in the spatial direction using the `offsets' parameter in IMCOMBINE to bring all the frames to the  position of the first frame.   For  LAE25972, ten 120-second frames were median combined.  For LAE7745 and LAE6559, the seven 240-second frames were median combined. 

\begin{figure*}
\centering
\begin{tabular}{cc}
\multicolumn{2}{c}{\includegraphics[bb=100 50 404 310,scale=0.6]{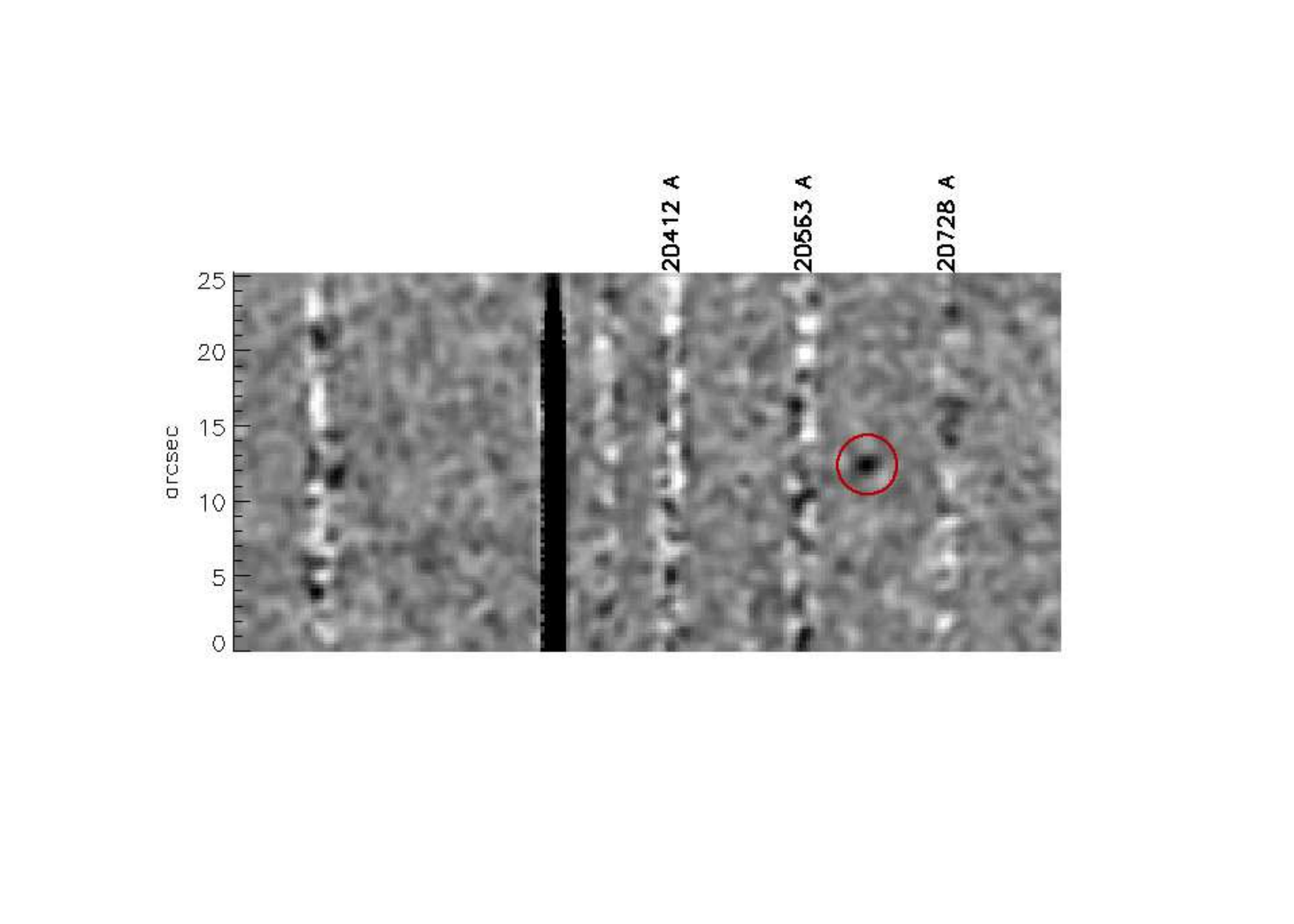} }\\
\includegraphics[bb=100 50 404 310,scale=0.6]{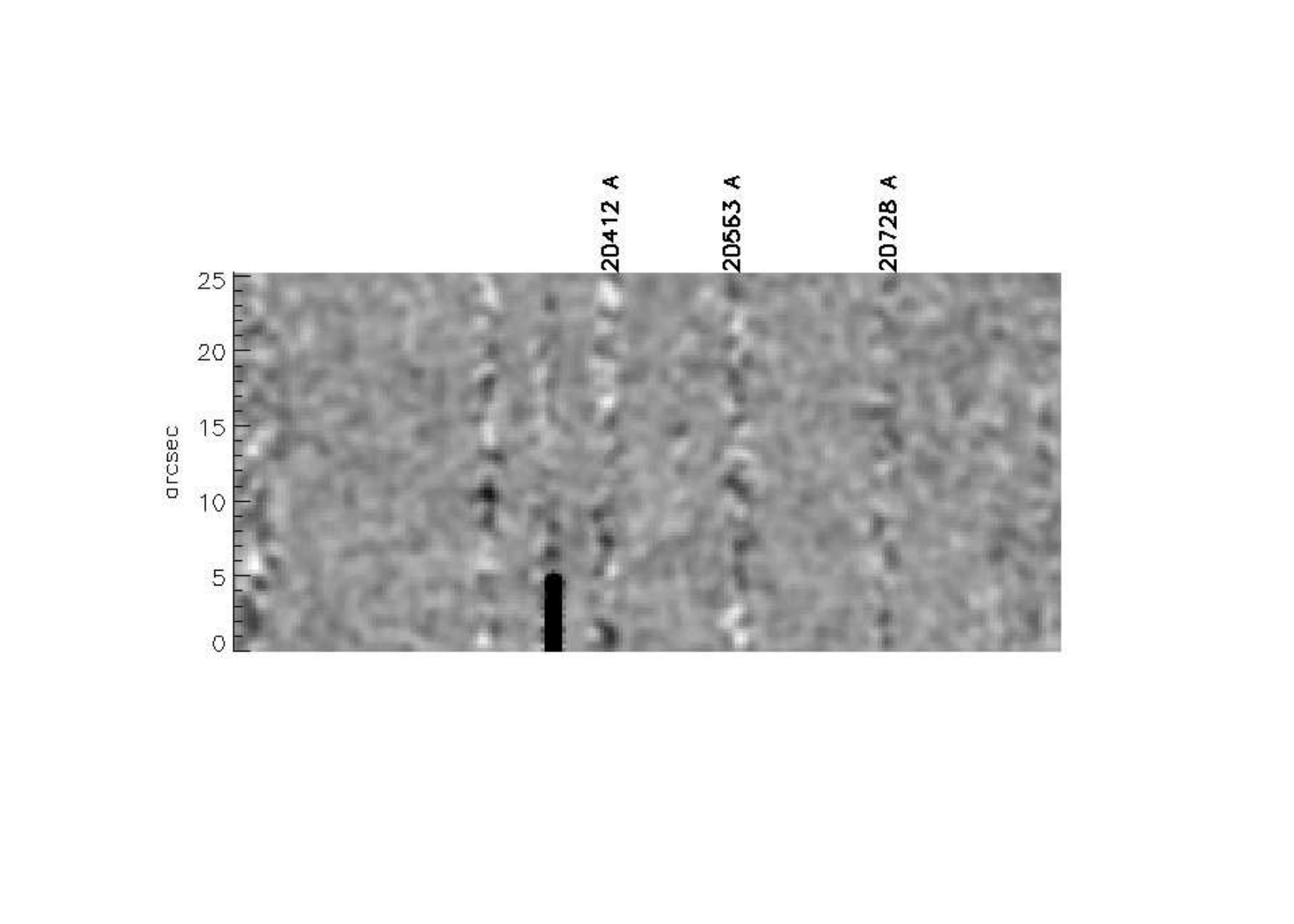} & \includegraphics[bb=100 50 404 310,scale=0.6]{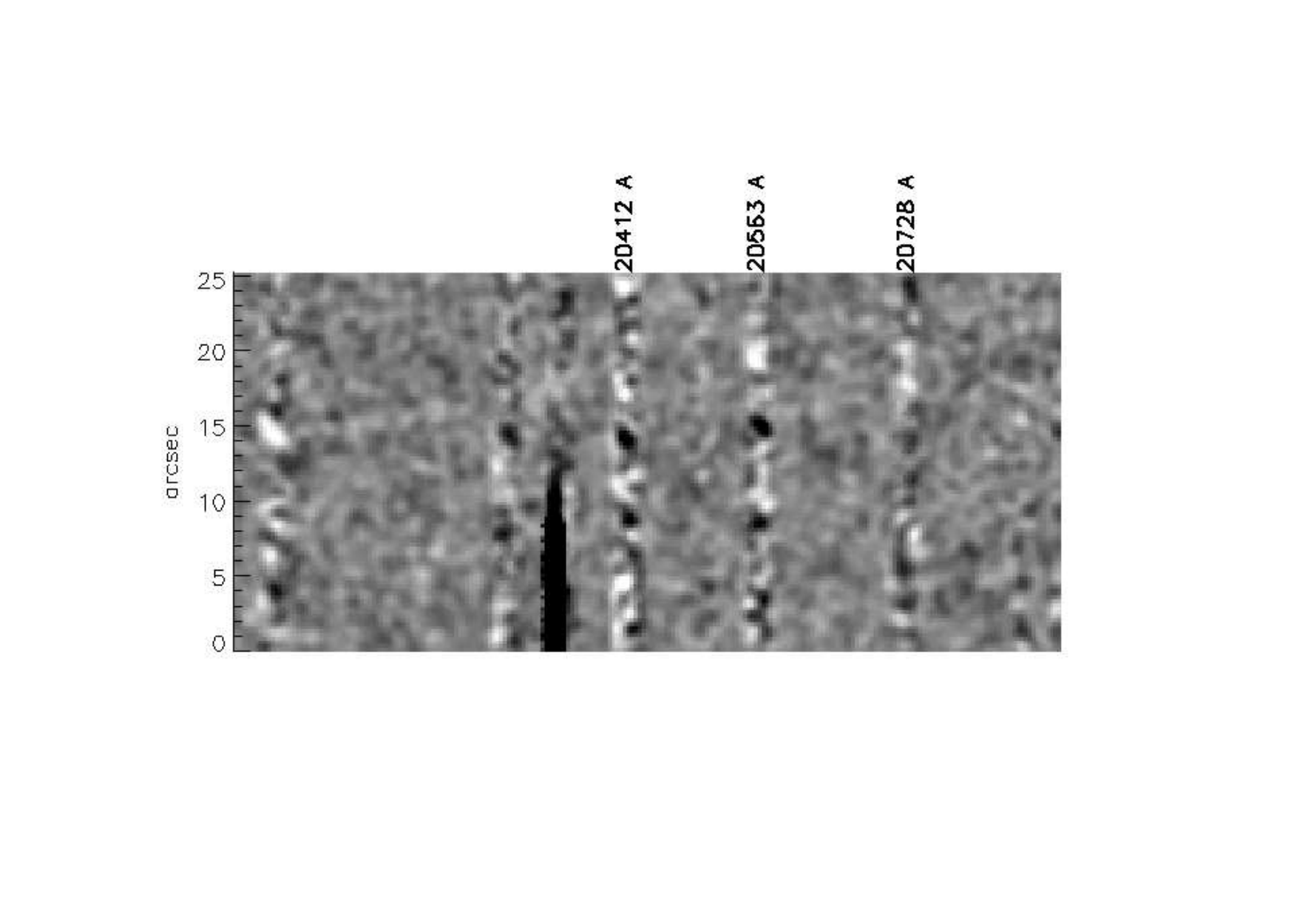} \\
\end{tabular}
\caption{2D NIR LUCIFER spectra of LAE7745 (top), LAE25972 (bottom left), LAE6559 (bottom right).  Images have smoothed with 3-pixel Gaussian kernel.  \oiii\ detection (5008.24 \AA) in LAE7745 highlighted in red circle.  A bad column in the detector is seen just left of center in each frame.}
\label{fig:twod}
\end{figure*}

Only one object, LAE7745, shows a detection in the reduced 2-D (Figure \ref{fig:twod}).  The detection corresponds to the expected spatial-direction location of the LAE based on its distance from the bright continuum object that shared the slit.  The detection also corresponds to the approximate expected dispersion-direction location of an \oiii\ detection based on the \lya\ redshift of z $\sim$ 3.1. Given that this detection appears at both the expected spatial and dispersion locations  gives strong credibility to this being a real detection of \oiii\ and not an errant cosmic ray.  In addition, while the detection can't be seen in a single exposure, it can be seen faintly when a single exposure is subtracted from a nodded subsequent exposure.  The other two objects show no detections and are shown in the bottom panel of Figure \ref{fig:twod}.  Possible reasons for non-detections are insufficient integration time for faint lines and emission lines located under OH skylines.  We argue in Section \ref{sec:comp} that insufficient integration time is a likely culprit for these two non-detections.

\subsubsection{1D Reduction of LUCIFER Spectra}
The 1D spectra were created following  a similar reduction process to that outlined in Mc11. We utilized the DOSLIT routine in IRAF \citep{val93}.  Because a bright continuum source  shared the slit with each LAE we were able to create a trace for extraction from the bright object.  The trace was then shifted along the spatial axis to extract the LAE spectrum, whose continuum emission is undetectably faint in individual exposures and therefore cannot be traced.  DOSLIT was performed on the median combined, sky subtracted 2D spectra from NIRSPEC\_REDUCE.  Wavelength calibration was done using night sky OH lines.  The average RMS uncertainty from wavelength calibration for LAE7745 $\sim$ 0.66 \AA.  Residual bright night sky lines were interpolated over using the SKYINTERP task from the WMKONSPEC package originally designed for Keck NIRSPEC reduction\footnote{http://www2.keck.hawaii.edu/inst/nirspec/wmkonspec.html}.  

Flux calibration proceeded, as in MC11, using the bright continuum sources that shared the slit with our LAEs. LAE7745 was calibrated using SDSS J095930.35+015646.6.  We flux calibrated the spectrum of the bright continuum star spectrum using an appropriate Pickles model spectrum \citep{pic98}, scaled in flux to match the object's apparent \textit{V} magnitude of the object that shared the slit.  
We determined the appropriate Pickles model spectrum to use by determining the star's spectral type from SDSS \textit{u-g} and \textit{g-r} colors \citep{fuk11}.  The SDSS \textit{u}, \textit{g} and \textit{r} magnitudes for this determination come from SDSS data release seven.  The \textit{V} magnitude of the observed of the calibration star was determined from its SDSS colors and the Lupton 2005 color transformation from SDSS \textit{g-r} color to \textit{V} magnitude\footnote{http://www.sdss.org/dr6/algorithms/sdssUBVRITransform.html}.   The sensitivity curve for calibration comes from dividing the scaled-down Pickles model by the bright continuum star's stellar spectrum in counts.  The raw LAE spectrum was multiplied by the sensitivity curve to produce a final flux-calibrated NIR LAE spectrum.  This method ought to account for slit losses automatically, provided that slit losses are the same for both the on-slit continuum source and the LAE.

\subsection{NIRSPEC Data and Reduction}
Two of our five additional NIR observations were made at the Keck~II telescope using NIRSPEC. Observations were made on 30 January 2010 and 1 February 2010.  We used the 42x0.76 arcsecond slit and the low-resolution mode of NIRSPEC for these observations.  For LAE42795 we obtained nine 360-second frames of \textit{K} band spectroscopy, using the blocking filter NIRSPEC-7 and seven 600-second frames of \textit{H} band
spectroscopy, using the blocking filter NIRSPEC-5.  The \textit{K} band spectra show a very broad \oiii\ emission line.  In Section \ref{sec:agn} we discuss our interpretation of this broad line as evidence of AGN activity.  For LAE27910 we obtained five 360-second frames of \textit{K} band spectroscopy using the NIRSPEC-7 filter. We saw no evidence of \oiii\ or any other optical emission lines in LAE27910.  In addition, LAE6559 was observed with  NIRSPEC in addition to LUCIFER, but yielded no detections with either instrument.  The Keck observations for LAE6559 consisted of five 360-second frames of \textit{K} band spectroscopy using the NIRSPEC-7 filter.

We reduced the NIRSPEC data again using the NIRSPEC reduction package \citep{bec06}.
For this reduction, the spectra were first  flat-fielded, and then corrected for dark current using a constant
value. The sky was then subtracted again using the optimal sky-subtraction technique of \citet{kel03}.

In order to correct for the distortion in both the x and y-direction,  we use the IRAF tasks XDISTCOR and YDISTCOR in the
WMKONSPEC package specifically developed for the NIRSPEC data reduction.  All pixels affected by cosmic rays 
are identified using the IRAF task CRMEDIAN, and these affected pixels are replaced by average counts calculated from
 neighboring pixels. We then average-combined each individual spectra using IRAF task imcombine, for each of the sources.
We detected no optical emission lines in any object except the likely AGN (LAE42795).

\section{RESULTS FROM OPTICAL AND NIR SPECTROSCOPY}\label{sec:res1}

\subsection{\lya\ Line Fluxes and Asymmetries}\label{sec:lyafit}
\begin{figure*}
\centering
\includegraphics[scale=0.45]{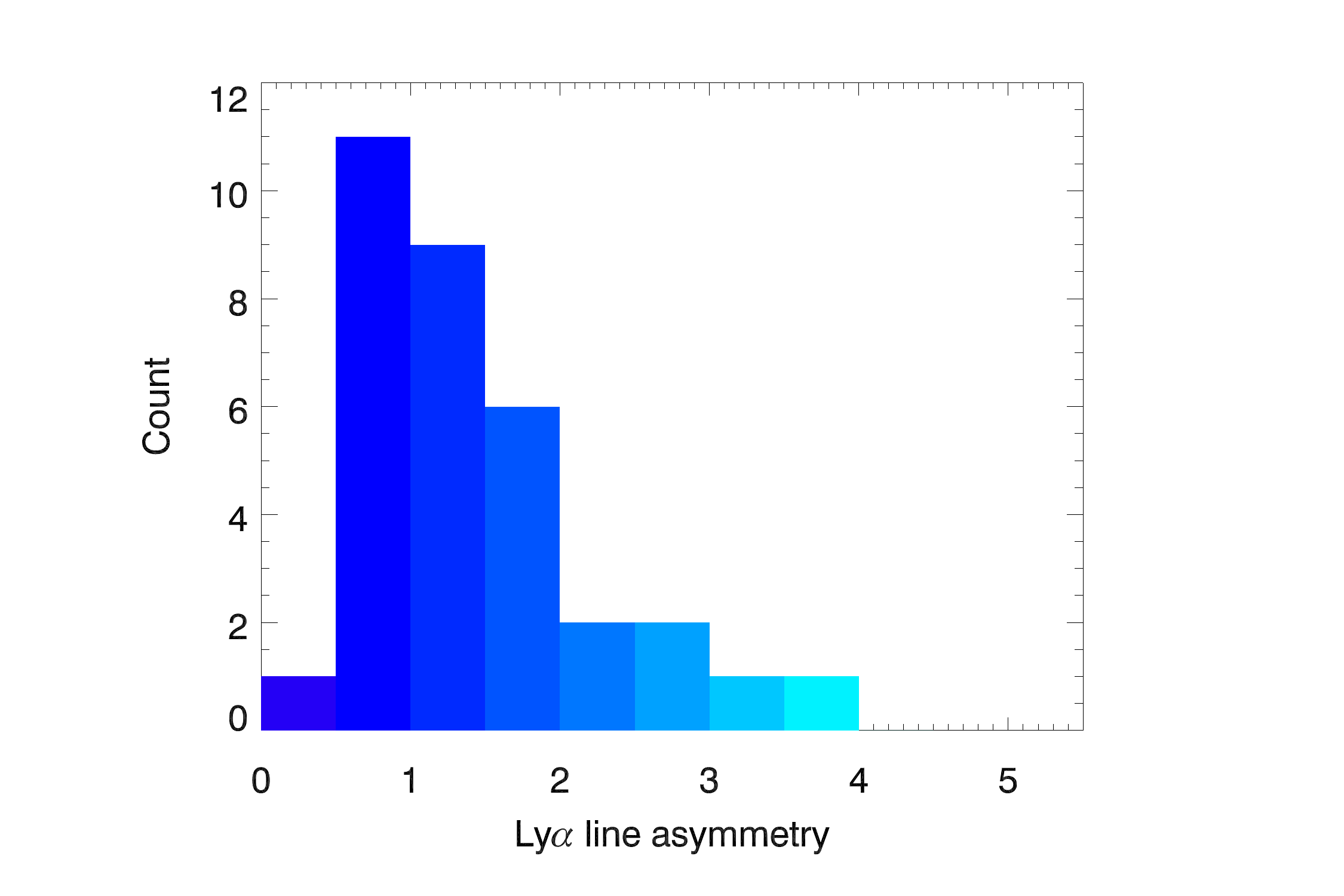}
\caption{Histogram of \lya\ line asymmetry in sample of 33 z $\sim$ 3.1 LAEs.}
\label{fig:lasym}
\end{figure*}

We are able to measure \lya\ line fluxes in our sample by fitting an asymmetric Gaussian to each line detected in our optical spectroscopy data.  A more detailed description of this process is found in Mc11.  To summarize, each \lya\ line is fit with an asymmetric Gaussian using a modified version of the ARM\_ASYMGAUSSFIT IDL routine developed by Andrew Marble\footnote{ http://hubble.as.arizona.edu/idl/arm/}.  The purpose of using a fitting routine that allows for, but does not require, an asymmetric solution is that it allows the red and blue sides of the \lya\ line to be fit with different sigmas.  In cases where the red-wing of the \lya\ is elongated and/or the blueside of the line is sharply truncated, this asymmetric fitting procedure will find a good fit that captures these characteristics.  Asymmetric line profiles are observed for high-z LAEs \citep{rho03,daw04,kash06} because the blue side of the line will be preferentially absorbed by intervening neutral hydrogen.  In addition, it has been shown that asymmetric \lya\ lines can also be produced by \lya\ radiative transfer through expanding shells, a model meant to represent outflows from star-bursting galaxies \citep[e.g.,][]{ver06,ver08}. 
The \lya\ line flux is determined from the area under the asymmetric Gaussian.  The average \lya\ line flux of our entire confirmed sample is 17.4 $\pm$ 0.9 $\times 10^{-17}$ \lf.  We quantify the asymmetry of the fitted \lya\ lines as a$_{rb}$, which comes directly from our asymmetric fitting process; where a$_{rb}$ is the ratio of the red side best-fit sigma to the blue side best-fit sigma, or a$_{rb}= \sigma_{red} / \sigma_{blue}$. From this definition, when a$_{rb}$ is $>$ 1.0, the line is considered asymmetric in the expected direction for \lya, i.e. with a larger red-side sigma.  When a$_{rb}$ is $<$ 1.0 the line is also asymmetric but with a larger blue-side sigma, and when a$_{rb}$ $=$ 1 the line is symmetric.   The average asymmetry, using this measure, of our entire confirmed sample of LAEs is 1.4 $\pm$ 0.2, indicating that, as a whole, our sample of LAEs does have asymmetric \lya\ lines. A histogram of \lya\ line asymmetries is shown in Figure \ref{fig:lasym}. 

\subsection{New \oiii\ Detection}\label{sec:detoiii}
As mentioned above, this paper presents one new \oiii\ detection in a z $\sim$ 3.1 LAE, excluding an \oiii\ detection in a likely AGN.  Our new measurement was made in the same manner as the \oiii\ line flux measurements in Mc11.  Namely, we fit the \oiii\ line with a symmetric Gaussian plus constant, using the IDL routine MPFITEXPR.  The area under the best-fit Gaussian determines the line flux of the \oiii\ line, the central wavelength of the fit determines the the systemic redshift of the galaxy, and the constant term is the continuum level.  We report an error on these measured quantities determined from 1000 Monte-Carlo simulations.  In these simulations, the actual 1D spectrum was modified at each point by a Gaussian random amount proportional to the error at that point, and then a Gaussian was fit to this modified data in each simulation.  The standard deviation of the 1000 iterations for  each quantity represents 1$\sigma$.  The errors we report are three times this.   Following this procedure, LAE7745 has an \oiii\ line flux of 13.7 $\pm$ 1.8 $\times 10^{-17}$ \lf.  This is in addition to the our two previous detections reported in Mc11, where line fluxes of 7.0 $\pm$ 0.3 $\times 10^{-17}$ and 35.5 $\pm$ 1.2 $\times 10^{-17}$ \lf\ were reported for LAE27878 and LAE40844, respectively.  The other characteristics of the best-fit Gaussian for LAE7745 are a central wavelength of  20636.7 $\pm$ 1.3 \AA, FWHM = 19.7 $\pm$ 0.12 \AA, and a constant term consistent with zero (0.9 $\pm$ 1.3 $\times$ 10$^{-18}$ \fluxl).  

We do not detect the second \oiii\ line at rest-frame 4960 \AA\ in LAE7745.  We placed an 3$\sigma$ upper limit on this line by adding a mock Gaussian emission line to the spectra to represent the 4960 \AA\ line and testing the level to which we could recover it, a procedure similar to that in Finkelstein (2011b). The sigma of the Gaussian was fixed to 12.7 \AA\ which is the best fit sigma from the 5008.24 \AA\ line.  The line center was fixed using the redshift of the 5008.24 \AA\ line as well.  We then measured the mock line by fitting a symmetric Gaussian using MPFITEXPR, as we would for an actual \oiii\ detection.  The noise on the measurement was determined from 1000 Monte Carlo iterations, where the flux was modified each time by a random amount proportional to the error bars.  We repeated this measurement with decreasing line fluxes until the signal to noise ratio (SNR) dropped below 5$\sigma$.  The line flux where the 5$\sigma$ threshold was crossed became our 5$\sigma$ value from which we were able to determine $\sigma$ and therefore a 3$\sigma$ line flux detection limit.  From this procedure we determine a 3$\sigma$ upper limit for this line of 6.9 $\times$ 10$^{-18}$ \lf.

\subsection{AGN in the Sample}\label{sec:agn}
 LAE25972 was not well fit with any of our star-forming SED models, leading to consideration that this object may instead be a \lya-selected AGN, especially since this object also had the largest \lya\ line flux in our sample. This object does not have an X-ray counterpart in the Chandra COSMOS Survey Point Source Catalog \citep{elv10} but the catalog may be too shallow to rule out a faint X-ray counterpart (limiting depth $=$ 5.7 $\times$ 10$^{-16}$ \lf, corresponding to X-ray luminosity of 4.9 $\times$ 10$^{43}$ ergs s$^{-1}$ at z $=$ 3.1, assuming a X-ray photon index $\Gamma$ = 2.0).  This object does, however, have a number of other AGN signatures based on the strength of its \lya\ line. For example, the \lya\ line flux of 7.8 $\times$ 10$^{-16}$ \lf\ corresponds to a \lya\ luminosity of 6.75 $\times$ 10$^{43}$ erg s$^{-1}$.  This \lya\ luminosity is larger than five of the six \lya-selected AGN at z $=$ 3.1 -- 3.7 discussed in \citet{ouch08}.  A comparison to \citet{zh10} yields a similar conclusion - namely \citet{zh10} found that all \lya\ detected objects with \lya\ luminosity $\ge$ 1.8 $\times$ 10$^{43}$ were AGN.  They investigated seven \lya-selected AGN from z $=$ 3.1 -- 4.5 to reach this conclusion.  Given the diagnostics from \citet{ouch08,zh10} we conclude that this object is likely an AGN. 

LAE42795 also does not have an  X-ray counterpart in the Chandra COSMOS Survey Point Source Catalog, but it does have a very strong, broad \oiii\ detection.   We interpret this as strong evidence for AGN activity in this object.  This interpretation is supported by a possible detection of the C\,\textsc{IV} 1549\AA\ line in the MMT optical spectrum at $\sim$ 6415.4 \AA\ which agrees with the \lya\ and \oiii-derived redshifts for this object.  Finally, the \lya\ line in this object is also broad.  The red side of the best-fit asymmetric Gaussian has $\sigma$ = 11.3 \AA, which would correspond to a FWHM of $\sim$ 1585 \kms.  The left side of the best-fit asymmetric Gaussian has $\sigma$ = 5.4 \AA\ which would correspond to a FWHM $\sim$ 756 \kms.  If we take the average of these FWHM values as the appropriate FWHM for the asymmetric Gaussian then velocity-width of this \lya\ line is $\sim$ 1170 \kms.

We exclude both of these likely AGN from our SED fitting results below, and they are excluded anywhere average characteristics of the LAEs are reported, so that these averages only reflect the characteristics of (33) typical star-forming LAEs in our sample.  

\subsection{\lya\ - \oiii\ Velocity Offsets}
Using the new \oiii\ detection, we are also able to determine a velocity offset between the \lya\ and \oiii\ lines as we did in Mc11.  The \oiii\ line defines the systemic velocity of the galaxy, and the \lya\ line, subject to resonant scattering from neutral hydrogen both in the galaxy and in the IGM as well as dust attenuation, is shifted red-ward.  We find a velocity offset between \oiii\ and \lya\ in LAE7745 of 52 $\pm$ 25.2 \kms, after correction for the Earth's motion. We follow the same procedure we reported previously in Mc11 to make this new measurement; the velocity offset is determined based by comparing the central wavelength of \oiii\ and \lya\ - where the central wavelength is determined by the best-fit asymmetric (for \lya) and symmetric (for \oiii) Gaussians.  The offset between the \oiii\ and \lya\ lines is illustrated in Figure \ref{fig:offset} below, mirroring the plots are shown in Figure 2 in Mc11.   The measurement reported here is in addition to the velocity offsets of  125 $\pm$ 17.3 and 342 $\pm$ 18.3 \kms\ we previously reported for LAE27878 and LAE40844, respectively, making the new measurement the smallest velocity offset we have seen. This result is suggestive of a wide distribution of velocity offsets in LAEs at z $\sim$ 3.1 - suggesting there is not a single characteristic velocity offset but rather a distribution.  This diversity of observed velocity offsets is supported by the velocity offsets presented in other samples of z $\sim$ 2-3 galaxies where velocity offsets between \lya\ and rest-frame optical emission lines are derived (e.g. Finkelstein et al. 2011a, Kulas et al., 2012, Chonis et al. 2013).  This result is not surprising considering the generally diverse physical characteristics (age, mass, star formation history etc.) of the sample that we find from SED fitting in section \ref{sec:sed}.  If these observed velocity offsets are due to star burst driven winds, one would expect galaxies with  diverse characteristics to drive different winds.  We also note that this result is still consistent with the various models \citep[e.g.][]{ver06, ver08, stei10} discussed in Mc11 as possible matches to our observations.  

While detections of \oiii\ in the NIR for high-z LAEs are still fairly novel, making this an exciting result and one that shed light on the kinematics of LAEs and \lya\ escape, we would like to have a better ability to predict which LAEs will yield \oiii\ detections, since only three of our six observations have yielded \oiii\ detections so far (excluding the two AGN).  A new approach that may help tackle this challenge is discussed in Section \ref{sec:sed} below.  In addition, more detections are needed to really understand the full distribution of these velocity offsets and how they correlate with other characteristics of LAEs.  This distribution is something we cannot characterize yet with only small samples currently available, but this is becoming more approachable as more NIR instruments come on-line, particularly those with multi-object capabilities.

\begin{figure*}
\includegraphics[scale=0.65]{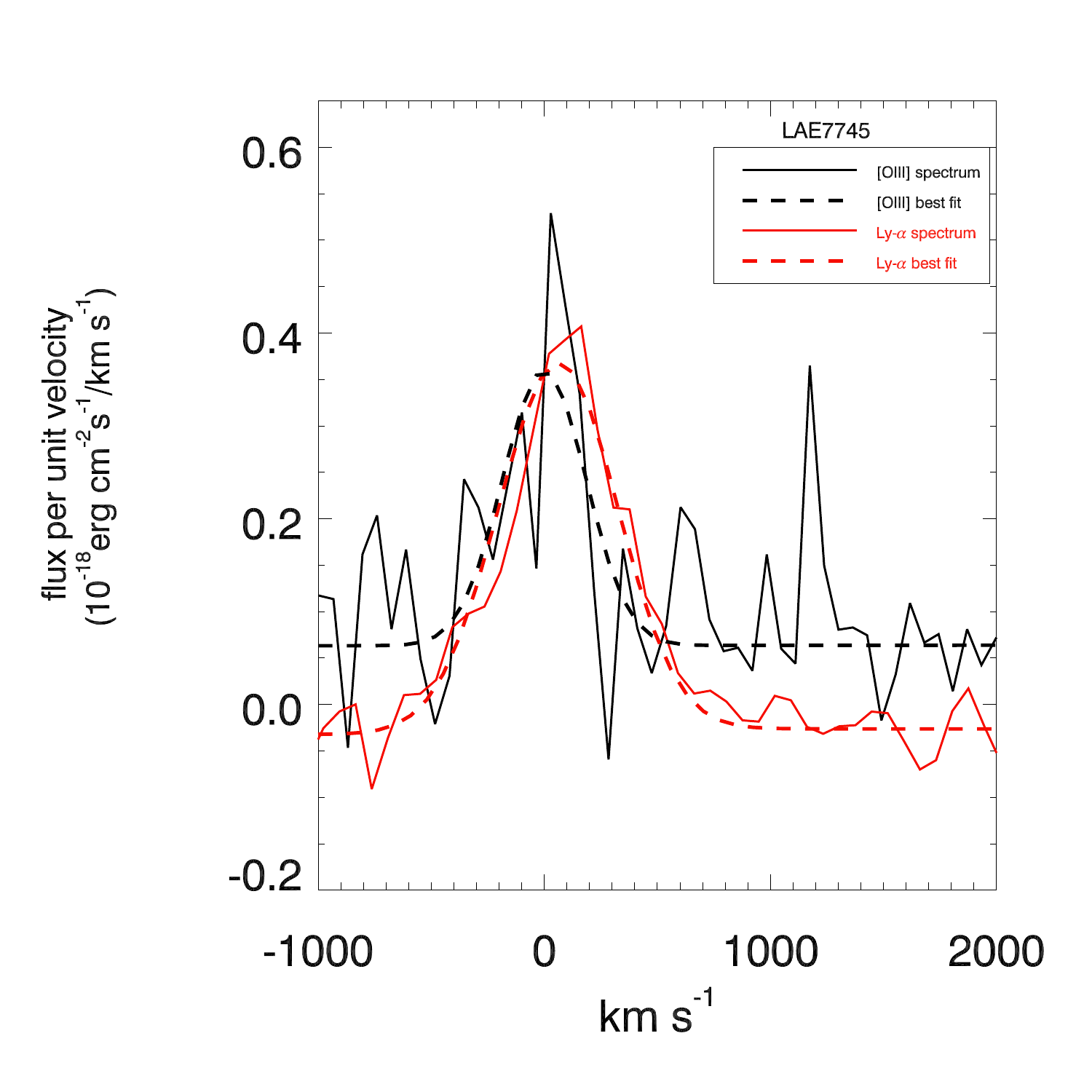}
\caption{ Velocity offset between \oiii\ and \lya\ as detected in one new LAE observed in 2011 with Hectospec and LUCIFER.  \oiii\ spectrum is in black, \lya\ spectrum is in red.  \lya\ line is offset from \oiii\ by 52 \kms.  See Mc11, Figure 2, for two previously observed LAEs with velocity offsets of 342 and 125 \kms.}\label{fig:offset}
\end{figure*}

\section{CONSTRAINING PHYSICAL PARAMETERS WITH SED FITTING}\label{sec:sed}
To constrain the physical properties of our 33 (non-AGN) LAEs we generated stellar population model spectra  produced using the updated models of  Bruzual \& Charlot (2003) which will henceforth be referred to as CB11.  This latest version includes contributions from TP-AGB stars and allows for exponentially increasing star formation histories.   We use a Salpeter IMF.  We created model spectra with an extensive grid of ages, metallicities, star formation histories, and dust extinction values.  We also present an additional fitted parameter, a line flux contribution to the \kfil\ band, which is discussed in more detail below.  For z $\sim$ 3.1 LAEs, the 5008.240 \AA\ \oiii\ line is redshifted into the \kfil\ filter so this is the line to which we assign the additional flux.   We report a single line flux for the \oiii\ line, but one can consider that this emission is really split between the two lines in the 4960.295/5008.240 \AA\ \oiii\ doublet (with a ratio of $\sim$ 1 to 3 in the 4960.295 and 5008.240 lines, respectively).  Technically, the 4862.683 \AA\ H$\beta$ line also falls in the \kfil\ filter for a z $\sim$ 3.1 galaxy, and the 3727.092/3729.875 \AA\ \oii\ lines could fall in the \textit{H} filter. However, our LUCIFER observations have covered the full \textit{H} and \kfil\ wavelength range and we have only detected \oiii.   Hence, for this work we attribute all the additional line flux in the \kfil\ band to \oiii.  We note, however, that our method does not rule out that this emission comes from multiple lines and it could be divided among \oiii\ and H$\beta$ a posteriori.  Because we have not yet detected \oii\ we do not alter the \textit{H} band flux for this line, but this could easily be added to future analyses if future observations indicate that it is warranted.  It is also worth noting that most SED results to date have indicated LAEs are relatively metal poor \citep[e.g.][]{fink11a,nak13} and hence we can expect the \oiii\ line to be much brighter and contribute much more to the broadband flux than the \oii\ lines.

Ages for our models vary on an irregular grid of 48 values, from 2 Myr to 2 Gyrs (approximately the age of the universe at z = 3.1).  Dust extinction, \textit{E(B-V)}, is allowed to assume 31 regular values to produce 0-6.6 magnitudes of dust extinction (A$_{1200}$).  Dust attenuation is applied to our models using the Calzetti formulation (2000).  Metallicity is allowed to assume 5 values from 0.005 - 1.0 \zsol.  We chose only exponential star formation rates, investigating both exponentially increasing and exponentially decreasing rates.  Star formation history e-folding time, $\tau$, can assume 6  positive values from $\tau$ = 0.0001 - 4.0 Gyr.  This essentially creates one template of instantaneous star formation (when $\tau$ = 0.0001 Gyr which is much younger than the age of the O/B star) and one template with continuous star formation (when $\tau$ = 4 Gyr which is longer than the age of the universe at z=3.1) with four templates of exponentially decaying star formation in between ($\tau$ = 0.001, 0.01, 0.1, 1 Gyr).  We add to this two negative e-folding times ($\tau$ =  -0.1, -1 Gyr) to explore exponentially increasing models due to recent results  \citep{mar10,fin11,pap11} that have indicated high-z LAEs may be better fit with exponentially increasing star formation rates.  This brings our total number of possible tau values to eight.   Redshifts were fixed for each object, depending on the redshift of the \lya\ line, as this should be close to the correct redshift  depending on the possible velocity offset of \lya\ from systemic (even with our largest detected offset of 342 \kms in MC11, $\delta$z between \lya\ and \oiii\ is $<$ 0.005).  Our full grid contains 1.116 $\times$ 10$^{6}$ models, probing a very large parameter space.

\subsubsection{Photometry for \csq\ Minimization}\label{sec:phot}
For our SED fits we used  model and observed photometry in the \textit{B}, \rfil, \ifil\, \zfil, \textit{J}, \textit{H} and \kfil\ bands and IRAC 3.6 $\mu$m bands. We use photometry from the COSMOS Intermediate and Broad Band Photometry Catalog catalog \citep{cap07} for the \textit{B}, \rfil, \ifil\, \zfil\ bands (3\arcsec\ aperture photometry). We do not use the \ufil\ filter because this is the dropout band for z $\sim$ 3.1 LAEs.  The \gfil\ band is not used since the redshifted \lya\ line is centered in this filter. The \textit{V} band is excluded for a similar reason, as the filter has transmission between 55-65 per cent at the location of the \lya\ line.  The IRAC 3.6 \um\ data come from the S-COSMOS IRAC 4-channel Photometry Catalog\footnote{http://irsa.ipac.caltech.edu/data/COSMOS/tables/scosmos/} available on the NASA/IPAC archive.  We use the 2.9\arcsec\ aperture fluxes from this catalog \citep{sand07}.  Six LAEs have IRAC 3.6 \um\ detections in this catalog. For uncrowded objects with no IRAC 3.6 \um\ detection, we use the 3$\sigma$ depth  (0.54 $\mu$Jy) of the IRAC 3.6 \um\ image for the observed data point.  This is the case for 15 LAEs.  In the 12 cases where neither a detection nor an upper limit could be used, the $\chi^2$ minimization process does not use an IRAC 3.6 \um\ point. 29 of the LAEs in our sample are covered by the deep UltraVISTA Survey in the COSMOS field \citep{mcc12} and we used these new \textit{J}, \textit{H}, and \kfil\ images (Data Release 1) for our NIR photometry.  The photometry for each object was measured using SExtractor. SExtractor detections were forced at the desired coordinates (coordinates taken from the COSMOS catalog) by creating images with bright, fake sources at the correct coordinates and running SExtractor in dual-image mode with these fake images as the detection images and the \textit{J}/\textit{H}/\kfil\ images as the measurement images. For the four LAEs not covered by UltraVISTA, we extracted \textit{J}, \textit{H} and \kfil\ photometry from earlier publicly available COSMOS images.  We used the CFHT \textit{H} and \kfil\ band images \citep{mcc10} and for \textit{J} we used the UKIRT \textit{J} images \citep{cap07}. Again, SExtractor detections were forced at the desired coordinates as described above, in 3\arcsec\ apertures, error bars UltraVISTA and COSMOS NIR data are taken directly from SExtractor.


\subsection{SED Models}
The CB11 code creates model spectra in units \lsol\ $A^{-1}$ for $\lambda$ = 91 - 3.6$\times$ 10$^8$ \AA.  This output was converted to flux density (F$_{\nu}$) using the following two conversions \citep{pap01}:
\begin{equation}
L_{\nu} = \frac{10^8 \lambda_0^2 l_{\lambda} L_{\odot}}{cM_{gal}}10^{-0.4[E(B-V)k^{'}(\lambda)]}
\end{equation}

\begin{equation}
\label{eq:fnu}
F_{\nu}=\frac{(1+z)L_{\nu}}{4\pi d_L^2}e^{-\tau_{IGM}}
\end{equation}

$l_{\lambda}$ is the CB11 output in units \lsol $A^{-1}$, 10$^8$ converts from \AA\ to $\mu$m, [E(B-V)$k^{'}(\lambda)$] is wavelength dependent and calculated from Calzetti (2000), $\lambda_0^2$ is wavelength in the galaxies rest-frame, M$_{gal}$ is the total mass in the stellar population at a given age (which results in the spectrum being normalized to 1 \msol), z is the redshift of the model, fixed to z = 3.1, $\tau_{IGM}$ is wavelength-dependent IGM absorption from \citet{mad95} and d$_L$ is the luminosity distance for z = 3.1.  After application of Equation \ref{eq:fnu} and convolution of the flux through each filter, we have an individual flux density value for each filter (\textit{B}, \rfil, \ifil, \zfil, \textit{J}, \textit{H}, \kfil\, IRAC 3.6 \um). 

It is at this point that we add our new parameter, \oiii\ line flux.  We modify (amplify) the flux in the \kfil\ band to mimic how \oiii\ can contribute flux in this filter.  The modification looks like
\begin{equation}
\label{eq:fmod}
f_{total} = f_k + f_{[OIII]} 
\end{equation} 
where $f_{k}$ is the unmodified model flux density in the \kfil\ band.  $f_{[OIII]i}$ is the \oiii\ line flux in the band, and $f_{tot}$ is the total flux density in the \kfil\ band after those of two fluxes are combined.  $f_{[OIII]}$ is defined as
\begin{equation}
\label{eq:foiii}
f_{[OIII]} = x f_k
\end{equation}
where x takes on 15 uniform values from 0 - 1.5, meaning there are 15 possible \oiii\ fluxes that could be fit.  When x=0 this means there is no additional line flux from \oiii\ added to \kfil\ band, and this result is chosen as the best fit for some of our LAES (see \ref{sec:res2}).   This method essentially allows for additional line flux in the \kfil\ band, but allows the underlying spectrum to still be a younger/less massive galaxy, which would not necessarily be the case if an artificially large \kfil\ band flux forced an older and more massive solution to be fit.  \citet{sd09} pointed out the importance of including some treatment of nebular emission lines when fitting starbursting galaxies, when they found that ages in a sample of z $\sim$ 6 galaxies could be overestimated by as much as four times and mass by as much 1.5 times when nebular emission lines were not accounted for.  Some treatment of nebular emission lines is certainly warranted, but we advocate for a simple methodology (Equation \ref{eq:fmod}) for accounting for nebular emission.  This methodology only requires that a single additional parameter be added to our fitting process, avoiding a complex recipe of adding a large number of lines to our spectra - and this single parameter can be accounted for across all possible star formation histories and metallicities.  As \citet{nil11} and \citet{nak12} have noted, accounting for such detailed nebular emission line recipes across multiple star formation histories and metallicities can be too complex, limiting the parameter space that can be probed.   We avoid this by dealing with only a single parameter that is matched to what we have actually observed - i.e. we have observed some of these LAEs in the NIR and only detected \oiii, so this is the only line/parameter we are adding.  Additionally, we can directly compare our NIR observations to the predictions from our model fitting process (\ref{sec:comp}).

Finally, mass is a fitted parameter, calculated from minimizing the $\chi^{2}$ in Equation \ref{eq:chi} for each model.  This means that for each model there is a single best-fit mass solution 
found by minimizing $\chi^2$ with respect to mass:
\begin{equation}\label{eq:chi}
\chi^2 = \sum_{i} [\frac{f_{\nu,i}^{obs} - Mf_{\nu,i}}{\sigma_{i}}]^2
\end{equation}
Here the subscript `i' represents each filter where the model and observed photometry are compared.

\subsection{Allowed Fits}\label{sec:allow}
Some of our LAEs are best fit, strictly via $\chi^2$ minimization, with old stellar populations.  These fits require careful consideration because older stellar populations may not be able to produce enough ionizing photons to produce the \lya\ lines we have measured (with optical spectroscopy) in these objects.  We therefore consider some additional constraints on these objects to see if these old best-fit solutions are, in fact, realistic, physically-motivated solutions or if they ought to be ruled out in favor of younger, dustier solutions.

The CB11 code produces a parameter, NLyc, that is the log rate of ionizing photons (s$^{-1}$) produced at each age of the model for a given metallicity.  Assuming case B recombination, where two of every three of these ionizing photons produces a \lya\ photon, we can turn this production rate into a \lya\ line strength at each model age.  This allows us to test if the best fit age for a given object is able to produce, at a minimum,  the \lya\ line we have measured for that object with spectroscopy.  We do not subject this \lya\ line to attenuation by dust and/or the IGM, as we are simply testing if, at a minimum, the model stellar population could intrinsically produce enough ionizing photons to begin with, before any attenuation. 

The actual mechanism for this calculation is as follows:
\begin{equation}\label{eq:ionize}
\textrm{\lya\ line flux} = \frac{2}{3}\frac{10^{\textrm{NLy$\alpha$}}[h\nu_{Lyc}]}{4\pi d_L^2}\frac{\textrm{M}_{m}}{M_{gal}}
\end{equation}
where $\frac{2}{3}$ is the coefficient for \lya\ for case B recombination, NLyc is the log production rate of ionizing photons, h$\nu_{\lya}$ is the energy of a \lya\ photon, and d$_L$ is the luminosity distance at z = 3.1 \citep{wr06}.  M$_{gal}$ is the total mass in the stellar population at a given age and M$_{m}$ is the best fit mass (in \msol) for the model under consideration, so that the final term $\frac{\textrm{M}_{m}}{M_{gal}}$ scales the model stellar population from its normalized, $<$ 1\msol\ mass, to the appropriate galactic size stellar mass.

Only models (i.e combinations of metallicity, age, star formation history, dust, and mass) that can produce, at a minimum, enough ionizing photons to power the \lya\ line we observe are considered `allowed' fits.  With this information, we find the model with the smallest $\chi^2$ from among only these `allowed' possibilities.  Henceforth, we'll refer to this as the best allowed-fit for each object.  Consequently, the best allowed-fit solution is not always the model with the absolute smallest $\chi^2$

An example of this calculation, for LAE40844, is shown in Figure \ref{fig:ionize}.  Figure \ref{fig:ionize} shows the strength of the \lya\ line (solid curve) as function of stellar populations of increasing age for constant mass, metallicity, and $\tau$.    This particular figure is constructed using the best allowed-fit model for LAE40844, where metallicity is 0.2 \zsol, $\tau$ is 0.001 Gyrs, and mass is 2.9 $\times$ 10$^9$ \msol. The maximum age this combination of mass, metallicity, and $\tau$ can have and still produce the amount of \lya\ flux we have observed is shown as a black vertical line.  The best allowed-fit age is shown as a red vertical line.  This diagnostic shows why this combination of mass, tau, metallicity, and age is an allowed solution for LAE40844 - namely that the model age is to the left (younger) than the maximum age allowed that could  still produce the number of \lya\ photons we have observed from this object.  The observed \lya\ line strength is shown as a dashed line.

\begin{figure*}
\centering
\includegraphics[scale=0.5]{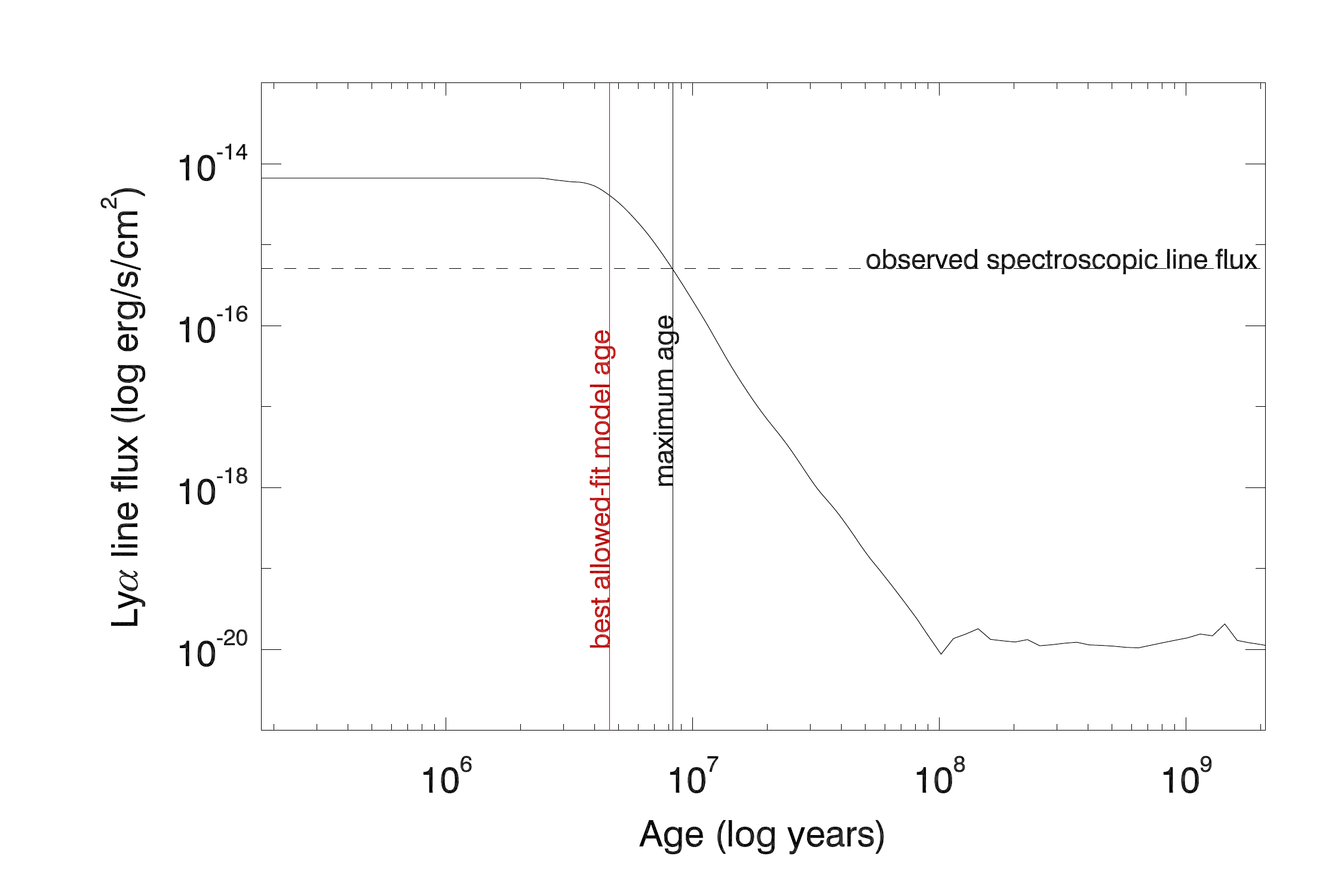}
\caption{Model \lya\ line flux (solid curve - from Equation \ref{eq:ionize}) that can be produced by stellar populations of increasing age, for a fixed mass, metallicity and star formation history.  This is the best allowed-fit model for LAE40844, where metallicity is 1\zsol, tau is 0.01 Gyrs, and mass is 2.17 $\times$ 10$^9$ \msol.  The horizontal dashed line shows the observed spectroscopic \lya\ line flux for LAE40844.  The age of the best allowed-fit model is the red vertical line, where only models younger than the black vertical  line can produce the observed \lya\ line flux.}\label{fig:ionize}
\end{figure*}

\section{RESULTS FROM SED FITTING}\label{sec:res2}
We find a diverse spread of physical characteristics from SED fitting for our sample of 33 LAEs.  Detailed descriptions of each physical characteristic are described in Sections \ref{sec:sfr} -- \ref{sec:oiiip} below.  The authors note that some of the diversity in physical characteristics is certainly real, some of it may arise from the iterative nature of our selection process that was used to build this sample (see Section \ref{sec:select}).  Model spectra are shown below in  Figures \ref{fig:tbl1} -- \ref{fig:tbl7} for all 33 LAEs.  Observed magnitudes are shown as red diamonds and error bars on observed photometry are also shown.  Red diamonds with a downward arrow instead of error bars indicates that an observed point was fainter than the 3$\sigma$ depth of that band.  Magnitudes from the model spectra are shown as blue diamonds.  For objects where the model included \oiii\ line flux in the \kfil\ band, you will note that both the red and blue diamonds lie above the black model spectrum.  This is expected as it means that an artificially large \kfil\ band flux from \oiii\ line flux pollution in this band is not dominating/skewing the best fit results.  Results for fitted parameters for each object are shown in Tables \ref{sedtbl1} and \ref{sedtbl2} along with 68 per cent confidence intervals for each parameter.  Plots of age versus mass, \oiii\ line flux versus age, metallicity versus age and \textit{E(B-V)} versus age are shown in Figures \ref{fig:dist1} and \ref{fig:dist2} for the population of 33 LAEs.

\subsection{Goodness of Fits}
Our median  reduced $\chi^2$ is 7.0.  The best fit object has a reduced $\chi^2$ of 0.9 and the worst fit object has a reduced $\chi^2$ of 48.3.  This particular object is one that may have multiple components in the HST image (see Section \ref{sec:final}), which may contribute to the large $\chi^2$ value.  We also remind the reader that the model chosen as the best fit is not always the smallest $\chi^2$ solution for each LAE, but rather, the model with smallest $\chi^2$ from among those models that can produce enough ionizing photons (best allowed-fit). For objects with IRAC 3.6 \um\ photometry (meaning either a detection or the limit was used) there are two degrees of freedom.  For objects with no IRAC 3.6 \um\ data, there is one degree of freedom.  These values comes from leveraging 8 bands (B, r, i, z, \textit{J}, \textit{H}, \kfil\, IRAC 3.6 \um) or 7 bands when no data are available for the IRAC 3.6 \um\ band, against six fitted parameters (age, mass, metallicity, dust, tau, [OIII]). 

We demonstrated how well constrained the fits are for each LAE with Monte Carlo (MC) simulations of each individual object.  We ran 1000 MC simulations for each object.  In each of the 1000 iterations, we modified the observed fluxes in each band by a Gaussian random amount proportional to the error bar in that band and then we determined the best allowed-fit model for the altered photometry in the same manner as described above.  Density plots showing the distribution of MC solutions around the best fit are shown in Figures \ref{fig:tbl1} -- \ref{fig:tbl7} for age, predicted \oiii\ line flux, and dust.  Similar plots for additional fitted parameters are included in Appendix A (online only).  Contours encompassing approximately 68 per cent and 95 per cent of the MC results are also shown on each plot.  In addition, Tables \ref{sedtbl1} and \ref{sedtbl2} list these 68 per cent confidence ranges for each fitted parameter.  This range was calculated by sorting (from smallest to largest) the 1000 MC solutions for a given parameter, and finding the spread given by the central 680 solutions in the sorted array.  The error bars reported below on any model predictions are derived from this 68 per cent confidence range.

\subsection{Star Formation History Results}\label{sec:sfr}
While some recent literature \citep{mar10,fin11,pap11} has suggested that, on average, high-z star-forming galaxies may be better fit with exponentially increasing star formation rates, in fitting 33 individual LAEs we find that only four galaxies in our sample are best fit with an exponentially increasing star formation rate.  Instead, we find that the majority of the sample is best fit with a single instantaneous burst (48 per cent, $\tau$ = 0.0001 Gyr) or exponentially decreasing star formation rates (39 per cent, $\tau$ = 0.001 -- 1.0 Gyr).  No LAEs in our sample are best fit with constant star formation rates ($\tau$ = 4 Gyr).

\subsection{ Age Results}\label{sec:age}
We report star formation rate weighted ages, age$_{\mathrm{SFR}}$, for each galaxy both here and in Tables \ref{sedtbl1} and \ref{sedtbl2}. Star formation weighted ages better represent the age of the bulk of the stars and are therefore more informative than directly quoting the ages of the models.  Equation \ref{eq:sfwdec} shows the derivation of this weighted age for exponentially decreasing star formation rates \citep{ra11}
\begin{equation}
\label{eq:sfwdec}
\langle age  \rangle_{\mathrm{SFR}} = \frac{\int_0^t(t-t') e^{-t'/\tau} \mathrm{d}t'}{\int_0^t e^{-t'/\tau} \mathrm{d}t'} = \frac{\tau e^{-\frac{t}{\tau}} - \tau + t}{1 - e^{-\frac{t}{\tau}}}
\end{equation}
where t is the age output from the model (i.e. time since star formation began) and tau is the e-folding time of the star formation rate as output from the model.  We also derived the same type of star formation weighted age for the case of exponentially increasing star formation, with the expression shown in Equation \ref{eq:sfwinc}.
\begin{equation}
\label{eq:sfwinc}
\langle age  \rangle_{\mathrm{SFR}} = \frac{\int_0^t(t-t') e^{t'/\tau} \mathrm{d}t'}{\int_0^t e^{t'/\tau} \mathrm{d}t'} = \frac{\tau - \tau e^{\frac{-t}{\tau}} - te^{\frac{-t}{\tau}} }{ 1 - e^{\frac{-t}{\tau}} }
\end{equation}

Our median age$_{\mathrm{SFR}}$ is 4.5 $\times$ 10$^{6}$ years, with age$_{\mathrm{SFR}}$ results spanning 1.4 $\times$ 10$^{6}$  -- 4.6 $\times$ 10$^{8}$ year.  The median size of the 68 per cent confidence ranges calculated for each object is 3.2 $\times$ 10$^{6}$ years. A majority of our sample, (85 per cent) have \ages\ $<$ 100 Myrs. Hence our sample of 33 galaxies fits with previously reported results (e.g. Gawiser et al. 2007, Pirzkal et al. 2007, Finkelstein 2009, Finkelstein 2011, Cowie et al. 2011) that LAEs have largely young to intermediate ages.  

\subsection{Stellar Mass Results}\label{sec:mass}
The median stellar mass in our sample is 6.9 $\times$ 10$^{8}$ \msol, the mean value is 5.4  $\times$ 10$^{9}$ \msol.  The most massive solution in our sample is 6.0 $\times$ 10$^{10}$ \msol\ and the smallest solution is 7.1 $\times$ 10$^{7}$ \msol.  
The large number of galaxies (14 of 33) that have masses $\ge$ 1 $\times$ 10$^{9}$ \msol\  is  a result of the wide-field and therefore shallower nature of our survey,  meaning we have selected LAEs from the brighter and more massive end of z $\sim$ 3.1 LAE population. We discuss this further in Section \ref{sec:comp2}.

\subsection{Dust Results}\label{sec:dust}
The median \textit{E(B-V)} value in our sample is 0.10, corresponding to less than a magnitude of extinction at $\lambda$ = 1200 \AA.  52 per cent of the sample has 68 per cent confidence ranges that include this median value.  The largest \textit{E(B-V)} value in the sample is 0.7.  
Only five objects are fit with absolutely no dust extinction, but an additional three objects have 68 per cent confidence ranges that include \textit{E(B-V)} = 0.  We also note that a total of 27 per cent of the sample has the smallest non-zero \textit{E(B-V)} solution, where \textit{E(B-V)} = 0.05.  These trends seem to indicate that overall, we are looking at a sample of galaxies that do not contain much dust.

\subsection{\oiii\ Line Fluxes Results}\label{sec:oiiip}
The main feature that distinguishes this work from previous SED fitting work with LAEs is the inclusion of an additional fitted parameter to account for \oiii\ line flux in the \kfil\ band (where the \kfil\ band encompasses the \oiii\ 5008.240 \AA\ for z $\sim$ 3.1 galaxies).   We chose to add this single line as this is the only rest-frame optical emission line we have detected in z $\sim$ 3.1 LAEs via NIR spectroscopy.  This puts us in a unique position to compare \oiii\ predictions from our models for these three objects with actual measurements in the same objects.  We also have three LAEs in which NIR observations yielded non-detections.  As for overall results of our  \oiii\ fitting approach, we find that  76 per cent of our sample is best fit with an \oiii\ line flux $>$ 0.  This means that 8 LAEs are best fit with no additional flux from the \oiii\ line contributing to the \kfil\ band.   The average best-fit \oiii\ line flux in our sample is 9.7 $\times$ 10$^{-17}$ \lf\ (calculated only among the 25 galaxies with non-zero solutions).


\begin{table*}
\centering
\setlength{\tabcolsep}{0.035in} 
\begin{tabular}{|l|c|c|c|c|c|c|c|c|c|c|c|c|c|}
Object & Mass & 68\% CR & Tau   & 68\% CR & Age$^{d}$ & 68\% CR &  Metal & 68\% CR &  \textit{E(B-V)} & 68\% CR & \oiii $^{c}$ & 68\% CR & $\chi_r^2$\\
 & (log) & & & & (log) & & & & & & & & \\
\hline
LAE\_J100049.56+021647.1$^{b}$    &       9.91  &       9.88-  &    1.0e-01 &   1.0e-01-  &       8.12  &       8.12-  &      0.020  &      0.020-   &      0.00  &       0.00-  &      0.00  &       0.00-  &      16.4\\
&    &      9.92  &    &    1.0e-01  &   &       8.12  &   &      0.020  &   &       0.05  &   &       0.00  &   \\
LAE\_J095859.33+014522.0  &       9.20  &       9.14-  &    1.0e-03 &   1.0e-04-  &       6.56  &       6.53-  &      1.000  &      1.000-   &      0.20  &       0.15-  &      0.90  &       0.53-  &      35.4\\
&    &      9.32  &    &    1.0e-03  &   &       6.71  &   &      1.000  &   &       0.20  &   &       1.18  &   \\
LAE\_J100212.99+020137.7  &       8.76  &       8.75-  &    1.0e-04 &   1.0e-04-  &       6.53  &       6.33-  &      1.000  &      1.000-   &      0.15  &       0.15-  &      0.00  &       0.00-  &      29.6\\
&    &      9.07  &    &    1.0e-03  &   &       6.59  &   &      1.000  &   &       0.20  &   &       0.00  &   \\
LAE\_J095929.41+020323.5$^{a}$   &       8.38  &       8.15-  &    1.0e-04 &   1.0e-04-  &       6.15  &       6.21-  &      1.000  &      0.005-   &      0.10  &       0.05-  &      0.23  &       0.00-  &       6.8\\
(LAE6559) &    &      8.51  &    &    1.0e-03  &   &       6.70  &   &      1.000  &   &       0.10  &   &       0.30  &   \\
LAE\_J095944.02+015618.8  &       9.73  &       9.30-  &    1.0e+00 &  -1.0e+00-  &       8.66  &       7.95-  &      0.005  &      0.005-   &      0.00  &       0.00-  &      0.00  &       0.00-  &       2.4\\
&    &      9.77  &    &    4.0e+00  &   &       8.78  &   &      0.200  &   &       0.05  &   &       0.07  &   \\
LAE\_J095930.52+015611.0  &       9.41  &       9.40-  &   -1.0e-01 &  -1.0e+00-  &       8.00  &       7.85-  &      0.200  &      0.005-   &      0.05  &       0.00-  &      1.47  &       1.37-  &      12.6\\
 (LAE7745)&    &      9.89  &    &    1.0e-01  &   &       8.78  &   &      0.400  &   &       0.05  &   &       1.59  &   \\
LAE\_J100217.05+015531.7$^{b}$  &      10.16  &       8.95-  &    1.0e+00 &   1.0e-04-  &       8.49  &       6.71-  &      1.000  &      1.000-   &      0.00  &       0.00-  &      0.00  &       0.00-  &      48.3\\
&    &     10.18  &    &    4.0e+00  &   &       8.54  &   &      1.000  &   &       0.10  &   &       0.43  &   \\
LAE\_J100157.87+021450.0$^{a}$   &       8.73  &       8.38-  &    1.0e-04 &   1.0e-04-  &       6.96  &       6.40-  &      1.000  &      0.020-   &      0.05  &       0.05-  &      0.28  &       0.16-  &       3.5\\
&    &      8.95  &    &    1.0e-03  &   &       6.98  &   &      1.000  &   &       0.15  &   &       0.52  &   \\
LAE\_J100124.36+021920.8  &       9.34  &       9.14-  &    1.0e-04 &   1.0e-04-  &       6.83  &       6.56-  &      0.200  &      0.200-   &      0.10  &       0.10-  &      3.71  &       3.63-  &       9.1\\
(LAE40844) &    &      9.34  &    &    1.0e-03  &   &       6.83  &   &      0.200  &   &       0.15  &   &       4.12  &   \\
LAE\_J095847.81+021218.2  &       8.26  &       8.27-  &    1.0e-04 &   1.0e-04-  &       6.59  &       6.48-  &      1.000  &      1.000-   &      0.00  &       0.00-  &      0.75  &       0.57-  &       2.7\\
&    &      8.65  &    &    1.0e-02  &   &       6.86  &   &      1.000  &   &       0.05  &   &       0.86  &   \\
LAE\_J095904.93+015355.4$^{a}$   &       8.24  &       8.03-  &    1.0e-03 &   1.0e-04-  &       6.70  &       6.40-  &      0.005  &      0.005-   &      0.10  &       0.05-  &      0.00  &       0.00-  &       4.1\\
&    &      8.54  &    &    1.0e-03  &   &       6.71  &   &      0.005  &   &       0.15  &   &       0.00  &   \\
LAE\_J095910.90+020631.6$^{b}$   &       8.44  &       8.18-  &    1.0e-02 &  -1.0e-01-  &       6.86  &       6.56-  &      1.000  &      0.200-   &      0.05  &       0.05-  &      0.21  &       0.00-  &       7.3\\
 (LAE14310)&    &      9.19  &    &    1.0e-01  &   &       8.00  &   &      1.000  &   &       0.05  &   &       0.33  &   \\
LAE\_J095921.06+022143.4  &       9.91  &       8.92-  &    1.0e-02 &   1.0e-04-  &       7.58  &       6.65-  &      0.005  &      0.005-   &      0.20  &       0.15-  &      0.27  &       0.13-  &       2.2\\
&    &      9.92  &    &    1.0e-02  &   &       7.58  &   &      1.000  &   &       0.20  &   &       0.51  &   \\
LAE\_J095948.47+022420.8$^{a}$    &       8.16  &       8.07-  &    1.0e-04 &   1.0e-04-  &       6.65  &       6.59-  &      0.005  &      0.005-   &      0.10  &       0.10-  &      0.57  &       0.44-  &       7.2\\
&    &      8.29  &    &    1.0e-03  &   &       6.65  &   &      0.005  &   &       0.15  &   &       0.69  &   \\
LAE\_J100019.07+022523.9  &       8.68  &       8.42-  &    1.0e-02 &  -1.0e-01-  &       6.39  &       6.42-  &      0.005  &      0.005-   &      0.15  &       0.10-  &      0.07  &       0.00-  &       2.4\\
 (LAE27878)&    &      9.37  &    &    1.0e-02  &   &       8.00  &   &      0.020  &   &       0.15  &   &       0.26  &   \\
LAE\_J100100.35+022834.7  &       8.84  &       8.59-  &   -1.0e-01 &  -1.0e-01-  &       6.47  &       6.46-  &      0.005  &      0.005-   &      0.15  &       0.10-  &      1.21  &       1.05-  &       7.0\\
&    &      8.86  &    &    1.0e-02  &   &       6.70  &   &      0.005  &   &       0.15  &   &       1.29  &   \\
LAE\_J100146.04+022949.0  &      10.78  &      10.69-  &    1.0e-01 &   1.0e-01-  &       8.49  &       8.49-  &      1.000  &      0.400-   &      0.10  &       0.10-  &      0.52  &       0.51-  &       6.7\\
&    &     10.78  &    &    1.0e-01  &   &       8.49  &   &      1.000  &   &       0.10  &   &       1.00  &   \\
LAE\_J095843.11+020312.3$^{b}$  &       8.32  &       8.30-  &    1.0e-04 &   1.0e-04-  &       6.59  &       6.56-  &      0.200  &      0.200-   &      0.05  &       0.05-  &      0.00  &       0.00-  &      18.1\\
&    &      8.36  &    &    1.0e-03  &   &       6.59  &   &      0.200  &   &       0.05  &   &       0.00  &   \\
LAE\_J100128.11+015804.7$^{a}$    &       8.91  &       8.71-  &    1.0e-02 &  -1.0e-01-  &       6.79  &       6.53-  &      0.400  &      0.005-   &      0.15  &       0.10-  &      1.27  &       1.16-  &      24.0\\
&    &      9.08  &    &    1.0e-02  &   &       7.07  &   &      1.000  &   &       0.20  &   &       1.37  &   \\
LAE\_J100017.84+022506.1  &       8.48  &       8.24-  &    1.0e-02 &   1.0e-04-  &       6.46  &       6.46-  &      0.005  &      0.005-   &      0.10  &       0.05-  &      0.60  &       0.30-  &       2.2\\
(LAE27910) &    &      8.51  &    &    1.0e-02  &   &       6.65  &   &      0.005  &   &       0.10  &   &       0.72  &   \\
LAE\_J095839.92+023531.3  &       9.01  &       8.74-  &    1.0e-04 &   1.0e-04-  &       6.34  &       5.42-  &      0.005  &      0.005-   &      0.20  &       0.15-  &      0.00  &       0.00-  &      35.2\\
&    &      9.08  &    &    1.0e-04  &   &       6.40  &   &      1.000  &   &       0.20  &   &       0.14  &   \\
LAE\_J095838.90+015858.2$^{a}$  &       7.85  &       7.84-  &    1.0e-03 &   1.0e-04-  &       6.56  &       6.53-  &      0.005  &      0.005-   &      0.10  &       0.10-  &      0.30  &       0.11-  &       7.9\\
&    &      8.09  &    &    1.0e-03  &   &       6.59  &   &      0.005  &   &       0.15  &   &       0.37  &   \\
LAE\_J100020.70+022927.0$^{a,b}$  &       8.08  &       8.02-  &    1.0e-04 &   1.0e-04-  &       6.59  &       6.49-  &      0.005  &      0.005-   &      0.05  &       0.05-  &      0.18  &       0.06-  &      10.9\\
&    &      8.13  &    &    1.0e-03  &   &       6.59  &   &      0.005  &   &       0.05  &   &       0.45  &   \\
LAE\_J095812.33+014737.6$^{a}$  &      10.35  &       9.39-  &    1.0e-01 &   1.0e-04-  &       7.95  &       6.56-  &      0.005  &      0.200-   &      0.15  &       0.15-  &      1.19  &       0.86-  &       5.2\\
&    &     10.07  &    &    1.0e-02  &   &       7.66  &   &      0.200  &   &       0.20  &   &       1.52  &   \\
LAE\_J095920.42+013917.1  &       8.85  &       8.83-  &    1.0e-04 &   1.0e-04-  &       6.71  &       6.65-  &      1.000  &      1.000-   &      0.05  &       0.05-  &      1.20  &       0.87-  &       4.6\\
&    &      8.94  &    &    1.0e-04  &   &       6.77  &   &      1.000  &   &       0.10  &   &       1.47  &   \\
LAE\_J095846.72+013706.1  &       8.44  &       8.26-  &    1.0e-04 &   1.0e-04-  &       6.47  &       6.34-  &      1.000  &      1.000-   &      0.10  &       0.05-  &      0.32  &       0.00-  &       4.6\\
&    &      8.73  &    &    1.0e-04  &   &       6.56  &   &      1.000  &   &       0.15  &   &       0.55  &   \\
\hline
\multicolumn{14}{l}{\textsuperscript{a}\footnotesize{Poor agreement between spectroscopic and photometric \lya\ line flux measurements}}\\
\multicolumn{14}{l}{\textsuperscript{b}\footnotesize{Possible multiple components in HST ACS image}}\\
\multicolumn{14}{l}{\textsuperscript{c}\footnotesize{Predicted \oiii\ line flux, Units are 10$^{-16}$ \lf.}}\\
\multicolumn{14}{l}{\textsuperscript{d}\footnotesize{Age is age$_{\mathrm{SFR}}$, i.e. star-formation weighted age.}}\\
\end{tabular}
\caption{Best allowed-fit parameters from SED fitting for each object in our sample, excluding AGN. Mass is in \msol, Tau is e-folding time for star formation, in Gyr, age$_{\mathrm{SFR}}$ is the star formation weighted age, in years, Metal is metallicity in \zsol, \textit{E(B-V)} is standard color excess from dust attenuation, \oiii\ is predicted \oiii\ line flux in \lf, $\chi_r$ is reduced chi square of the best allowed-fit model.  68 per cent confidence ranges (CR) are also given for each parameter. Continued in Table \ref{sedtbl2}. }
\label{sedtbl1}
\end{table*}

\begin{table*}\centering
\setlength{\tabcolsep}{0.035in} 
\begin{tabular}{|l|c|c|c|c|c|c|c|c|c|c|c|c|c|}
Object & Mass & 68\% CR & Tau   & 68\% CR & Age$^{d}$ & 68\% CR &  Metal & 68\% CR &  \textit{E(B-V)} & 68\% CR & \oiii $^{c}$ & 68\% CR & $\chi_r^2$\\
 & (log) & & & & (log) & & & & & & & & \\
\hline
LAE\_J095923.79+013045.6  &      10.55  &      10.55-  &    1.0e-04 &   1.0e-04-  &       6.65  &       6.40-  &      1.000  &      0.005-   &      0.70  &       0.65-  &      3.43  &       3.23-  &       4.1\\
&    &     11.37  &    &    1.0e-03  &   &       7.58  &   &      1.000  &   &       0.85  &   &       4.43  &   \\
LAE\_J100213.17+013226.8  &       8.04  &       8.00-  &    1.0e-04 &   1.0e-04-  &       6.59  &       6.53-  &      1.000  &      0.020-   &      0.05  &       0.00-  &      0.12  &       0.00-  &      30.4\\
&    &      9.92  &    &    1.0e+00  &   &       8.83  &   &      1.000  &   &       0.10  &   &       0.23  &   \\
LAE\_J095838.94+014107.9  &       8.18  &       8.01-  &    1.0e-04 &   1.0e-04-  &       6.59  &       6.59-  &      0.005  &      0.005-   &      0.05  &       0.00-  &      0.88  &       0.69-  &       0.9\\
&    &      9.50  &    &    1.0e-04  &   &       8.54  &   &      0.005  &   &       0.05  &   &       1.02  &   \\
LAE\_J095834.43+013845.6  &       8.29  &       8.28-  &    1.0e-04 &   1.0e-04-  &       6.53  &       6.53-  &      0.005  &      0.005-   &      0.10  &       0.10-  &      1.16  &       0.87-  &       5.2\\
&    &      8.58  &    &    1.0e-04  &   &       6.65  &   &      0.005  &   &       0.15  &   &       1.23  &   \\
LAE\_J100302.10+022406.7  &       9.29  &       9.29-  &    1.0e-04 &   1.0e-04-  &       6.53  &       6.47-  &      1.000  &      1.000-   &      0.20  &       0.20-  &      0.00  &       0.00-  &      10.0\\
&    &      9.50  &    &    1.0e-04  &   &       6.53  &   &      1.000  &   &       0.25  &   &       0.28  &   \\
LAE\_J100157.45+013556.2  &       9.14  &       8.62-  &   -1.0e-01 &  -1.0e-01-  &       7.50  &       6.53-  &      1.000  &      0.020-   &      0.05  &       0.05-  &      1.25  &       0.69-  &       3.9\\
&    &      9.45  &    &    1.0e-02  &   &       8.00  &   &      1.000  &   &       0.15  &   &       1.44  &   \\
LAE\_J100152.14+013533.2  &       9.85  &       8.45-  &   -1.0e+00 &  -1.0e+00-  &       8.54  &       6.53-  &      0.400  &      0.005-   &      0.00  &       0.00-  &      2.10  &       1.90-  &      15.8\\
&    &      9.88  &    &    1.0e+00  &   &       8.54  &   &      0.400  &   &       0.05  &   &       2.28  &   \\
\hline
\multicolumn{14}{l}{\textsuperscript{a}\footnotesize{Poor agreement between spectroscopic and photometric \lya\ line flux measurements}}\\
\multicolumn{14}{l}{\textsuperscript{b}\footnotesize{Possible multiple components in HST ACS image}}\\
\multicolumn{14}{l}{\textsuperscript{c}\footnotesize{Predicted \oiii\ line flux, Units are 10$^{-16}$ \lf.}}\\
\multicolumn{14}{l}{\textsuperscript{d}\footnotesize{Age is age$_{\mathrm{SFR}}$, i.e. star-formation weighted age.}}\\
\end{tabular}
\caption{Continued from Table \ref{sedtbl1}.  Best allowed-fit parameters from SED fitting for each object in our sample, excluding AGN. Mass is in \msol, Tau is e-folding time for star formation, in Gyr, age$_{\mathrm{SFR}}$ is the star formation weighted age, in years, Metal is metallicity in \zsol, \textit{E(B-V)} is standard color excess from dust attenuation, \oiii\ is predicted \oiii\ line flux in \lf, $\chi_r$ is reduced chi square of the best allowed-fit model.  68 per cent confidence ranges (CR) are also given for each parameter. }
\label{sedtbl2}
\end{table*}

\begin{figure*}
\centering
\includegraphics[scale=0.85, trim=0cm 5cm 0cm 0cm, clip]{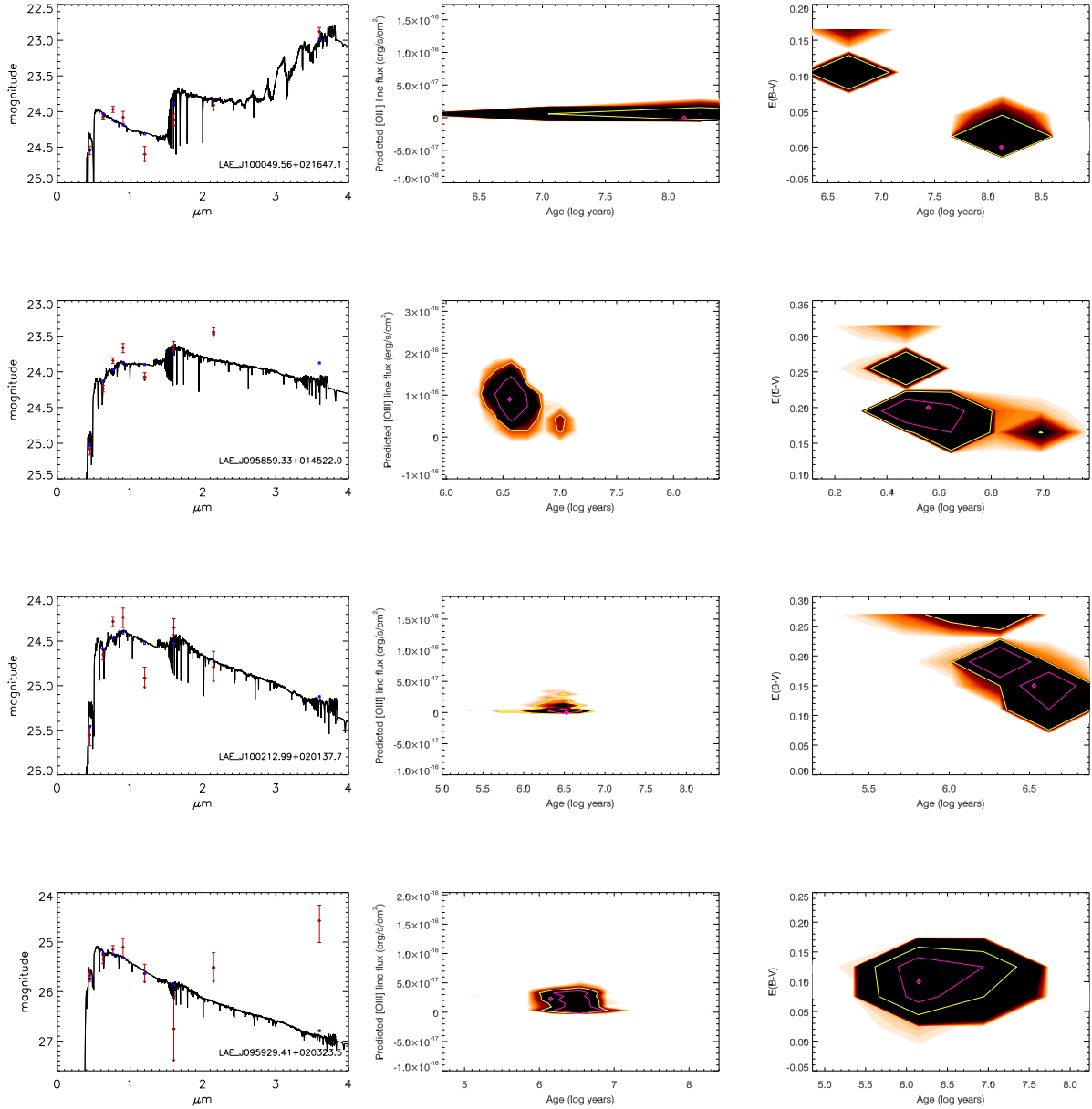}
\caption{The first column contains the best allowed-fit model spectra for the first four LAEs.  Model spectrum is black, model magnitudes are shown as blue squares. Observed magnitudes are shown  as red diamonds.  Red diamonds with a downward arrow instead of error bars indicates that an observed point was fainter than the 3$\sigma$ depth.  Plotted magnitudes are in \textit{B}, \textit{V}, \gfil, \rfil, \ifil\, \zfil, J, \textit{H} \kfil\, and IRAC 3.6\um, from left to right.  Large error bars in \textit{V} and \gfil\ bands are sometimes a consequence of subtracting the \lya\ line from these filters.  The second and third columns show density plots from our MC simulations.  Ages shown here are star formation weighted ages.  The best allowed-fit is shown as a magenta diamond. Contours encompassing $\sim$ 68 per cent and $\sim$ 95 per cent of the results are shown in magenta and yellow, respectively. The order of objects in Figure \ref{fig:tbl1} - Figure \ref{fig:tbl7} matches the order of objects in Tables \ref{sedtbl1} and \ref{sedtbl2}. AGN are excluded from both table and figures.}\label{fig:tbl1}
\end{figure*}

\begin{figure*}
\centering
\includegraphics[scale=0.85, trim=0cm 2cm 0cm 0cm, clip]{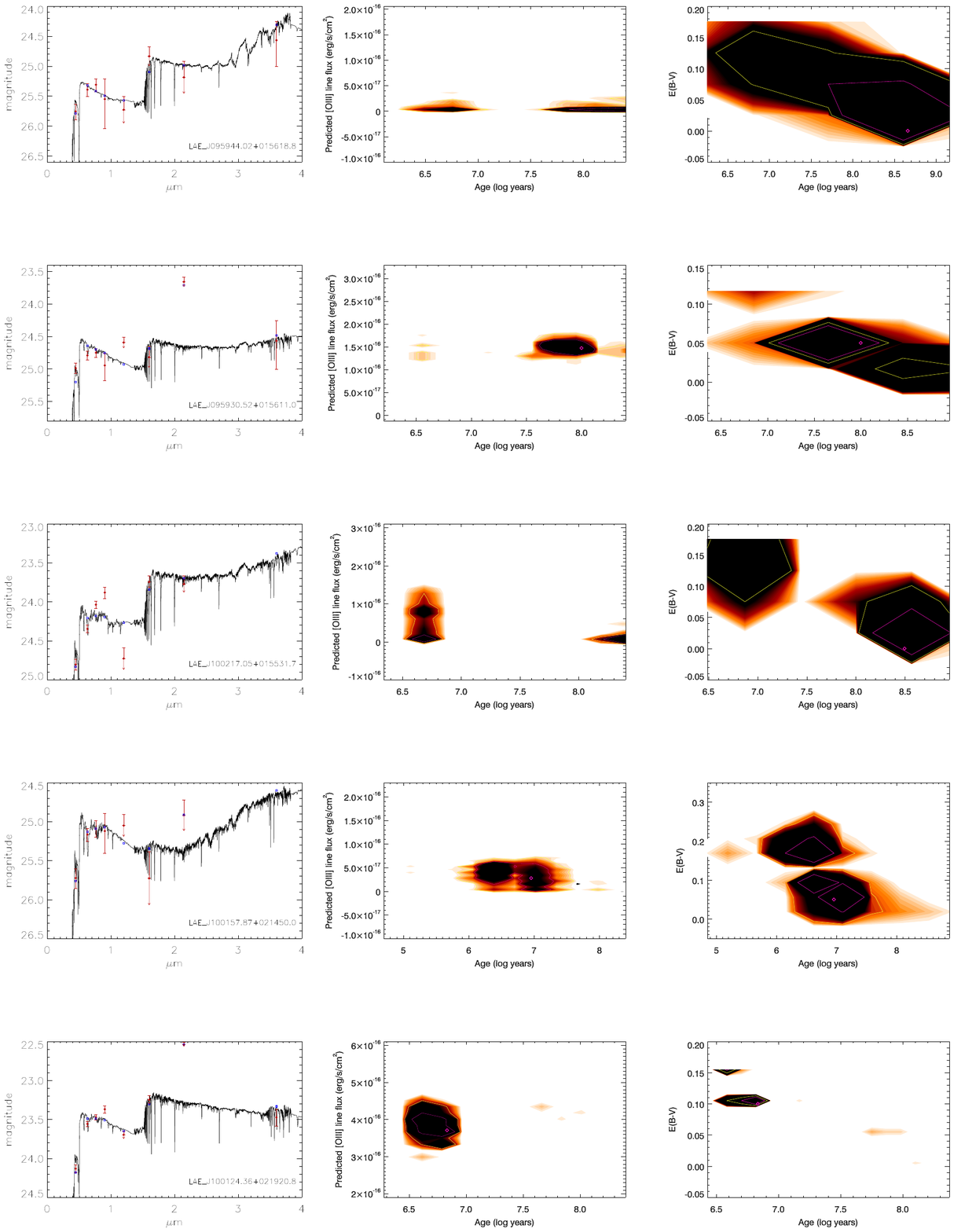}
\caption{Same as Figure \ref{fig:tbl1} for next 5 objects.}\label{fig:tbl2}
\end{figure*}

\begin{figure*}
\centering
\includegraphics[scale=0.85, trim=0cm 2cm 0cm 0cm, clip]{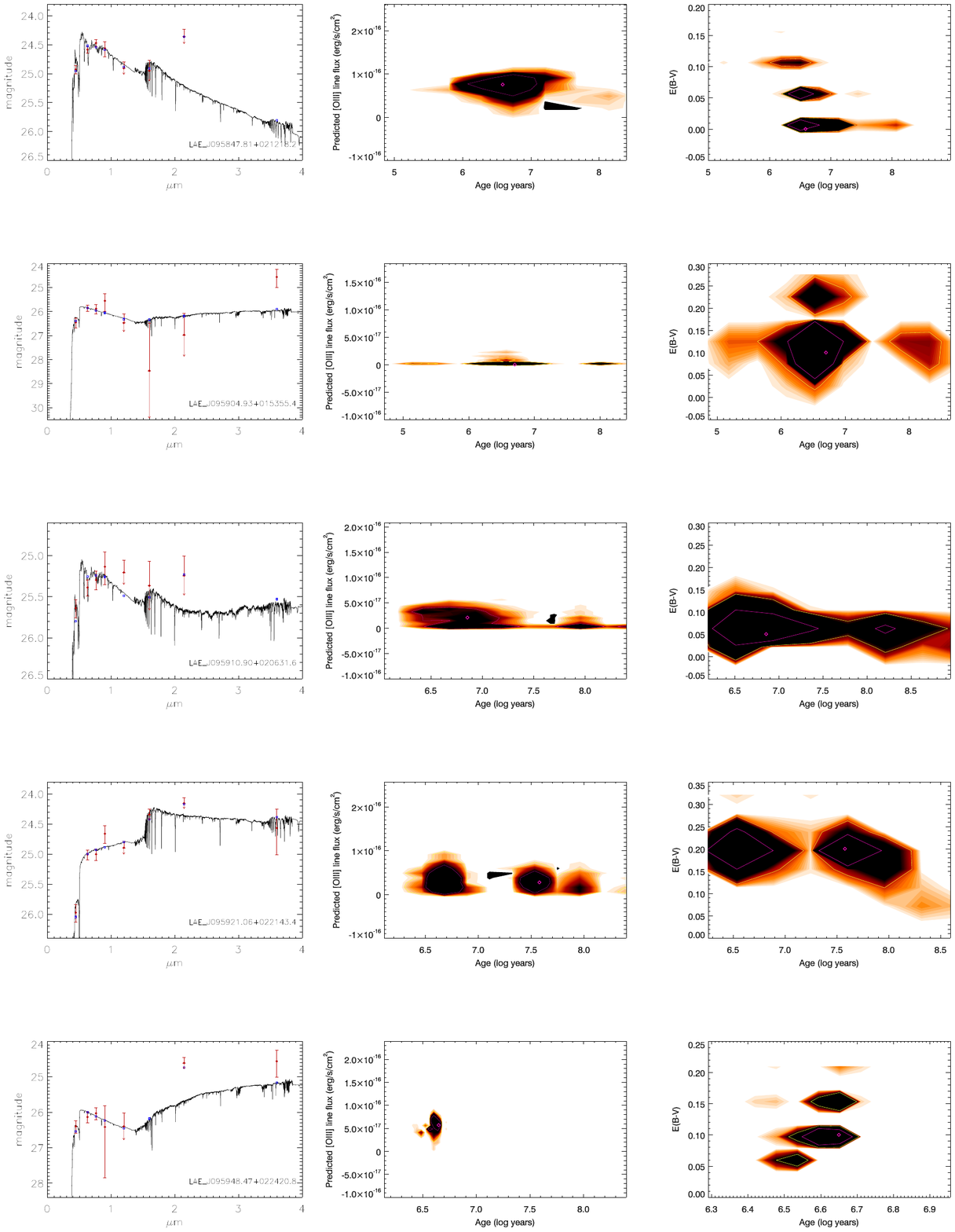}
\caption{Same as Figure \ref{fig:tbl1} for next 5 objects.}\label{fig:tbl3}
\end{figure*}

\begin{figure*}
\centering
\includegraphics[scale=0.85, trim=0cm 2cm 0cm 0cm, clip]{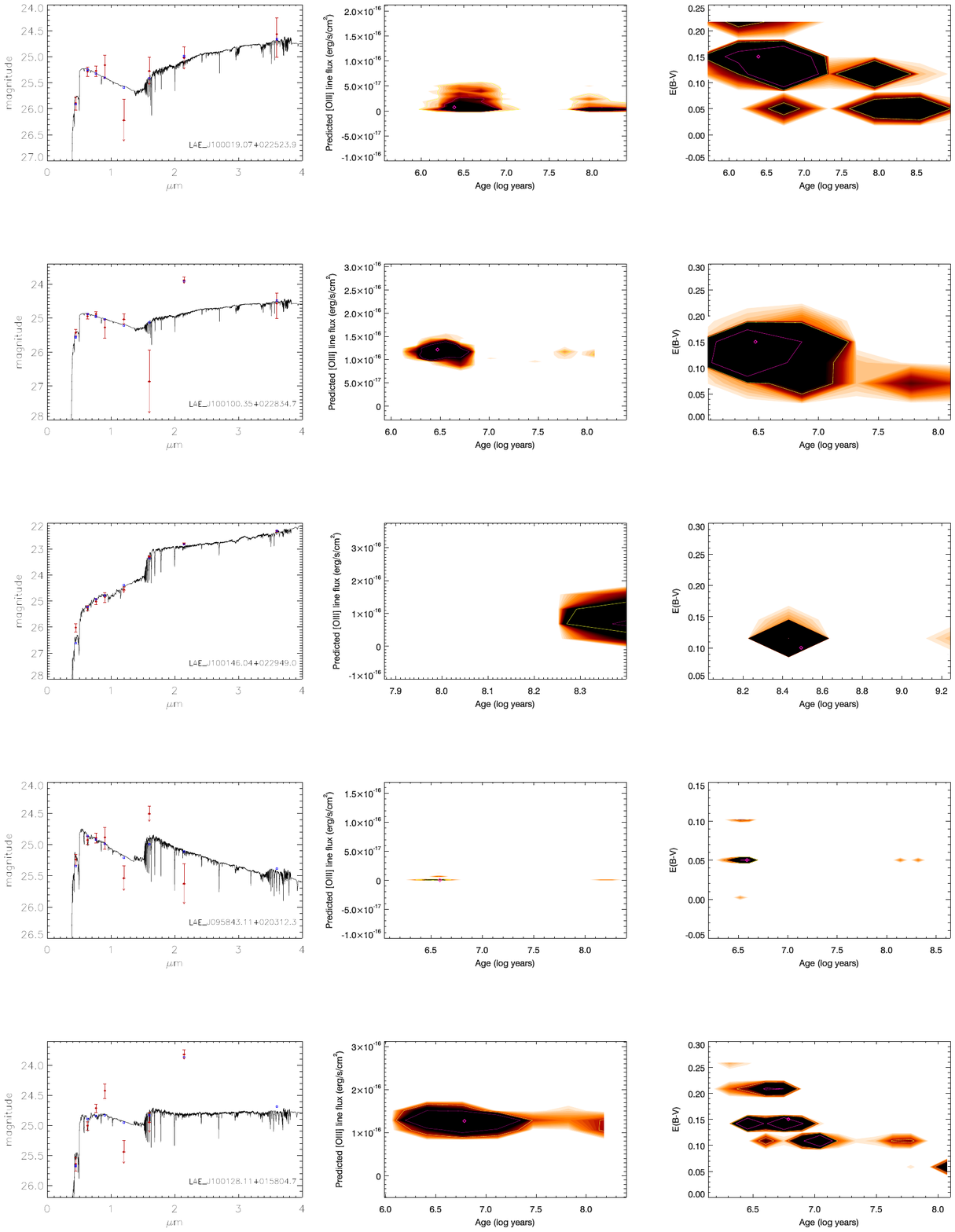}
\caption{Same as Figure \ref{fig:tbl1} for next 5 objects.}\label{fig:tbl4}
\end{figure*}

\begin{figure*}
\centering
\includegraphics[scale=0.85, trim=0cm 2cm 0cm 0cm, clip]{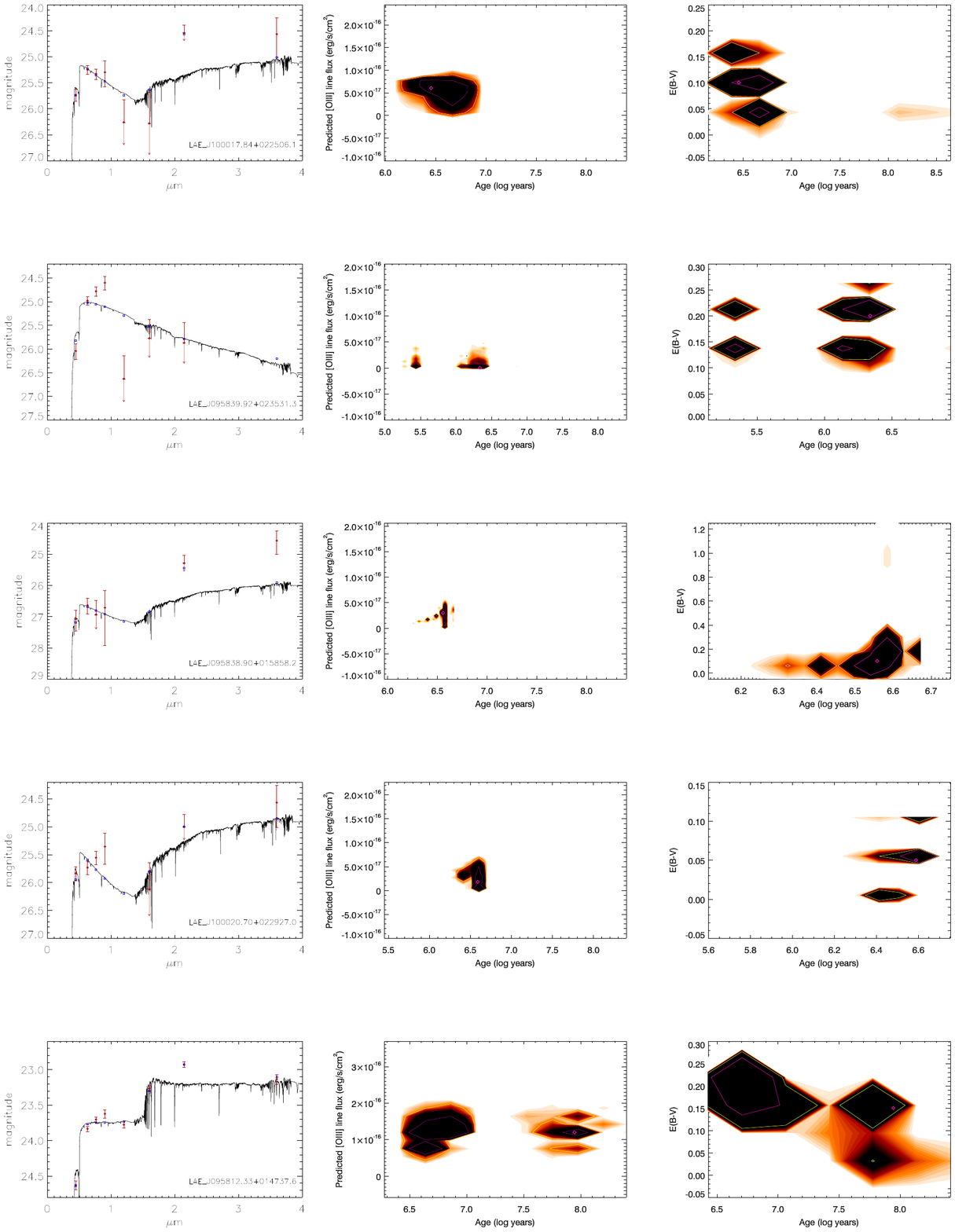}
\caption{Same as Figure \ref{fig:tbl1} for next 5 objects.}\label{fig:tbl5}
\end{figure*}

\begin{figure*}
\centering
\includegraphics[scale=0.85, trim=0cm 2cm 0cm 0cm, clip]{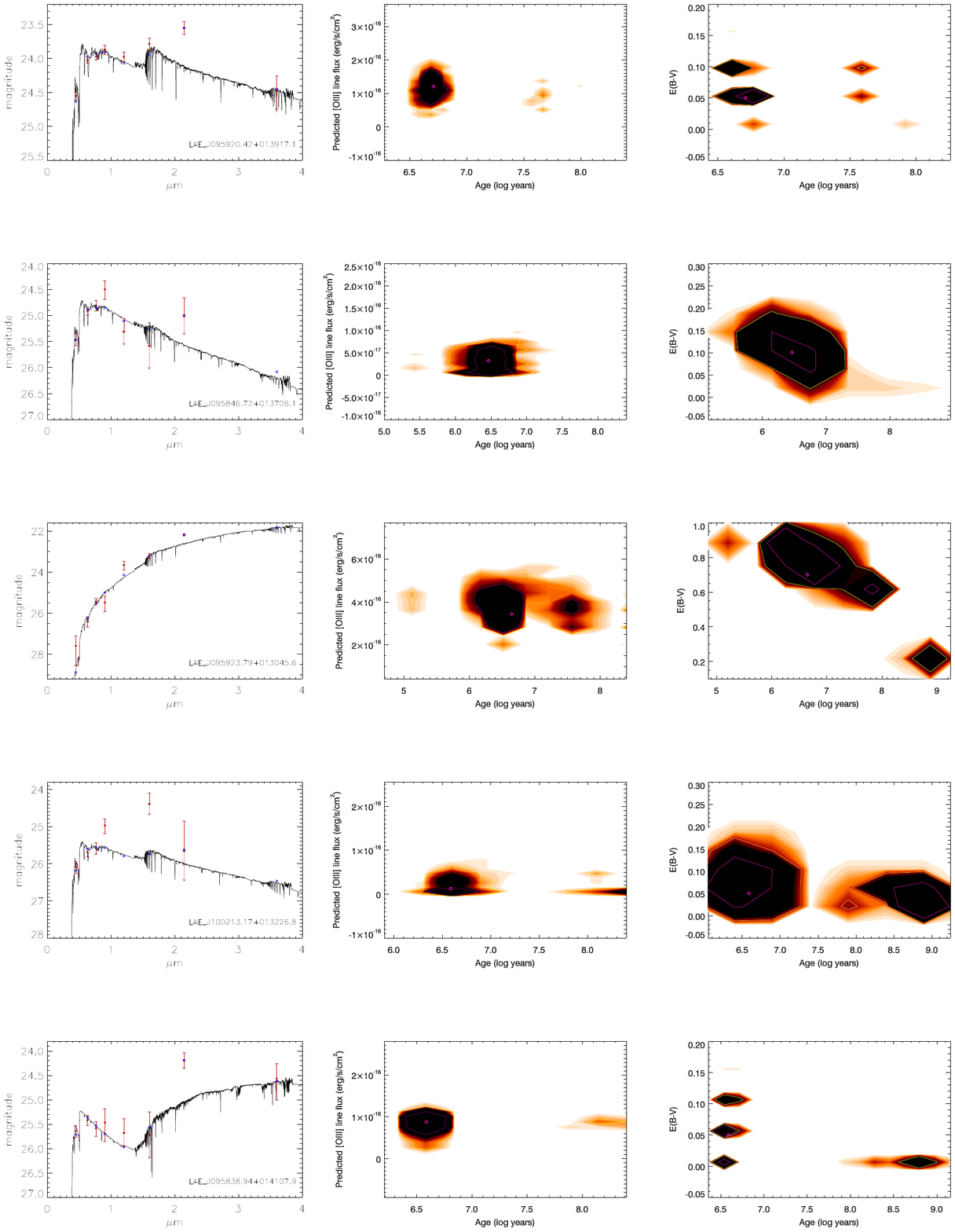}
\caption{Same as Figure \ref{fig:tbl1} for next 5 objects.}\label{fig:tbl6}
\end{figure*}

\begin{figure*}
\centering
\includegraphics[scale=0.85, trim=0cm 6cm 0cm 0cm, clip]{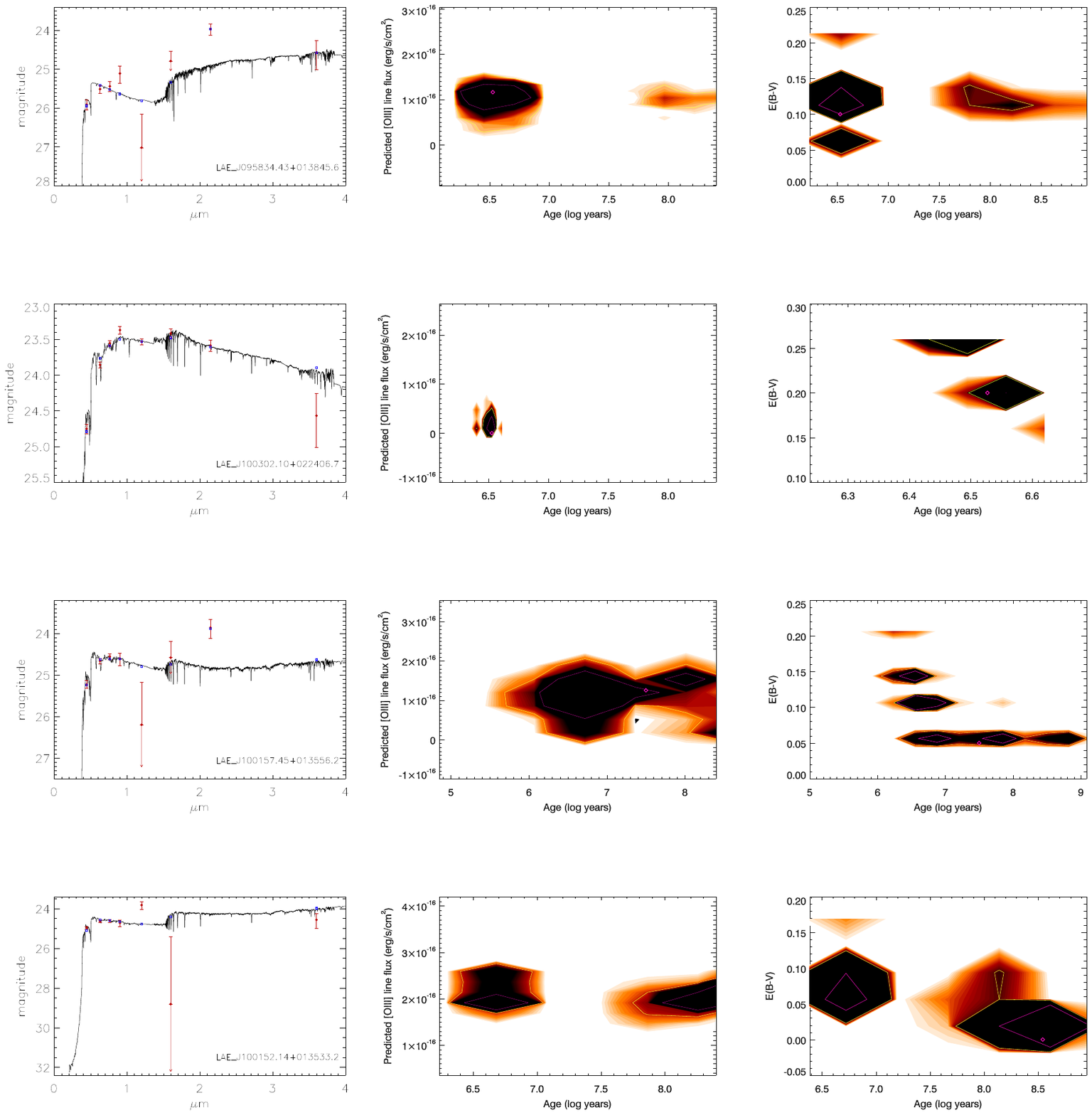}
\caption{Same as Figure \ref{fig:tbl1} for next 4 objects.}\label{fig:tbl7}
\end{figure*}

\begin{figure*}
\centering
\includegraphics[scale=0.6]{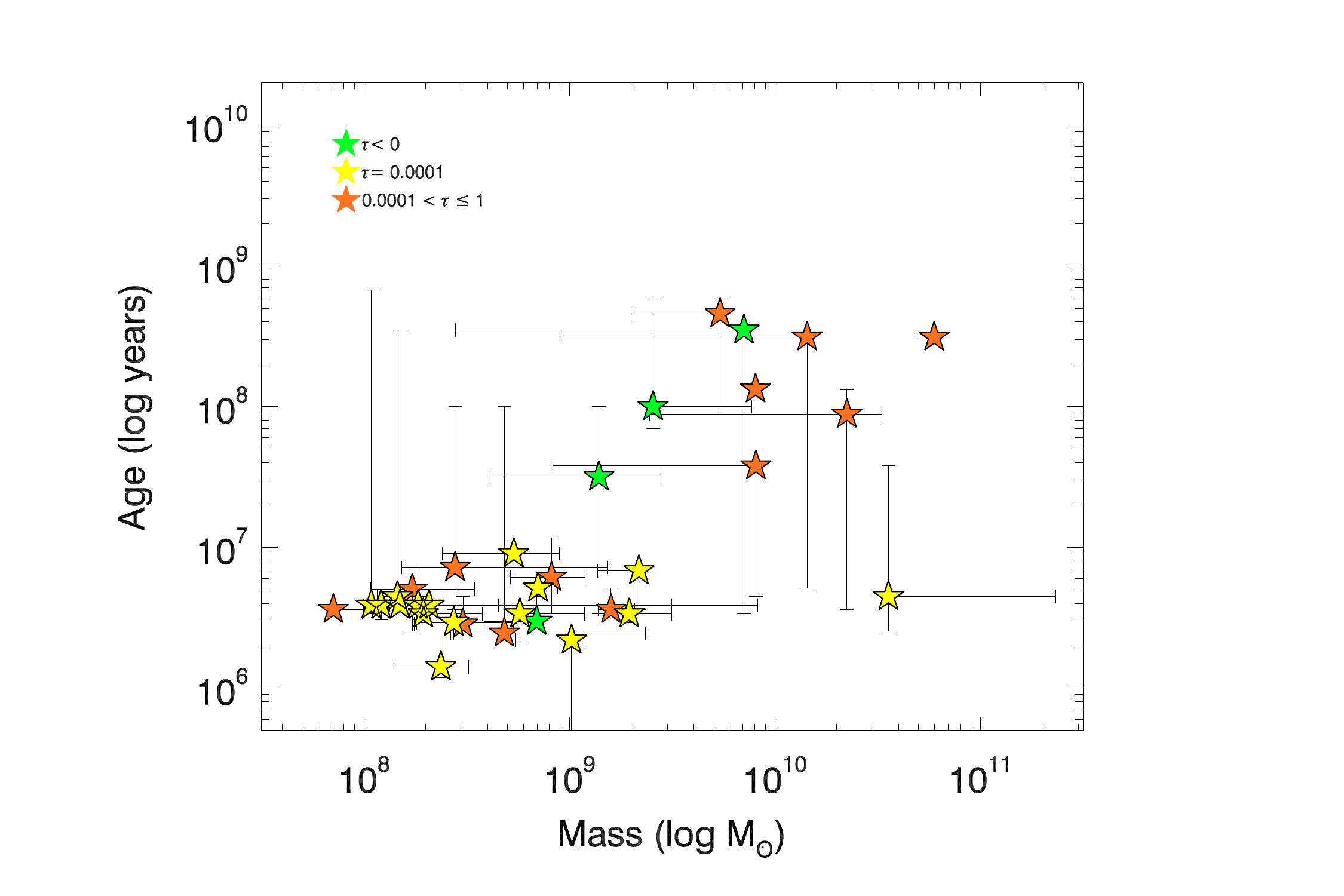}\caption{The distribution of age$_{\mathrm{SFR}}$ versus mass for various tau values for 33 LAEs.  Green stars indicate models with exponentially increasing star formation rates ($\tau < 0$), yellow stars are fits with a single instantaneous burst ($\tau_{\mathrm{SFR}}$ = 0.0001 Gyr), and orange stars are those with exponentially decaying star formation rates.  Since mass and age parameters are correlated this plot is mainly meant to illustrate and confirm the distribution of $\tau_{\mathrm{SFR}}$ with these parameters, showing that the oldest and most massive LAEs are those fit with increasing star formation rates, the youngest and least massive galaxies are fit with instantaneous star formation histories, and those LAEs with exponentially decaying star formation rates lie between those two populations.} 
\label{fig:dist1}
\end{figure*}

\begin{figure*}
\centering
\includegraphics[scale=1.25,trim=4cm 18cm 0cm 1cm,clip]{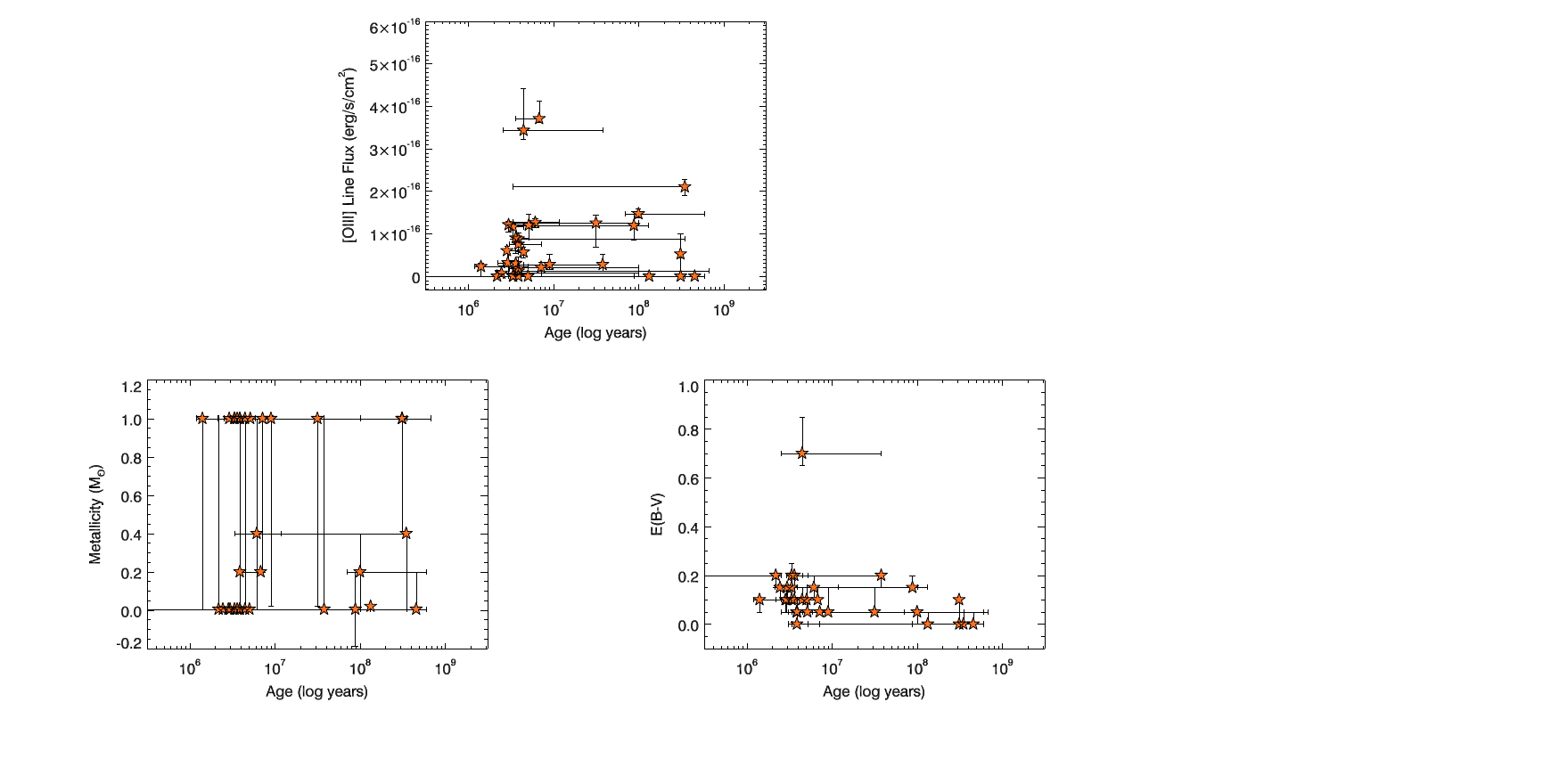}\caption{Top plot shows distribution of \oiii\ line flux versus age.  Bottom panel shows similar distributions for metallicity versus age (left) and \textit{E(B-V)} versus age (right). All ages plotted here are star-formation weighted ages.}
\label{fig:dist2}
\end{figure*}

\section{DISCUSSION}\label{sec:disc}

\subsection{Comparison of Predicted to Observed \oiii\ Line Flux}\label{sec:comp}
We have observed six non-AGN objects with NIR spectroscopy to look for \oiii\ and other rest-frame optical nebular emission lines.  As discussed previously, we have \oiii\ detections for three of these objects.  Comparing \oiii\ line flux predictions from our model predictions to the the actually observed line fluxes we find that all three of the galaxies with observed \oiii\ lines select models with \oiii\ lines.  Our best prediction is for LAE40844, in which we observed a line flux of 3.6 $\pm$ 0.1 $\times$ 10$^{-16}$ \lf\ and our model predicted  3.7$^{+0.41}_{-0.08}$ $\times$ 10$^{-16}$ \lf; an error between the observation and prediction of less than 3 per cent.  The observed \oiii\ line flux contributes 53 per cent contributes of the observed flux in the \kfil\ for this object; the predicted \oiii\ line flux contribute 54 per cent. This prediction also lies within the 1$\sigma$ error bar on the observed \oiii\ line.  In LAE7745, the per cent difference between the observed line flux (1.4 $\times$ 10$^{-16}$ \lf) and the predicted \oiii\ line bflux (1.5$^{+0.12}_{-0.1}$ $\times$ 10$^{-16}$ \lf) is also small, at $\sim$ 7 per cent.  In this case the observed \oiii\ line contributes $\sim$ 61 per cent of the flux in the \kfil\ filter, where the model prediction is that $\sim$ 65 per cent of the flux in the \kfil\ ifilter comes from \oiii.  The model prediction for LAE27878 provides the worst agreement.  The model prediction is only 0.7$^{+1.9}_{-0.7}$ $\times$ 10$^{-17}$, while the observed line flux in this object is 7 $\pm$ 0.3 $\times$ 10$^{-17}$.  In this object, the bobserved \oiii\ line contributes $\sim$ 100 per cent of the flux in the \kfil\ filter, where the model only predicts a contribution of 10 per cent.  The agreement is not good, but it is worth noting that LAE27878 has the smallest \oiii\ line flux of the three line fluxes we have measured to date, and the model correspondingly assigns the smallest predicted line flux of the three to this object as well.

As for the three LAEs in which we detected no \oiii\ line flux (LAE14310, LAE6559, and LAE27910), our models predict very little \oiii\ emission (2.1$^{+1.2}_{-2.1}$ $\times$ 10$^{-17}$, 2.3$^{+0.7}_{-2.3}$ $\times$ 10$^{-17}$, 6.0$^{+1.2}_{-3.0}$ $\times$ 10$^{-17}$ \lf, respectively).  This corresponds to predicted contributions to the flux in the \kfil\ filter of $\sim$ 34 per cent, 50 per cent, and 50 per cent, respectively).  While the agreement between observations and predictions does not initially seem very good for these objects, we note that the 68 per cent confidence ranges for the \oiii\ line flux predictions (reported in Tables \ref{sedtbl1} and \ref{sedtbl2}) in LAE14310 and LAE6559 include zero (i.e. no \oiii\ line flux). Additionally, while these objects did not have \oiii\ detections in our LUCIFER or NIRSPEC data, the predicted line fluxes are quite modest. The predicted fluxes for LAE14310 and LAE6559 are the fourth and fifth faintest predicted line fluxes among the 25 models with predicted line flux $\ne$ 0.  We derive a 3$\sigma$ line flux limit from the 28 minute LUCIFER spectrum of LAE6559 of $\sim$ 1.4 $\times$ 10$^{-16}$ \lf.  So the predicted model line flux of 2.3 $\times$ 10$^{-17}$ \lf\ is well below what we would have been able to observe in this object.  Given that this same object was also observed with NIRSPEC using a similar 30 minute integration which also yielded no detection, we argue that this upper limit should also approximate the upper limit for LAE27910, which was also observed for 30 minutes with NIRSPEC and which also yielded no detection.  Comparison of the model prediction for LAE27910 (6.0 $\times$ 10$^{-17}$ \lf) and this approximate upper limit (1.4 $\times$ 10$^{-16}$ \lf) once again shows that even if the galaxy produced the predicted \oiii\ flux we would see it as a non-detection given our modest integration time.  For LAE14310, which had a nosier NIR spectrum, we derive a 3$\sigma$ upper limit of $\sim$ 2.8  $\times$ 10$^{-16}$ \lf, again well above the line flux predicted for this object of 2.1 $\times$ 10$^{-17}$ \lf. Most importantly these upper limits tell us there is really no big disagreement between our observed non-detections and our model predictions of a very faint \oiii\ line.  

To compute the 3$\sigma$ line flux upper limits quoted above, we added and recovered mock Gaussian emission lines following the procedure described in Section \ref{sec:detoiii}. For the three \oiii\ line flux upper limits calculated here we fixed the sigma of the Gaussian to 5.52\AA, or the $\sigma$ from our faintest \oiii\ detection (LAE27878).   Because it is impossible to know a priori exactly how much the \lya\ line is offset from the \oiii\ line, we had to repeat these calculations, fixing the mock line at different wavelengths to recreate different velocity offsets.  We found the 3$\sigma$ line flux detection limit at 11 different wavelengths for each object, corresponding to velocity offsets of 0 -- 500 \kms, in increments of 50 \kms.  This range of velocity offsets was chosen to encompass the magnitude of \lya\ - \oiii\ velocity offsets we have observed of 52 - 342 \kms.  The 3$\sigma$ line flux detection limits at each of these 11 locations were then averaged to give an approximate upper limit for the entire wavelength range. 

We contend that in light of the discussion put forth above, the SED modeling discussed in this paper has done a reasonable job of matching our observations, but there is still room for improvement.  It is possible that attributing some of the model line flux to the H$\beta$ line, instead of solely to the \oiii\ would provide an even better match between the observed line fluxes and observed \oiii\ line fluxes.  This can be explored in future work and is beyond the scope of this paper. We also assert that additional spectroscopic observations of LAEs in the NIR are needed, yielding both detections and non-detections, to better quantify exactly how successful this approach can be, beyond what we can say with a sample of only six LAEs with \oiii\ detections/non-detections.  Perhaps most importantly, the predictions of \oiii\ flux that we have made from the new SED fitting approach in this paper should allow us to select the LAEs that are mostly likely to yield \oiii\ detections in future NIR spectroscopic observations.  Based on our comparisons of predicted \oiii\ line fluxes to observed line fluxes in the three objects that had \oiii\ detections - it seems likely that objects with strong \oiii\ line fluxes predicted would be our best bet for NIR followup observations. This is a testable hypothesis and should allow us to more efficiently use telescope time and more carefully plan appropriate integration times for each object.   

\subsection{Effects of Including \oiii\ Emission}
As has been pointed out by \citet{sd09} and others, inclusion or exclusion of nebular emission lines during SED fitting can significantly alter the results obtained, specifically masses and ages.  To investigate how our additional \oiii\ parameter affects our best fit solutions, we compare the best allowed-fit solutions with and without \oiii\ emission.  We focus our discussion here on the three objects for which we have \oiii\ measurements, and repeat the same fitting procedure described above, but with the \oiii\ line flux contribution to the \kfil\ band fixed to zero.  Unsurprisingly, the object most affected by removing the \oiii\ parameter is LAE40844.  This is unsurprising as this was the LAE with the largest of the three observed \oiii\ fluxes, and was also fit with largest \oiii\ flux solution among the entire LAE sample.  For LAE40844, the best allowed-fit mass increases from 2.2$^{+0.0}_{-0.8}$ $\times$ 10$^{9}$ \msol\ (\oiii\ included) to 3.5$^{+0.05}_{-0.2}$ $\times$ 10$^{9}$ \msol when \oiii\ emission is not included.  So the best allowed-fit mass solution in this object increases 1.6 times when \oiii\ is not properly accounted for.  Perhaps most tellingly, the reduced $\chi^2$ value increases from 9.1 to 145.9 when the \oiii\ contribution is removed, indicating that the fit without an \oiii\ contribution is quite poor. This increase in mass is in excellent agreement with those reported in \citet{sd09}.  Figure \ref{fig:compare} illustrates the difference between the models when  \oiii\ flux is and is not included.

\begin{figure*}
\centering
\begin{tabular}{c}
\includegraphics[scale=0.4]{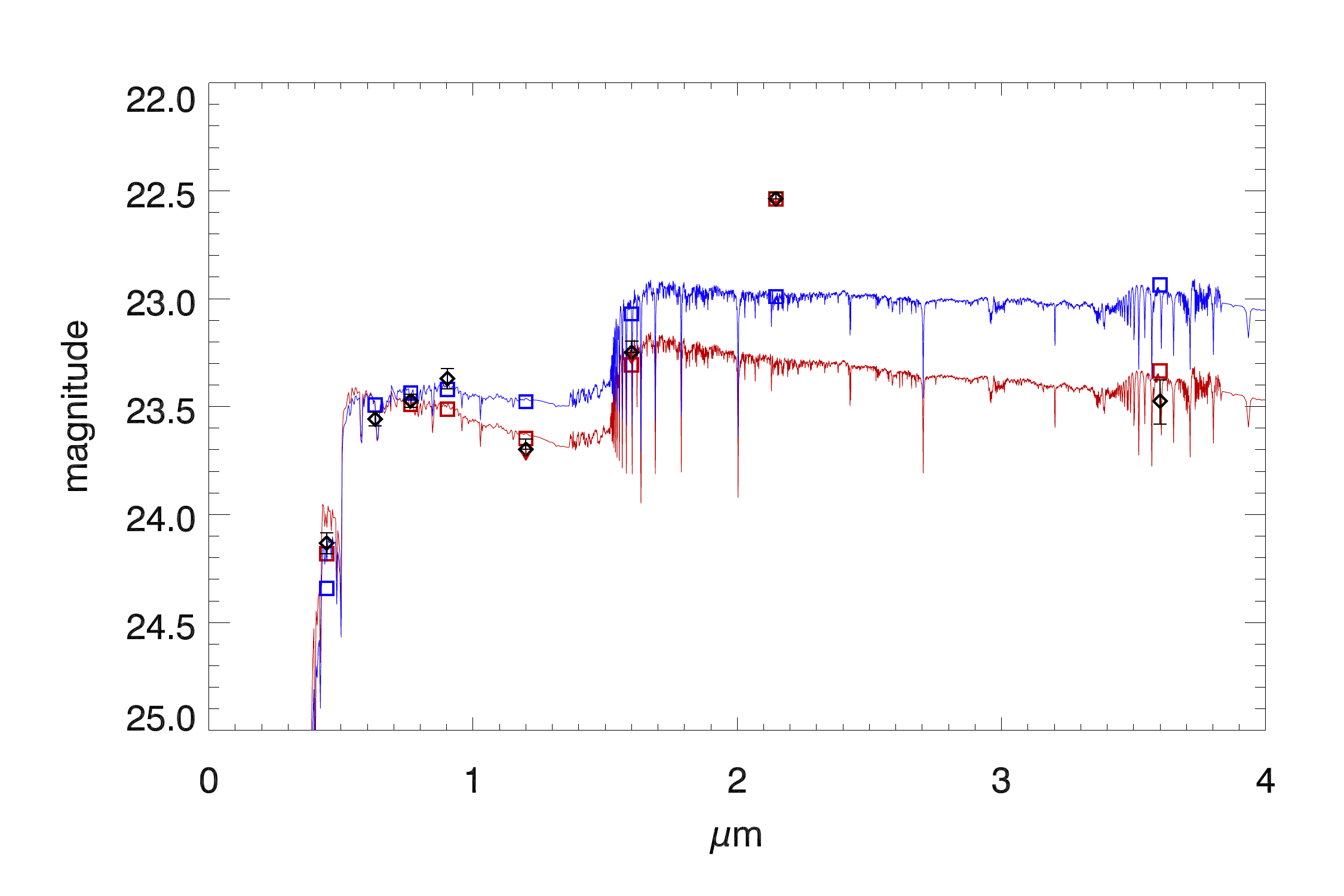}
\end{tabular}
\caption{Observed magnitudes are in black. The best allowed-fit solution with an \oiii\ contribution is shown with red spectrum and red squares, the best fit solution with with no \oiii\ line flux is shown with blue spectrum and blue squares.  In addition to yielding a more massive solution, the blue spectrum is a much poorer fit.}\label{fig:compare}
\end{figure*}
LAE27878 and LAE7745 are the other two objects with measured \oiii\ fluxes.  LAE27878 has a very small best allowed-fit \oiii\ solution, and the results are indistinguishable when \oiii\ is fixed at 0.  The case of LAE7745 is not quite as clear as those of LAE27878 and LAE40844.  LAE7745 has a negative best allowed-fit tau parameter when \oiii\ flux is considered.  It has not been previously investigated how age and mass solutions behave when you exclude nebular emission in objects fit with negative tau value. For this object, we find that when  \oiii\ flux is fixed to zero, the best-allowed fit solution instead chooses a positive tau.  Subsequently, the best allowed-fit age and mass in this object decrease, rather than increase (in all comparisons in this section we are comparing model ages, not star formation weighted ages which are dependent on tau).  But, in spite of these decreases, the reduced $\chi^2$ value still increases significantly when \oiii\ is excluded from 12.6 to 33.2.  

We present the overall trends for the changes in age, mass and reduced $\chi^2$ for the entire sample of 33 LAEs in Figure \ref{fig:sedcomp} when \oiii\ flux is and is not included in the fitting process.  The histograms in Figure \ref{fig:sedcomp} present the per cent difference between the solution with \oiii\ and the solution without \oiii.   In all three histograms a positive per cent difference means the solution without \oiii\ was larger, a negative per cent difference means the solution with \oiii\ was larger.  The two most definitive trends are seen in the histograms for mass and reduced $\chi^2$.  Overall, the solutions without \oiii\ are on average more massive, as seen by the fact that most of the per cent differences in this panel are positive.  19 solutions become more massive, 8 stay the same, and only 6 get less massive.  The  reduced $\chi^2$ results are even more clear, every $\chi^2$ value gets larger or stays the same, none get smaller.  More precisely, 25 solutions get bigger and 8 stay the same.  The trend in the age results (model ages, not star formation weighted ages) is not quite as definitive, 16 solutions do not change, while 6 get older and 11 get younger.  So overall we can say that when \oiii\ contributions to the \kfil\ band are not included, the sample becomes more massive and less well fit.  The magnitude of these effects, however, can vary significantly from object to object within a sample.  

\begin{figure*}
\centering
\begin{tabular}{c}
\includegraphics[scale=0.35]{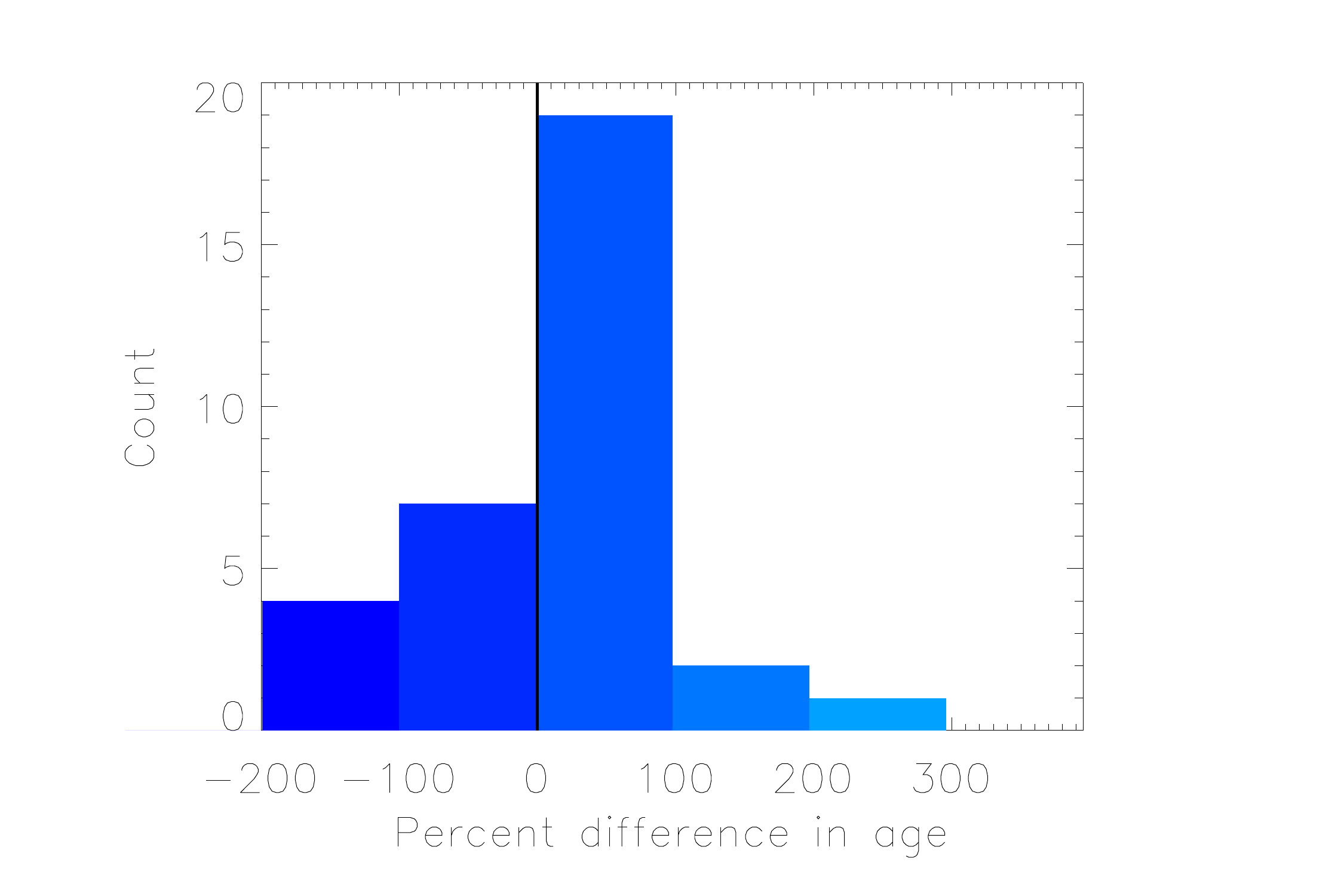}\\
\includegraphics[scale=0.35]{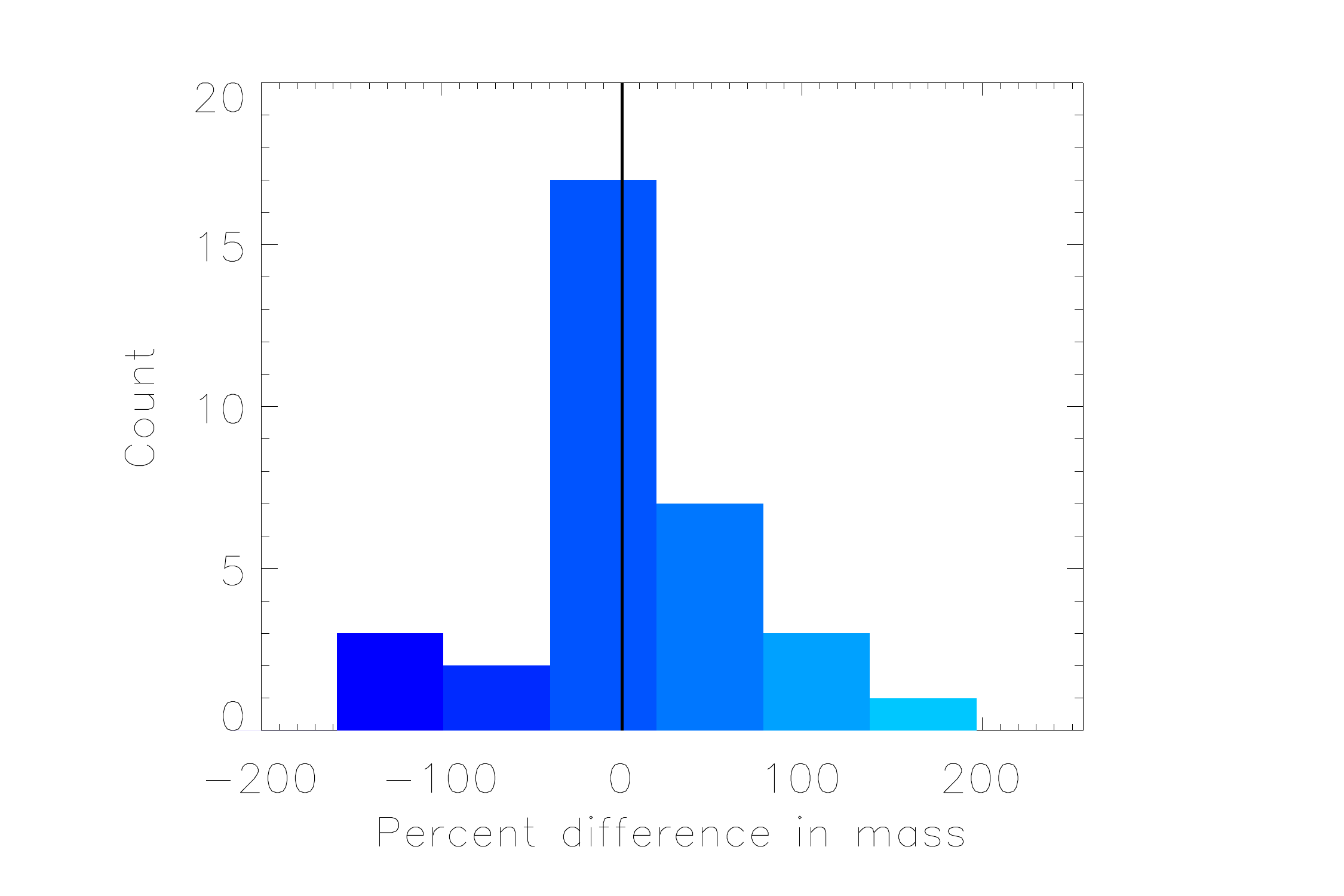}\\
\includegraphics[scale=0.35]{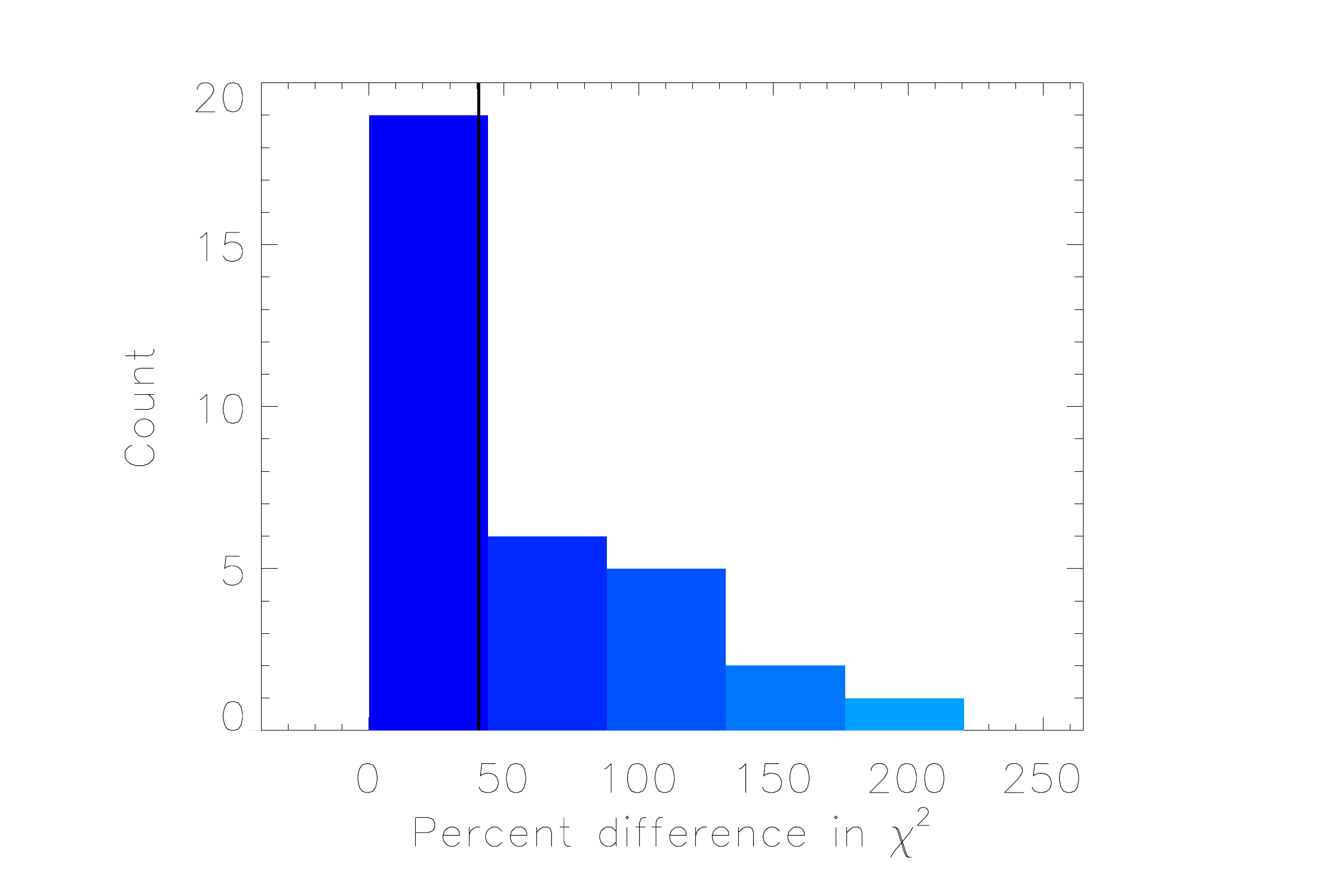}\\
\end{tabular}
\caption{Per Cent difference between best allowed-fit solutions in full LAE sample, when \oiii\ line flux contributions are and are not included in the fitting process. A positive per cent difference means the solution without \oiii\ was larger, a negative per cent difference means the solution with \oiii\ was larger.  Black vertical line indicates median per cent difference for sample.  Overall, when \oiii\ contributions to the \kfil\ band are not included, the sample becomes more massive and less well fit, even when increasing star formation histories are allowed with some less definitive changes in ages be expected as well - see text for further details}\label{fig:sedcomp}
\end{figure*}

\subsection{Comparison of Physical Characteristics to Other Samples}\label{sec:comp2}
Table \ref{comp} shows best-fit age and mass results from the majority of recent papers on SED fitting of LAEs from z $\sim$ 0.3 -- 6.6, including the results of this paper.  We note which models were used in each paper and whether nebular emission lines were included.  The reader should also consider that star formation histories and metallicities are sometimes treated differently from paper to paper (i.e. in some cases these are fixed parameters, in others they are free).  Focusing specifically on the z $\sim$ 3.1 samples detailed in Table \ref{comp}, we find the results vary substantially from sample to sample.  We find that our sample of 33 individually fit LAEs has, on average, a systematically more massive solution than all the stacked samples at z $\sim$ 3.1, even in the samples where nebular emission lines were not treated during the fitting process. 

There are a number of systematic differences between the samples that may indicate that our results do not necessarily contradict the other works to which we are comparing, but rather we may be probing different subsamples of LAEs. For instance, while our LAEs are spectroscopically confirmed, our selection criteria (Section \ref{sec:select} in some cases differ substantially from other authors and this may contribute to some of the differences in derived physical characteristics we have observed.  Also, as we alluded to earlier, given the wide-field and correspondingly shallow nature of our narrowband survey, we have selected a subset of bright LAEs, brighter than many surveys to which we can compare in  Table \ref{comp}.  L$^{\ast}$ for z $\sim$ 3.1 LAEs is $\sim$ 5.75 $\times$ 10$^{42}$ erg s$^{-1}$ \citep{cia12}.  The majority of our sample is above this luminosity, as illustrated in Figure \ref{fig:lstar}, where our average L$_{\lya}$ luminosity is $\sim$ 1.50 $\times$ 10$^{43}$ erg s$^{-1}$.  This is in contrast, for instance, to the z $\sim$ 2.1 and 3.1 LAEs selected from the deep MUSYC survey \citep{gaw07,lai08,gua11} where the area surveyed was much smaller but the 5$\sigma$ narrowband depth reached magnitudes of 25.4 and 25.1 for z $\sim$ 3.1 and z $\sim$ 2.1, respectively.  We have analyzed the effect of L$_{\lya}$ on the SED-derived masses in Figure \ref{fig:lstar} in a subset of samples from z $\sim$ 0.3 -- 3.1 from Table \ref{comp} where L$_{\lya}$ information readily available.  L$^{\ast}$ for z $\sim$ 0.3 is taken from \citet{cow10}, and L$^{\ast}$ at z $\sim$ 2.1 comes from \citet{cia12}.  This preliminary analysis indicates that individually fit LAEs have larger masses than the masses derived from stacked analysis.  Also the stacked LAEs from the deeper MUSYC data have smaller masses than those LAEs in our wide-field survey.  

\begin{figure*}
\begin{tabular}{ll}
\includegraphics[bb = 100 0 548 432,scale=0.45]{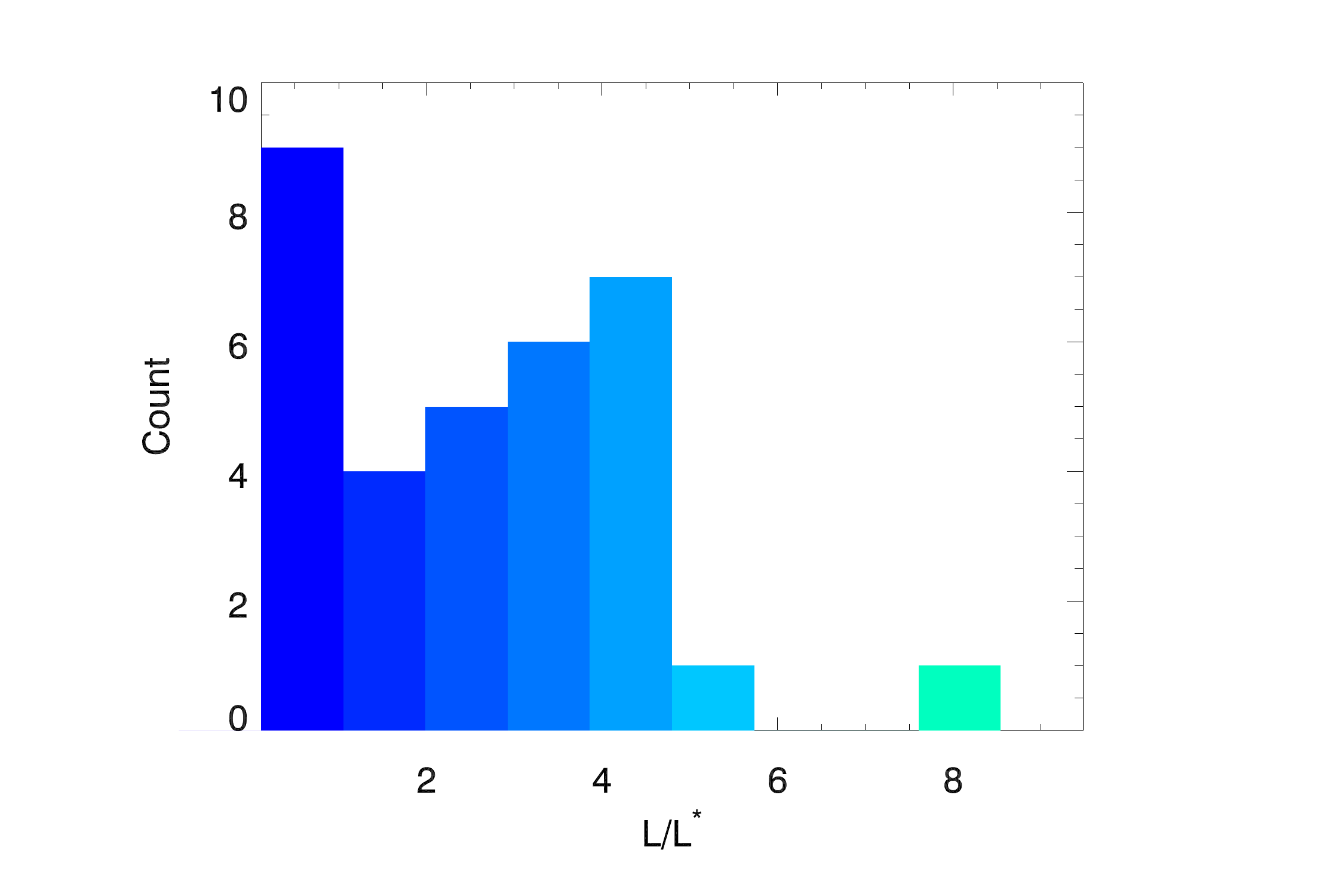} & \includegraphics[bb=50 0 548 432,scale=0.45]{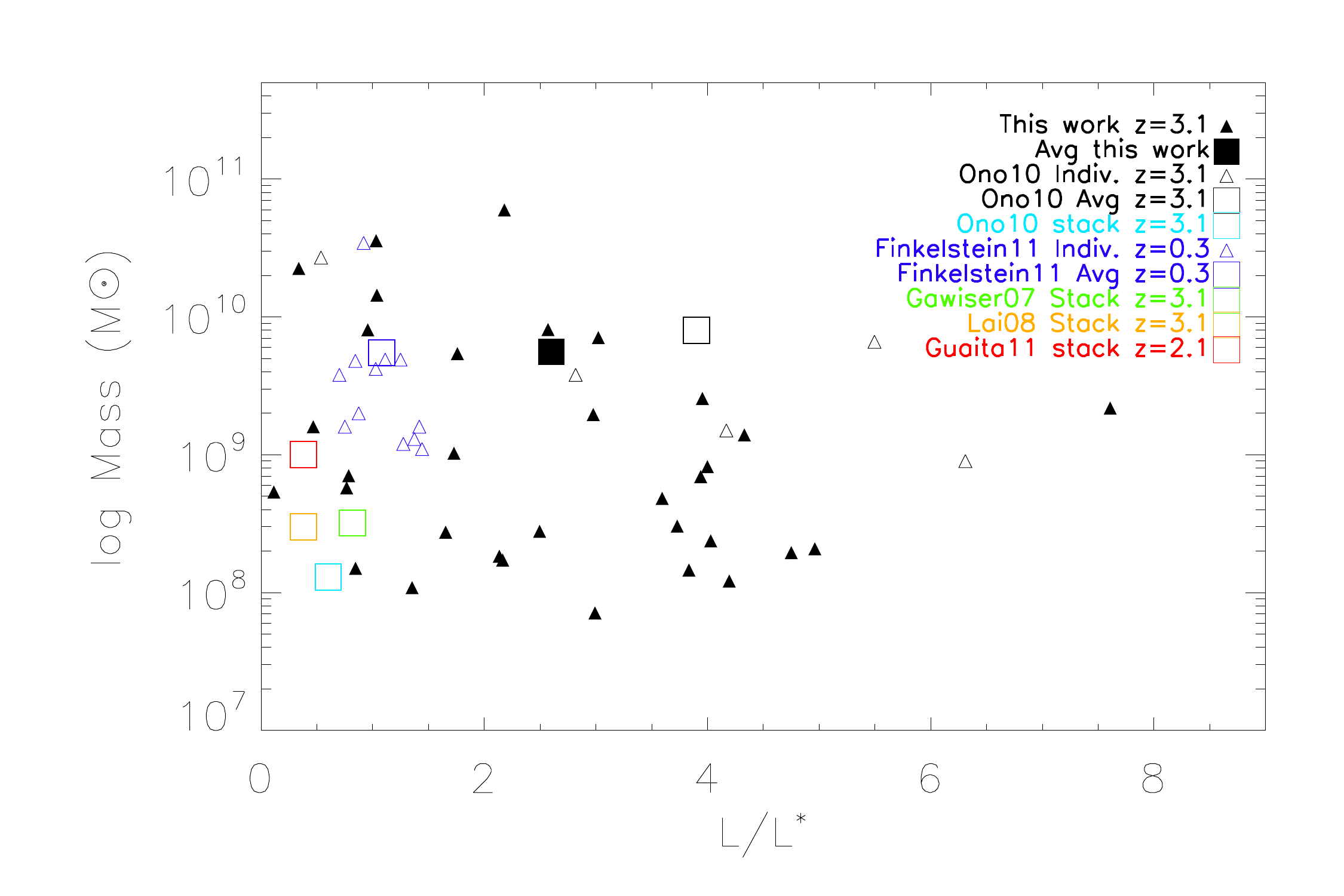}\\
\end{tabular}
\caption{Left shows histogram of L/L$^{\ast}$, where L$^{\ast}$ is from \citet{cia12}. Right shows derived masses as a function of the ratio of L/L$^{\ast}$. Individual results from this paper are shown as small black triangles, an average value from this work is indicated with a large filled black square.  12 individual z $\sim$ 0.3 LAEs from \citet{fink11b} are shown as blue triangles, a large blue square indicates the average value from this sample.  Small open black triangles are 5 individual z $\sim$ 3.1 LAEs from \citet{ono10a} and the large open black square shows the average value of this sample.  The large cyan square is the result from the 200 stacked z $\sim$ LAEs in \citet{ono10a}.  The green and orange squares are the same stack of 52 LAEs at z $\sim$ 3.1 from \citet{lai08,gaw07} fit with different star formation histories.  The red square is the stack of 216 z $\sim$ 2.1 LAEs from \citet{gua11}.}
\label{fig:lstar}
\end{figure*}

We cannot make similarly broad statements about any systematic offset when comparing our age results to the stacked age results at z $\sim$ 3.1. \citet{acq12} have an older average while \citet{gaw06, gaw07} have younger solutions (where both authors stack their samples) than our average from individually fit LAEs.   \citet{nil07} find a solution more than five times older than our average, with constant star formation assumed, but various metallicities allowed.  The \citet{acq12} fitting procedure assumes constant star formation, but metallicity is allowed to vary; the \citet{gaw06} sample was also fit with a constant star formation but with metallicity fixed to solar. \citet{gaw07} used a two-burst scenario for their star formation history, and metallicity was allowed to vary.  \citet{lai08} on the other hand, in spite of being a stacked sample, with no treatment of nebular emission lines and assuming constant star formation and solar metallicity, finds a very similar average age of 160 Myr compared to our average of 151 Myr.  We note, however, that our full range of age solutions (1.1 -- 1100 Myr) does encompass the all the average ages put forth by other authors for their stacked samples.  We acknowledge that the variety of methods used by different authors can make direct comparison somewhat difficult, but it is worth trying to catalog the various results and compare to the extent we are able.

\cite{ono10a} presents the only other sample of individually fit LAEs at z $\sim$ 3.1, albeit in a sample of only five objects, to which we can compare.  The fitting procedure of \citet{ono10a} includes an assumed metallicity of Z = 0.2 \zsol\, the star formation history can be constant or decreasing exponentially, and no treatment of nebular lines is included.  In spite of these differences we find good agreement between their ranges for both mass and age and those that we have presented for our sample.  They find (as we do) a large range of ages, 4.8--407 Myr (1.5 -- 1800 Myr), and masses, 9.3$\times$10$^{8}$ -- 2.7$\times$10$^{10}$ \msol\ (7.1$\times$10$^{7}$ -- 6$\times$10$^{10}$ \msol).  The fact that this individually fit sample is the only one that matches our results well may lend further credence to the idea that stacked analyses may not be capturing the diversity that we have found in the LAE population at this redshift.  While \citet{var13} find a moderately diverse sample of properties in their 20 individually fit z $\sim$ 2.1 LAEs, the spread we find in both our age and mass results are significantly larger.   We also note that there is broad agreement between our individually fit LAEs at z $\sim$ 3.1 and the 40 individually fit LAEs of \citet{cow11} at z $\sim$ 0.3 and 12 individually fit LAEs from \citet{fink11b} .  The age and mass spread of the samples is quite similar, except, of course, the fact that there are older possible ages allowed for galaxies in the z $\sim$ 0.3 universe compared to the z $\sim$ 3.1 universe.  Such agreement between samples far removed from one another in cosmic time could suggest that \lya\ selection techniques are capturing similar objects, at similar states of evolution, regardless of the redshift sampled.  The broad agreement may also be a result of the similar L$_{\lya}$ space probed by the z $\sim$ 0.3 sample and our sample, as the z $\sim$ 0.3 sample has a rather large average  L/L$^{\ast}$ value of $\sim$ 1.6 (using L$^{\ast}$ from \citet{cow10}).

\begin{table*}
\begin{tabular}{|l|l|l|l|l|l|}
Author & Redshift & Sample & Models & Neb. Em.$^{1}$ & Results\\
\hline
Acquaviva 2011 & z $\sim$ 2.1 & 216 stacked & CB11 & yes & 50 Myr, 3$\times$10$^8$ \msol\\
                      & z $\sim$ 3.1 & 70 stacked & CB11 & yes & 1000 Myr, 1.5$\times$10$^9$ \msol, \\
Cowie 2011 & z $\sim$ 0.3 & 40 individual & BC03 & yes & 10--10000 Myr, 10$^7$--10$^{11}$ \msol\\
Finkelstein 2007 & z $\sim$ 4.5 & 98 stacked$^{2}$ & BC03 & \lya\ & 1--40 Myr, 0.68--16.2$\times$10$^8$ \msol\\
Finkelstein 2009 & z $\sim$ 4.5  & 14 individual & BC03 &  \lya, H$\alpha$ & 3-500 Myr, 1.6 $\times$10$^8$--5.0$\times$10$^{10}$ \msol \\
Finkelstein 2011 & z $\sim$ 0.3 & 12 individual &  BC07 & yes & 60 -- 9000 Myr, 1.1 $\times$10$^9$ -- 3.4 $\times$ 10$^{10}$ \msol \\
Gawiser 2006 & z $\sim$ 3.1 & 40 stacked & BC03 & no & 90 Myr, 5$\times$10$^8$ \msol \\
Gawiser 2007 & z $\sim$ 3.1 & 52 stacked & BC03 & no & 20 Myr, 1$\times$10$^9$ \msol \\
Guaita 2011 & z $\sim$ 2.1 & 216 stacked & CB10 & no & 10 Myr, 3.2$\times$10$^8$ \msol \\
Lai 2008 & z $\sim$ 3.1 & 76 stacked & BC03 & only \lya & 160 Myr, 3$\times$10$^8$ \msol \\
McLinden 2014 (this paper) & z $\sim$ 3.1 & 33 individual & CB11 & yes, see Sec. \ref{sec:sed} & 1.5-1800 Myr, 7.1$\times$10$^{7}$ -- 6$\times$10$^{10}$ \msol \\ 
Nakajima 2012$^{5}$ & z $\sim$ 2.2 & 304 stacked & BC03 & yes & 12.6 Myr, 3$\times$10$^8$ \msol \\
                                                  & z $\sim$ 2.2 & 55 stacked & BC03 & yes & 8.3 Myr, 5$\times$10$^8$ \msol \\
Nilsson 2007 & z $\sim$ 3.15 & 23 stacked & BC03 & no & 830 Myr$^{3}$, 8$\times$10$^8$ \msol \\
Nilsson 2011 & z $\sim$ 2.3 & 40 stacked & NisseFit$^{4}$ & yes & 440 Myr, 2.5$\times$10$^{10}$ \msol \\
                    & z $\sim$ 2.3  & 40 individual &  NisseFit$^{4}$ & yes & 1000 Myr, 1.7$\times$10$^{10}$ \msol \\
Ono 2010a & z $\sim$ 3.1 & 200 stacked & BC03 & no & 65 Myr, 1.3$\times$10$^8$ \msol \\
                & z $\sim$ 3.1 & 5 individual & BC03 & no & 4.8--407 Myr , 0.93--27$\times$10$^9$ \msol \\
                & z $\sim$ 3.7 & 61 stacked & BC03 & no & 5.8 Myr, 3.2$\times$10$^8$ \msol \\
                & z $\sim$ 3.7 & 6 individual & BC03 & no & 1.4--900 Myr, 3.9--51$\times$10$^9$ \msol \\
Ono 2010b & z $\sim$ 5.7 & 165 stacked & BC03 & yes & 3 Myr, 3$\times$10$^7$ \msol \\
                & z $\sim$ 6.6 & 91 stacked & BC03 & yes & 1 Myr, 1$\times$10$^8$ \msol \\
Pirzkal 2007 & z $\sim$ 4--5.7 & 9 individual & BC03 & no & 0.5 -- 20 Myr, 5$\times$10$^6$ -- 18$\times$10$^8$ \msol \\
Vargas 2013 & z $\sim$ 2.1 & 20 individual & BC03 & yes & 4 -- 470 Myr, 2.3$\times$10$^7$ -- 8.5 $\times$10$^9$ \msol\\ 
\hline
\multicolumn{6}{l}{\textsuperscript{1}\footnotesize{Was nebular emission accounted for?}}\\
\multicolumn{6}{l}{\textsuperscript{2}\footnotesize{Divided into 6 subsamples}}\\
\multicolumn{6}{l}{\textsuperscript{3}\footnotesize{Author notes this is poorly constrained}}\\
\multicolumn{6}{l}{\textsuperscript{4}\footnotesize{Based on BC03}}\\
\multicolumn{6}{l}{\textsuperscript{5}\footnotesize{Two different stacks for two different fields at z $\sim$ 2.2}}\\
\end{tabular}
\caption{Comparison to SED fitting in the literature.  Note that ages presented here are direct model ages, not star formation weighted ages for easier comparison to other samples.}
\label{comp}
\end{table*}

\section{CONCLUSIONS}
We have presented one new \oiii\ detection in a z $\sim$ 3.1 LAE.  Combining this new detection with the two we presented in Mc11, we are able to present a total of three measurements of the velocity offset between \lya\ and \oiii\ in these z $\sim$ 3.1 LAEs, ranging from 52 -- 342 \kms.  This new result is still consistent with the outflow models explored in Mc11.  

In addition to the new \oiii\ detection, we have put forth a simple method to account for nebular emission in high-z starbursting galaxies, motivated by our three \oiii\ measurements.  We have individually fit 33 z $\sim$ 3.1 LAEs using this powerful yet simple method to account for nebular emission line contributions to galaxy SEDs.  From these fits we find constraints on age, mass, dust content, metallicity, star formation history, and \oiii\ line flux.  We find that our sample has quite diverse characteristics, but some generalizations can be made.  For instance, a majority of the galaxies are fit with a single instantaneous burst or exponentially decreasing star formation history.  As a whole, the sample has only moderate amounts of dust, and sub-solar metallicity.  Mass and age solutions vary widely, but median values of 4.5 $\times$ 10$^{6}$ years (full range is 1.4 $\times$ 10$^{6}$ -- 4.6 $\times$ 10$^{8}$ years)  and 6.9 $\times$ 10$^{8}$ \msol\ (full range is 7.1 $\times$ 10$^{7}$ -- 6 $\times$ 10$^{10}$ \msol)  are found.  Finally, most of the galaxies are best fit with an \oiii\ line contributing additional flux to the \kfil\ band, with an average flux of 7.3 $\times$ 10$^{-17}$ \lf\ (or an average of 9.7 $\times$ 10$^{-17}$ \lf\ among the 25 galaxies with non-zero line flux solutions, ranging from 7.0 $\times$ 10$^{-18}$ -- 3.7 $\times$ 10$^{-16}$ \lf).

The \oiii\ line strength predictions from our new SED fitting methodology have reasonably matched the observations of the \oiii\ line in the six objects for which we can make this comparison.   These predictions gives us confidence that these results can be used to select the LAEs mostly likely to yield \oiii\ detections in future NIR observations and aids in planning adequate integration times for the most efficient use of such future NIR observing time.  Further observations of LAEs in the NIR will allow us to fill in the distribution of velocity offsets found in LAEs at this redshift, and will allow us to further test the validity of \oiii\ line strength predictions from our SED fitting process.  In the meantime we have, with this work, provided a comprehensive picture of LAE characteristics in a large sample of individually examined objects.

\section*{Acknowledgments}
We thank NOAO for loaning the KPNO \oiii\ filter for use on the Bok
Telescope.  This work has been supported by NASA (program N067NS),
and by the National Science Foundation through NSF grant AST-0808165.  Additional thanks to Dr. Seth Cohen for many
helpful discussions.

Some of the data presented herein were obtained at the W.M. Keck
Observatory, which is operated as a scientific partnership among the
California Institute of Technology, the University of California and
the National Aeronautics and Space Administration. The Observatory was
made possible by the generous financial support of the W.M. Keck
Foundation.

The LBT is an international collaboration among institutions in the United States, Italy and Germany. LBT Corporation partners are: The University of Arizona on behalf of the Arizona university system; Instituto Nazionale di Astrofisica, Italy; LBT Beteiligungsgesellschaft, Germany, representing the Max-Planck Society, the Astrophysical Institute Potsdam, and Heidelberg University; The Ohio State University, and The Research Corporation, on behalf of The University of Notre Dame, University of Minnesota and University of Virginia.

Observations reported here were obtained at the MMT Observatory, a joint facility of the University of Arizona and the Smithsonian Institution.

The authors wish to recognize and acknowledge the very significant
cultural role and reverence that the summit of Mauna Kea has always
had within the indigenous Hawaiian community.  We are most fortunate
to have the opportunity to conduct observations from this mountain.

\bibliographystyle{mn2e}
\bibliography{mclinden1215}

\begin{thebibliography}{60}
\expandafter\ifx\csname natexlab\endcsname\relax\def\natexlab#1{#1}\fi

\bibitem[{{Acquaviva} {et~al}\mbox{.}(2012){Acquaviva}, {Vargas}, {Gawiser}, \&
  {Guaita}}]{acq12}
{Acquaviva} V., {Vargas} C., {Gawiser} E., {Guaita} L., 2012, \apjl, 751, L26

\bibitem[{{Ageorges} {et~al}\mbox{.}(2010){Ageorges}, {Seifert}, {J{\"u}tte},
  {Knierim}, {Lehmitz}, {Germeroth}, {Buschkamp}, {Polsterer}, {Pasquali},
  {Naranjo}, {Gemperlein}, {Hill}, {Feiz}, {Hofmann}, {Laun}, {Lederer},
  {Lenzen}, {Mall}, {Mandel}, {M{\"u}ller}, {Quirrenbach}, {Sch{\"a}ffner},
  {Storz}, \& {Weiser}}]{ag10}
{Ageorges} N. {et~al.}, 2010, in Society of Photo-Optical Instrumentation
  Engineers (SPIE) Conference Series, Vol. 7735, Society of Photo-Optical
  Instrumentation Engineers (SPIE) Conference Series

\bibitem[{{Becker} {et~al}\mbox{.}(2006){Becker}, {Sargent}, {Rauch}, \&
  {Simcoe}}]{bec06}
{Becker} G.~D., {Sargent} W.~L.~W., {Rauch} M., {Simcoe} R.~A., 2006, \apj,
  640, 69

\bibitem[{{Boulade} {et~al}\mbox{.}(2003){Boulade}, {Charlot}, {Abbon}, {Aune},
  {Borgeaud}, {Carton}, {Carty}, {Da Costa}, {Deschamps}, {Desforge},
  {Eppell{\'e}}, {Gallais}, {Gosset}, {Granelli}, {Gros}, {de Kat}, {Loiseau},
  {Ritou}, {Rouss{\'e}}, {Starzynski}, {Vignal}, \& {Vigroux}}]{bo03}
{Boulade} O. {et~al.}, 2003, in Society of Photo-Optical Instrumentation
  Engineers (SPIE) Conference Series, Vol. 4841, Society of Photo-Optical
  Instrumentation Engineers (SPIE) Conference Series, {Iye} M., {Moorwood}
  A.~F.~M., eds., pp. 72--81

\bibitem[{{Capak} {et~al}\mbox{.}(2007){Capak}, {Aussel}, {Ajiki}, {McCracken},
  {Mobasher}, {Scoville}, {Shopbell}, {Taniguchi}, {Thompson}, {Tribiano},
  {Sasaki}, {Blain}, {Brusa}, {Carilli}, {Comastri}, {Carollo}, {Cassata},
  {Colbert}, {Ellis}, {Elvis}, {Giavalisco}, {Green}, {Guzzo}, {Hasinger},
  {Ilbert}, {Impey}, {Jahnke}, {Kartaltepe}, {Kneib}, {Koda}, {Koekemoer},
  {Komiyama}, {Leauthaud}, {Le Fevre}, {Lilly}, {Liu}, {Massey}, {Miyazaki},
  {Murayama}, {Nagao}, {Peacock}, {Pickles}, {Porciani}, {Renzini}, {Rhodes},
  {Rich}, {Salvato}, {Sanders}, {Scarlata}, {Schiminovich}, {Schinnerer},
  {Scodeggio}, {Sheth}, {Shioya}, {Tasca}, {Taylor}, {Yan}, \&
  {Zamorani}}]{cap07}
{Capak} P. {et~al.}, 2007, \apjs, 172, 99

\bibitem[{{Ciardullo} {et~al}\mbox{.}(2012){Ciardullo}, {Gronwall}, {Wolf},
  {McCathran}, {Bond}, {Gawiser}, {Guaita}, {Feldmeier}, {Treister}, {Padilla},
  {Francke}, {Matkovi{\'c}}, {Altmann}, \& {Herrera}}]{cia12}
{Ciardullo} R. {et~al.}, 2012, \apj, 744, 110

\bibitem[{{Cowie}, {Barger} \& {Hu}(2010){Cowie}, {Barger}, \& {Hu}}]{cow10}
{Cowie} L.~L., {Barger} A.~J., {Hu} E.~M., 2010, \apj, 711, 928

\bibitem[{{Cowie}, {Barger} \& {Hu}(2011){Cowie}, {Barger}, \& {Hu}}]{cow11}
{Cowie} L.~L., {Barger} A.~J., {Hu} E.~M., 2011, \apj, 738, 136

\bibitem[{{Dawson} {et~al}\mbox{.}(2004){Dawson}, {Rhoads}, {Malhotra},
  {Stern}, {Dey}, {Spinrad}, {Jannuzi}, {Wang}, \& {Landes}}]{daw04}
{Dawson} S. {et~al.}, 2004, \apj, 617, 707

\bibitem[{{Elvis} {et~al}\mbox{.}(2009){Elvis}, {Civano}, {Vignali},
  {Puccetti}, {Fiore}, {Cappelluti}, {Aldcroft}, {Fruscione}, {Zamorani},
  {Comastri}, {Brusa}, {Gilli}, {Miyaji}, {Damiani}, {Koekemoer}, {Finoguenov},
  {Brunner}, {Urry}, {Silverman}, {Mainieri}, {Hasinger}, {Griffiths},
  {Carollo}, {Hao}, {Guzzo}, {Blain}, {Calzetti}, {Carilli}, {Capak}, {Ettori},
  {Fabbiano}, {Impey}, {Lilly}, {Mobasher}, {Rich}, {Salvato}, {Sanders},
  {Schinnerer}, {Scoville}, {Shopbell}, {Taylor}, {Taniguchi}, \&
  {Volonteri}}]{elv10}
{Elvis} M. {et~al.}, 2009, \apjs, 184, 158

\bibitem[{{Fabricant} {et~al}\mbox{.}(2005){Fabricant}, {Fata}, {Roll},
  {Hertz}, {Caldwell}, {Gauron}, {Geary}, {McLeod}, {Szentgyorgyi}, {Zajac},
  {Kurtz}, {Barberis}, {Bergner}, {Brown}, {Conroy}, {Eng}, {Geller},
  {Goddard}, {Honsa}, {Mueller}, {Mink}, {Ordway}, {Tokarz}, {Woods}, {Wyatt},
  {Epps}, \& {Dell'Antonio}}]{fab05}
{Fabricant} D. {et~al.}, 2005, \pasp, 117, 1411

\bibitem[{{Finkelstein} {et~al}\mbox{.}(2011{\natexlab{a}}){Finkelstein},
  {Cohen}, {Moustakas}, {Malhotra}, {Rhoads}, \& {Papovich}}]{fink11b}
{Finkelstein} S.~L., {Cohen} S.~H., {Moustakas} J., {Malhotra} S., {Rhoads}
  J.~E., {Papovich} C., 2011{\natexlab{a}}, \apj, 733, 117

\bibitem[{{Finkelstein} {et~al}\mbox{.}(2011{\natexlab{b}}){Finkelstein},
  {Hill}, {Gebhardt}, {Adams}, {Blanc}, {Papovich}, {Ciardullo}, {Drory},
  {Gawiser}, {Gronwall}, {Schneider}, \& {Tran}}]{fink11a}
{Finkelstein} S.~L. {et~al.}, 2011{\natexlab{b}}, \apj, 729, 140

\bibitem[{{Finkelstein} {et~al}\mbox{.}(2009){Finkelstein}, {Rhoads},
  {Malhotra}, \& {Grogin}}]{fink09}
{Finkelstein} S.~L., {Rhoads} J.~E., {Malhotra} S., {Grogin} N., 2009, \apj,
  691, 465

\bibitem[{{Finkelstein} {et~al}\mbox{.}(2008){Finkelstein}, {Rhoads},
  {Malhotra}, {Grogin}, \& {Wang}}]{fink08}
{Finkelstein} S.~L., {Rhoads} J.~E., {Malhotra} S., {Grogin} N., {Wang} J.,
  2008, \apj, 678, 655

\bibitem[{{Finkelstein} {et~al}\mbox{.}(2007){Finkelstein}, {Rhoads},
  {Malhotra}, {Pirzkal}, \& {Wang}}]{fink07}
{Finkelstein} S.~L., {Rhoads} J.~E., {Malhotra} S., {Pirzkal} N., {Wang} J.,
  2007, \apj, 660, 1023

\bibitem[{{Finlator}, {Oppenheimer} \& {Dav{\'e}}(2011){Finlator},
  {Oppenheimer}, \& {Dav{\'e}}}]{fin11}
{Finlator} K., {Oppenheimer} B.~D., {Dav{\'e}} R., 2011, \mnras, 410, 1703

\bibitem[{{Fukugita} {et~al}\mbox{.}(2011){Fukugita}, {Yasuda}, {Doi}, {Gunn},
  \& {York}}]{fuk11}
{Fukugita} M., {Yasuda} N., {Doi} M., {Gunn} J.~E., {York} D.~G., 2011, \aj,
  141, 47

\bibitem[{{Gawiser} {et~al}\mbox{.}(2007){Gawiser}, {Francke}, {Lai},
  {Schawinski}, {Gronwall}, {Ciardullo}, {Quadri}, {Orsi}, {Barrientos},
  {Blanc}, {Fazio}, {Feldmeier}, {Huang}, {Infante}, {Lira}, {Padilla},
  {Taylor}, {Treister}, {Urry}, {van Dokkum}, \& {Virani}}]{gaw07}
{Gawiser} E. {et~al.}, 2007, \apj, 671, 278

\bibitem[{{Gawiser} {et~al}\mbox{.}(2006){Gawiser}, {van Dokkum}, {Gronwall},
  {Ciardullo}, {Blanc}, {Castander}, {Feldmeier}, {Francke}, {Franx},
  {Haberzettl}, {Herrera}, {Hickey}, {Infante}, {Lira}, {Maza}, {Quadri},
  {Richardson}, {Schawinski}, {Schirmer}, {Taylor}, {Treister}, {Urry}, \&
  {Virani}}]{gaw06}
{Gawiser} E. {et~al.}, 2006, \apjl, 642, L13

\bibitem[{{Guaita} {et~al}\mbox{.}(2011){Guaita}, {Acquaviva}, {Padilla},
  {Gawiser}, {Bond}, {Ciardullo}, {Treister}, {Kurczynski}, {Gronwall}, {Lira},
  \& {Schawinski}}]{gua11}
{Guaita} L. {et~al.}, 2011, \apj, 733, 114

\bibitem[{{Kashikawa} {et~al}\mbox{.}(2006){Kashikawa}, {Shimasaku}, {Malkan},
  {Doi}, {Matsuda}, {Ouchi}, {Taniguchi}, {Ly}, {Nagao}, {Iye}, {Motohara},
  {Murayama}, {Murozono}, {Nariai}, {Ohta}, {Okamura}, {Sasaki}, {Shioya}, \&
  {Umemura}}]{kash06}
{Kashikawa} N. {et~al.}, 2006, \apj, 648, 7

\bibitem[{{Kelson}(2003)}]{kel03}
{Kelson} D.~D., 2003, \pasp, 115, 688

\bibitem[{{Koekemoer} {et~al}\mbox{.}(2007){Koekemoer}, {Aussel}, {Calzetti},
  {Capak}, {Giavalisco}, {Kneib}, {Leauthaud}, {Le F{\`e}vre}, {McCracken},
  {Massey}, {Mobasher}, {Rhodes}, {Scoville}, \& {Shopbell}}]{koe07}
{Koekemoer} A.~M. {et~al.}, 2007, \apjs, 172, 196

\bibitem[{{Lai} {et~al}\mbox{.}(2007){Lai}, {Huang}, {Fazio}, {Cowie}, {Hu}, \&
  {Kakazu}}]{lai07}
{Lai} K., {Huang} J.-S., {Fazio} G., {Cowie} L.~L., {Hu} E.~M., {Kakazu} Y.,
  2007, \apj, 655, 704

\bibitem[{{Lai} {et~al}\mbox{.}(2008){Lai}, {Huang}, {Fazio}, {Gawiser},
  {Ciardullo}, {Damen}, {Franx}, {Gronwall}, {Labb{\'e}}, {Magdis}, \& {van
  Dokkum}}]{lai08}
{Lai} K. {et~al.}, 2008, \apj, 674, 70

\bibitem[{{Leauthaud} {et~al}\mbox{.}(2007){Leauthaud}, {Massey}, {Kneib},
  {Rhodes}, {Johnston}, {Capak}, {Heymans}, {Ellis}, {Koekemoer}, {Le
  F{\`e}vre}, {Mellier}, {R{\'e}fr{\'e}gier}, {Robin}, {Scoville}, {Tasca},
  {Taylor}, \& {Van Waerbeke}}]{leau07}
{Leauthaud} A. {et~al.}, 2007, \apjs, 172, 219

\bibitem[{{Madau}(1995)}]{mad95}
{Madau} P., 1995, \apj, 441, 18

\bibitem[{{Malhotra} {et~al}\mbox{.}(2012){Malhotra}, {Rhoads}, {Finkelstein},
  {Hathi}, {Nilsson}, {McLinden}, \& {Pirzkal}}]{mal12}
{Malhotra} S., {Rhoads} J.~E., {Finkelstein} S.~L., {Hathi} N., {Nilsson} K.,
  {McLinden} E., {Pirzkal} N., 2012, \apjl, 750, L36

\bibitem[{{Maraston} {et~al}\mbox{.}(2010){Maraston}, {Pforr}, {Renzini},
  {Daddi}, {Dickinson}, {Cimatti}, \& {Tonini}}]{mar10}
{Maraston} C., {Pforr} J., {Renzini} A., {Daddi} E., {Dickinson} M., {Cimatti}
  A., {Tonini} C., 2010, \mnras, 407, 830

\bibitem[{{Massey} {et~al}\mbox{.}(2010){Massey}, {Stoughton}, {Leauthaud},
  {Rhodes}, {Koekemoer}, {Ellis}, \& {Shaghoulian}}]{mas10}
{Massey} R., {Stoughton} C., {Leauthaud} A., {Rhodes} J., {Koekemoer} A.,
  {Ellis} R., {Shaghoulian} E., 2010, \mnras, 401, 371

\bibitem[{{McCracken} {et~al}\mbox{.}(2010){McCracken}, {Capak}, {Salvato},
  {Aussel}, {Thompson}, {Daddi}, {Sanders}, {Kneib}, {Willott}, {Mancini},
  {Renzini}, {Cook}, {Le F{\`e}vre}, {Ilbert}, {Kartaltepe}, {Koekemoer},
  {Mellier}, {Murayama}, {Scoville}, {Shioya}, \& {Tanaguchi}}]{mcc10}
{McCracken} H.~J. {et~al.}, 2010, \apj, 708, 202

\bibitem[{{McCracken} {et~al}\mbox{.}(2012){McCracken}, {Milvang-Jensen},
  {Dunlop}, {Franx}, {Fynbo}, {Le F{\`e}vre}, {Holt}, {Caputi}, {Goranova},
  {Buitrago}, {Emerson}, {Freudling}, {Hudelot}, {L{\'o}pez-Sanjuan},
  {Magnard}, {Mellier}, {M{\o}ller}, {Nilsson}, {Sutherland}, {Tasca}, \&
  {Zabl}}]{mcc12}
{McCracken} H.~J. {et~al.}, 2012, \aap, 544, A156

\bibitem[{{McLean} {et~al}\mbox{.}(1998){McLean}, {Becklin}, {Bendiksen},
  {Brims}, {Canfield}, {Figer}, {Graham}, {Hare}, {Lacayanga}, {Larkin},
  {Larson}, {Levenson}, {Magnone}, {Teplitz}, \& {Wong}}]{mc98}
{McLean} I.~S. {et~al.}, 1998, in Presented at the Society of Photo-Optical
  Instrumentation Engineers (SPIE) Conference, Vol. 3354, Society of
  Photo-Optical Instrumentation Engineers (SPIE) Conference Series,
  {A.~M.~Fowler}, ed., pp. 566--578

\bibitem[{{McLinden} {et~al}\mbox{.}(2011){McLinden}, {Finkelstein}, {Rhoads},
  {Malhotra}, {Hibon}, {Richardson}, {Cresci}, {Quirrenbach}, {Pasquali},
  {Bian}, {Fan}, \& {Woodward}}]{mcl11}
{McLinden} E.~M. {et~al.}, 2011, \apj, 730, 136

\bibitem[{{Nakajima} {et~al}\mbox{.}(2013){Nakajima}, {Ouchi}, {Shimasaku},
  {Hashimoto}, {Ono}, \& {Lee}}]{nak13}
{Nakajima} K., {Ouchi} M., {Shimasaku} K., {Hashimoto} T., {Ono} Y., {Lee}
  J.~C., 2013, \apj, 769, 3

\bibitem[{{Nakajima} {et~al}\mbox{.}(2012){Nakajima}, {Ouchi}, {Shimasaku},
  {Ono}, {Lee}, {Foucaud}, {Ly}, {Dale}, {Salim}, {Finn}, {Almaini}, \&
  {Okamura}}]{nak12}
{Nakajima} K. {et~al.}, 2012, \apj, 745, 12

\bibitem[{{Nilsson} {et~al}\mbox{.}(2007){Nilsson}, {M{\o}ller}, {M{\"o}ller},
  {Fynbo}, {Micha{\l}owski}, {Watson}, {Ledoux}, {Rosati}, {Pedersen}, \&
  {Grove}}]{nil07}
{Nilsson} K.~K. {et~al.}, 2007, \aap, 471, 71

\bibitem[{{Nilsson} {et~al}\mbox{.}(2011){Nilsson}, {{\"O}stlin}, {M{\o}ller},
  {M{\"o}ller-Nilsson}, {Tapken}, {Freudling}, \& {Fynbo}}]{nil11}
{Nilsson} K.~K., {{\"O}stlin} G., {M{\o}ller} P., {M{\"o}ller-Nilsson} O.,
  {Tapken} C., {Freudling} W., {Fynbo} J.~P.~U., 2011, \aap, 529, A9

\bibitem[{{Ono} {et~al}\mbox{.}(2010){Ono}, {Ouchi}, {Shimasaku}, {Akiyama},
  {Dunlop}, {Farrah}, {Lee}, {McLure}, {Okamura}, \& {Yoshida}}]{ono10a}
{Ono} Y. {et~al.}, 2010, \mnras, 402, 1580

\bibitem[{{Ouchi} {et~al}\mbox{.}(2008){Ouchi}, {Shimasaku}, {Akiyama},
  {Simpson}, {Saito}, {Ueda}, {Furusawa}, {Sekiguchi}, {Yamada}, {Kodama},
  {Kashikawa}, {Okamura}, {Iye}, {Takata}, {Yoshida}, \& {Yoshida}}]{ouch08}
{Ouchi} M. {et~al.}, 2008, \apjs, 176, 301

\bibitem[{{Papovich} {et~al}\mbox{.}(2006){Papovich}, {Cool}, {Eisenstein}, {Le
  Floc'h}, {Fan}, {Kennicutt}, {Smith}, {Rieke}, \& {Vestergaard}}]{pap06}
{Papovich} C. {et~al.}, 2006, \aj, 132, 231

\bibitem[{{Papovich}, {Dickinson} \& {Ferguson}(2001){Papovich}, {Dickinson},
  \& {Ferguson}}]{pap01}
{Papovich} C., {Dickinson} M., {Ferguson} H.~C., 2001, \apj, 559, 620

\bibitem[{{Papovich} {et~al}\mbox{.}(2011){Papovich}, {Finkelstein},
  {Ferguson}, {Lotz}, \& {Giavalisco}}]{pap11}
{Papovich} C., {Finkelstein} S.~L., {Ferguson} H.~C., {Lotz} J.~M.,
  {Giavalisco} M., 2011, \mnras, 412, 1123

\bibitem[{{Pickles}(1998)}]{pic98}
{Pickles} A.~J., 1998, \pasp, 110, 863

\bibitem[{{Pirzkal} {et~al}\mbox{.}(2007){Pirzkal}, {Malhotra}, {Rhoads}, \&
  {Xu}}]{pirz07}
{Pirzkal} N., {Malhotra} S., {Rhoads} J.~E., {Xu} C., 2007, \apj, 667, 49

\bibitem[{{Raichoor} {et~al}\mbox{.}(2011){Raichoor}, {Mei}, {Nakata},
  {Stanford}, {Holden}, {Rettura}, {Huertas-Company}, {Postman}, {Rosati},
  {Blakeslee}, {Demarco}, {Eisenhardt}, {Illingworth}, {Jee}, {Kodama},
  {Tanaka}, \& {White}}]{ra11}
{Raichoor} A. {et~al.}, 2011, \apj, 732, 12

\bibitem[{{Rhoads} {et~al}\mbox{.}(2003){Rhoads}, {Dey}, {Malhotra}, {Stern},
  {Spinrad}, {Jannuzi}, {Dawson}, {Brown}, \& {Landes}}]{rho03}
{Rhoads} J.~E. {et~al.}, 2003, \aj, 125, 1006

\bibitem[{{Sanders} {et~al}\mbox{.}(2007){Sanders}, {Salvato}, {Aussel},
  {Ilbert}, {Scoville}, {Surace}, {Frayer}, {Sheth}, {Helou}, {Brooke},
  {Bhattacharya}, {Yan}, {Kartaltepe}, {Barnes}, {Blain}, {Calzetti}, {Capak},
  {Carilli}, {Carollo}, {Comastri}, {Daddi}, {Ellis}, {Elvis}, {Fall},
  {Franceschini}, {Giavalisco}, {Hasinger}, {Impey}, {Koekemoer}, {Le
  F{\`e}vre}, {Lilly}, {Liu}, {McCracken}, {Mobasher}, {Renzini}, {Rich},
  {Schinnerer}, {Shopbell}, {Taniguchi}, {Thompson}, {Urry}, \&
  {Williams}}]{sand07}
{Sanders} D.~B. {et~al.}, 2007, \apjs, 172, 86

\bibitem[{{Schaerer} \& {de Barros}(2009)}]{sd09}
{Schaerer} D., {de Barros} S., 2009, \aap, 502, 423

\bibitem[{{Seifert} {et~al}\mbox{.}(2003){Seifert}, {Appenzeller},
  {Baumeister}, {Bizenberger}, {Bomans}, {Dettmar}, {Grimm}, {Herbst},
  {Hofmann}, {Juette}, {Laun}, {Lehmitz}, {Lemke}, {Lenzen}, {Mandel},
  {Polsterer}, {Rohloff}, {Schuetze}, {Seltmann}, {Thatte}, {Weiser}, \&
  {Xu}}]{sei03}
{Seifert} W. {et~al.}, 2003, in Society of Photo-Optical Instrumentation
  Engineers (SPIE) Conference Series, Vol. 4841, Society of Photo-Optical
  Instrumentation Engineers (SPIE) Conference Series, {Iye} M., {Moorwood}
  A.~F.~M., eds., pp. 962--973

\bibitem[{{Spergel} {et~al}\mbox{.}(2007){Spergel}, {Bean}, {Dor{\'e}},
  {Nolta}, {Bennett}, {Dunkley}, {Hinshaw}, {Jarosik}, {Komatsu}, {Page},
  {Peiris}, {Verde}, {Halpern}, {Hill}, {Kogut}, {Limon}, {Meyer}, {Odegard},
  {Tucker}, {Weiland}, {Wollack}, \& {Wright}}]{sper}
{Spergel} D.~N. {et~al.}, 2007, \apjs, 170, 377

\bibitem[{{Steidel} {et~al}\mbox{.}(2010){Steidel}, {Erb}, {Shapley},
  {Pettini}, {Reddy}, {Bogosavljevi{\'c}}, {Rudie}, \& {Rakic}}]{stei10}
{Steidel} C.~C., {Erb} D.~K., {Shapley} A.~E., {Pettini} M., {Reddy} N.,
  {Bogosavljevi{\'c}} M., {Rudie} G.~C., {Rakic} O., 2010, \apj, 717, 289

\bibitem[{{Valdes}(1993)}]{val93}
{Valdes} F., 1993, {Guide to the Kitt Peak Coude Slit Reduction Task DOSLIT}.
  Central Computer Services, NOAO

\bibitem[{{Vargas} {et~al}\mbox{.}(2013){Vargas}, {Bish}, {Acquaviva},
  {Gawiser}, {Finkelstein}, {Ciardullo}, {Ashby}, {Feldmeier}, {Ferguson},
  {Gronwall}, {Guaita}, {Hagen}, {Koekemoer}, {Kurczynski}, {Newman}, \&
  {Padilla}}]{var13}
{Vargas} C.~J. {et~al.}, 2013, ArXiv e-prints

\bibitem[{{Verhamme} {et~al}\mbox{.}(2008){Verhamme}, {Schaerer}, {Atek}, \&
  {Tapken}}]{ver08}
{Verhamme} A., {Schaerer} D., {Atek} H., {Tapken} C., 2008, \aap, 491, 89

\bibitem[{{Verhamme}, {Schaerer} \& {Maselli}(2006){Verhamme}, {Schaerer}, \&
  {Maselli}}]{ver06}
{Verhamme} A., {Schaerer} D., {Maselli} A., 2006, \aap, 460, 397

\bibitem[{{Williams} {et~al}\mbox{.}(2004){Williams}, {Olszewski}, {Lesser}, \&
  {Burge}}]{wil04}
{Williams} G.~G., {Olszewski} E., {Lesser} M.~P., {Burge} J.~H., 2004, in
  Society of Photo-Optical Instrumentation Engineers (SPIE) Conference Series,
  Vol. 5492, Society of Photo-Optical Instrumentation Engineers (SPIE)
  Conference Series, {Moorwood} A.~F.~M., {Iye} M., eds., pp. 787--798

\bibitem[{{Wright}(2006)}]{wr06}
{Wright} E.~L., 2006, \pasp, 118, 1711

\bibitem[{{Zheng} {et~al}\mbox{.}(2010){Zheng}, {Wang}, {Finkelstein},
  {Malhotra}, {Rhoads}, \& {Finkelstein}}]{zh10}
{Zheng} Z.~Y., {Wang} J.~X., {Finkelstein} S.~L., {Malhotra} S., {Rhoads}
  J.~E., {Finkelstein} K.~D., 2010, \apj, 718, 52

\end{thebibliography}

\appendix
\section{Additional Figures}\label{sec:appendix}
\begin{figure*}
\centering
\includegraphics[scale=0.8]{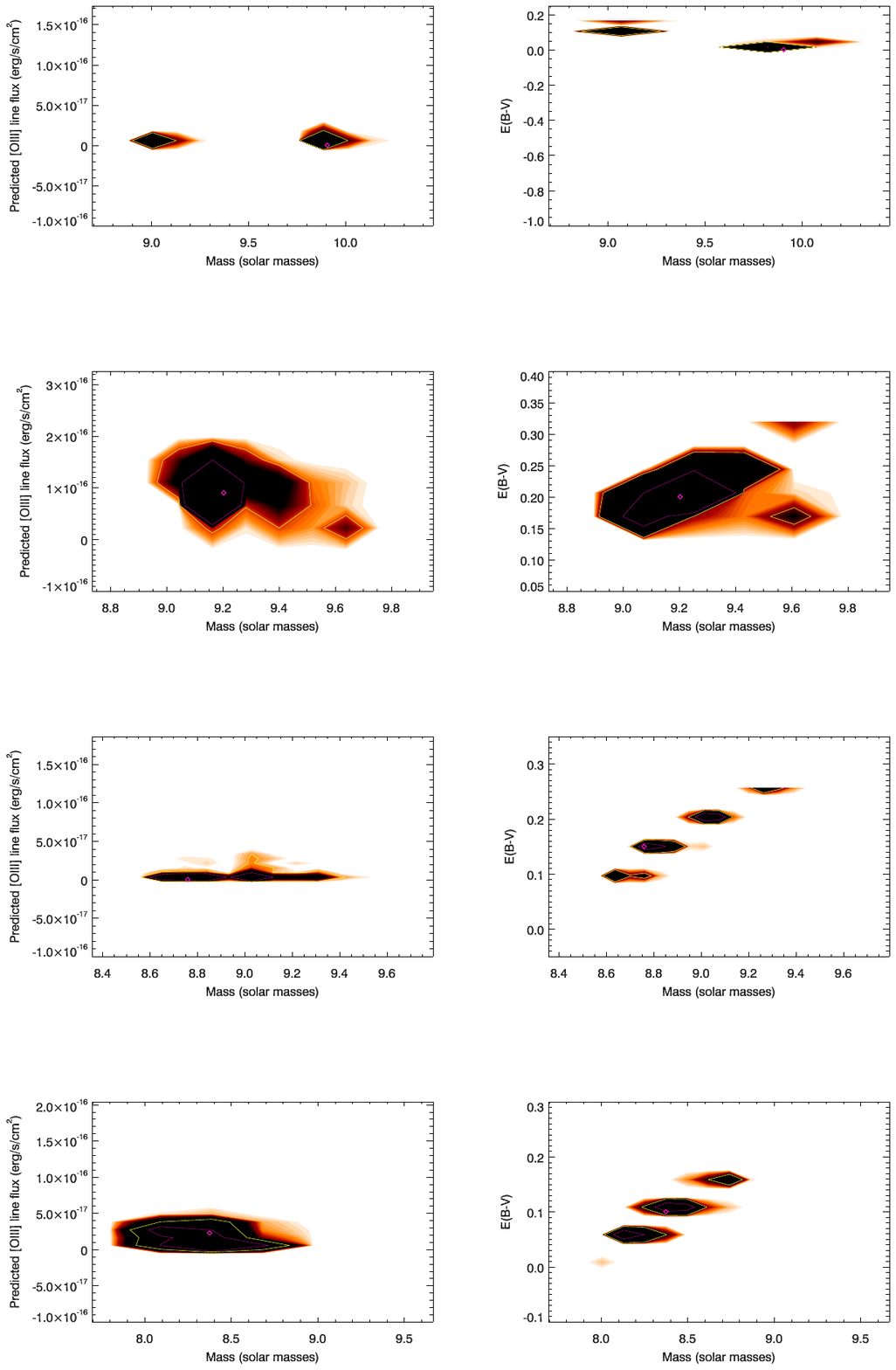}
\caption{Both columns show density plots from our MC simulations for additional parameters - left column is predicted \oiii\ line flux vs. mass (log) and right column is \textit{E(B-V)} vs. mass (log).  The best allowed-fit is shown as a magenta diamond. Contours encompassing $\sim$ 68 per cent and $\sim$ 95 per cent of the results are shown in magenta and yellow, respectively. The order of objects in Figure \ref{fig:tbl17} - Figure \ref{fig:tbl18} matches the order of objects in Tables \ref{sedtbl1} and \ref{sedtbl2}. AGN are excluded from both table and figures.}\label{fig:tbl17}
\end{figure*}

\begin{figure*}
\centering
\includegraphics[scale=0.8]{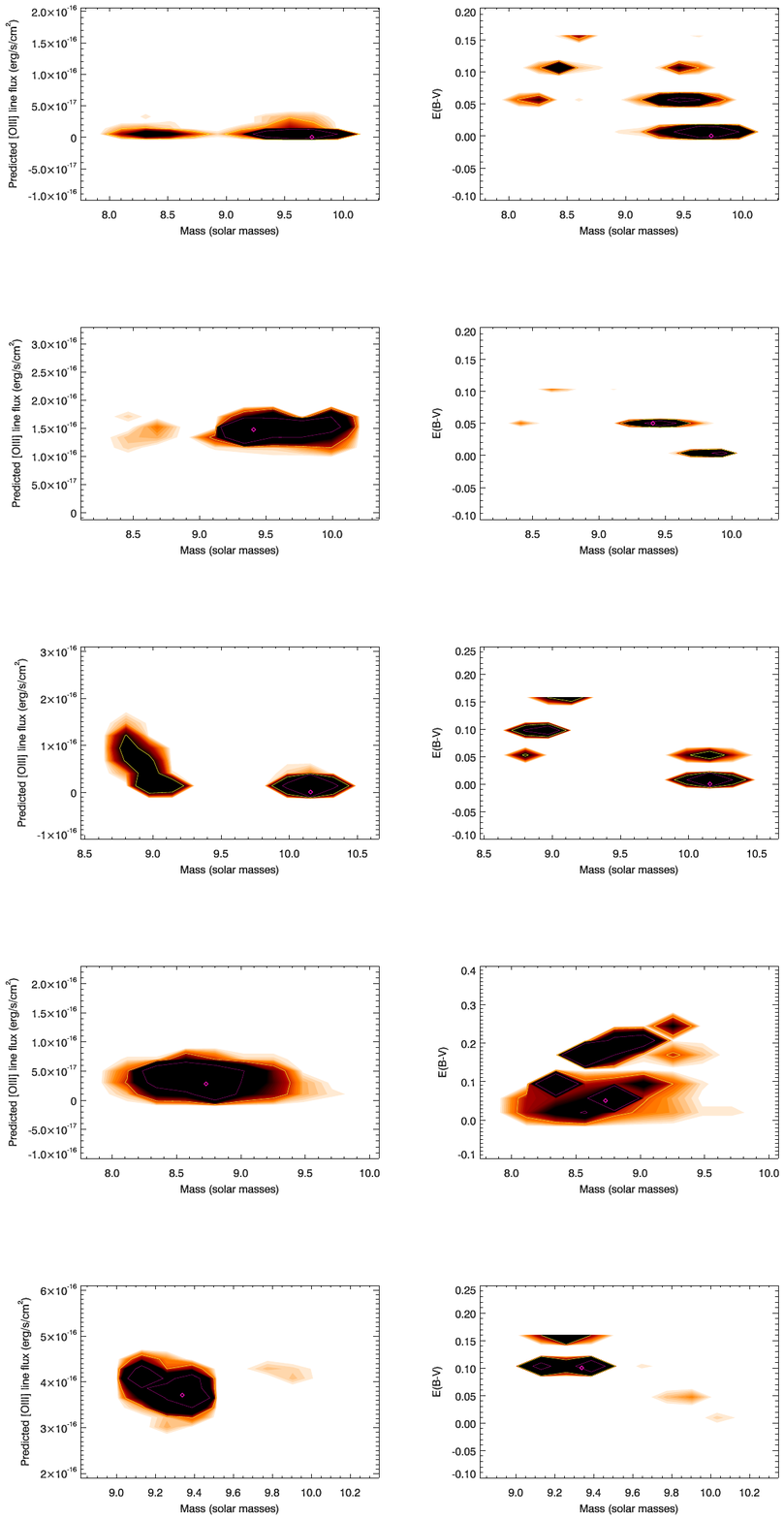}
\caption{Same as Figure \ref{fig:tbl17} for next 5 objects.}\label{fig:tbl18}
\end{figure*}

\begin{figure*}
\centering
\includegraphics[scale=0.8]{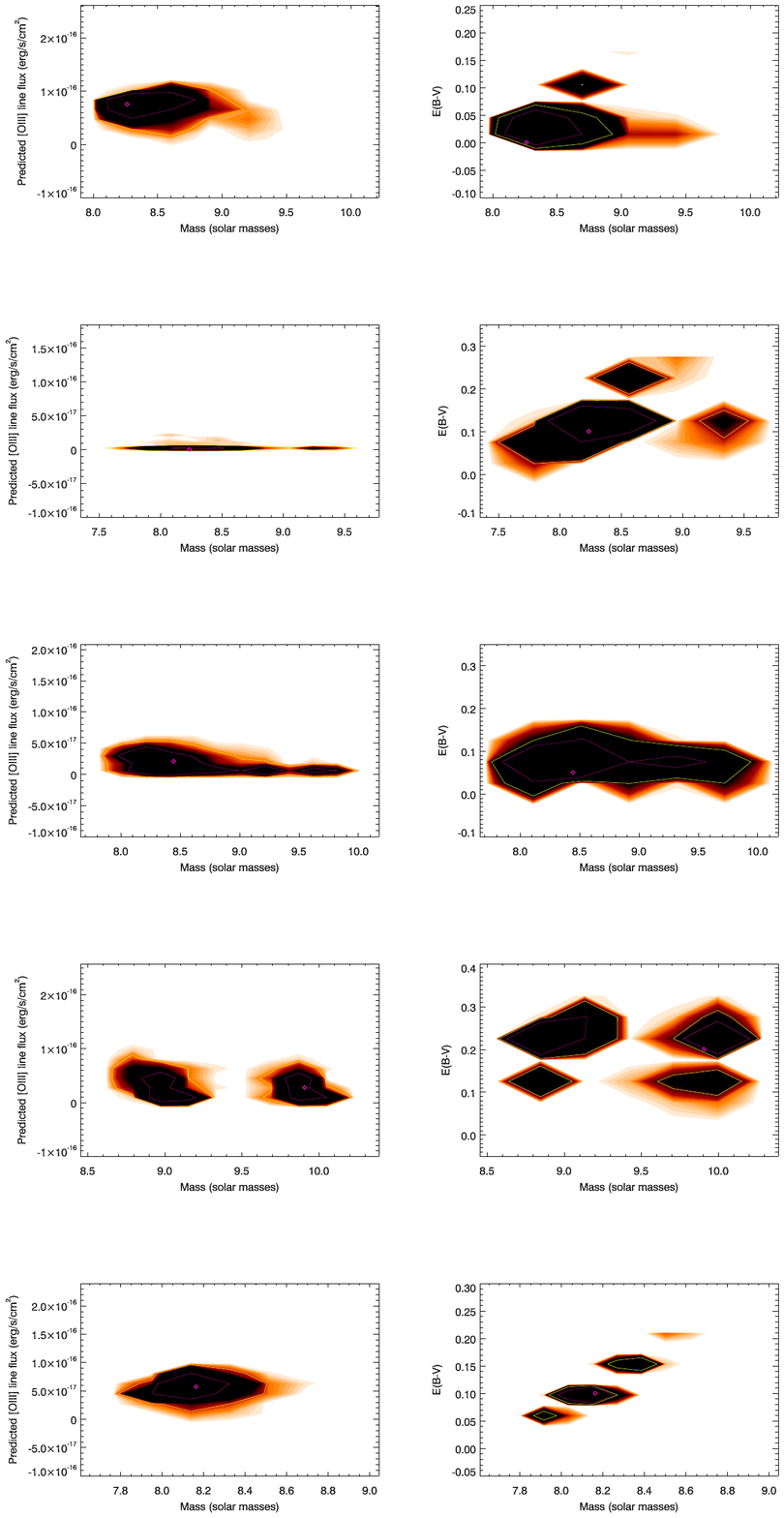}
\caption{Same as Figure \ref{fig:tbl17} for next 5 objects.}\label{fig:tbl19}
\end{figure*}

\begin{figure*}
\centering
\includegraphics[scale=0.8]{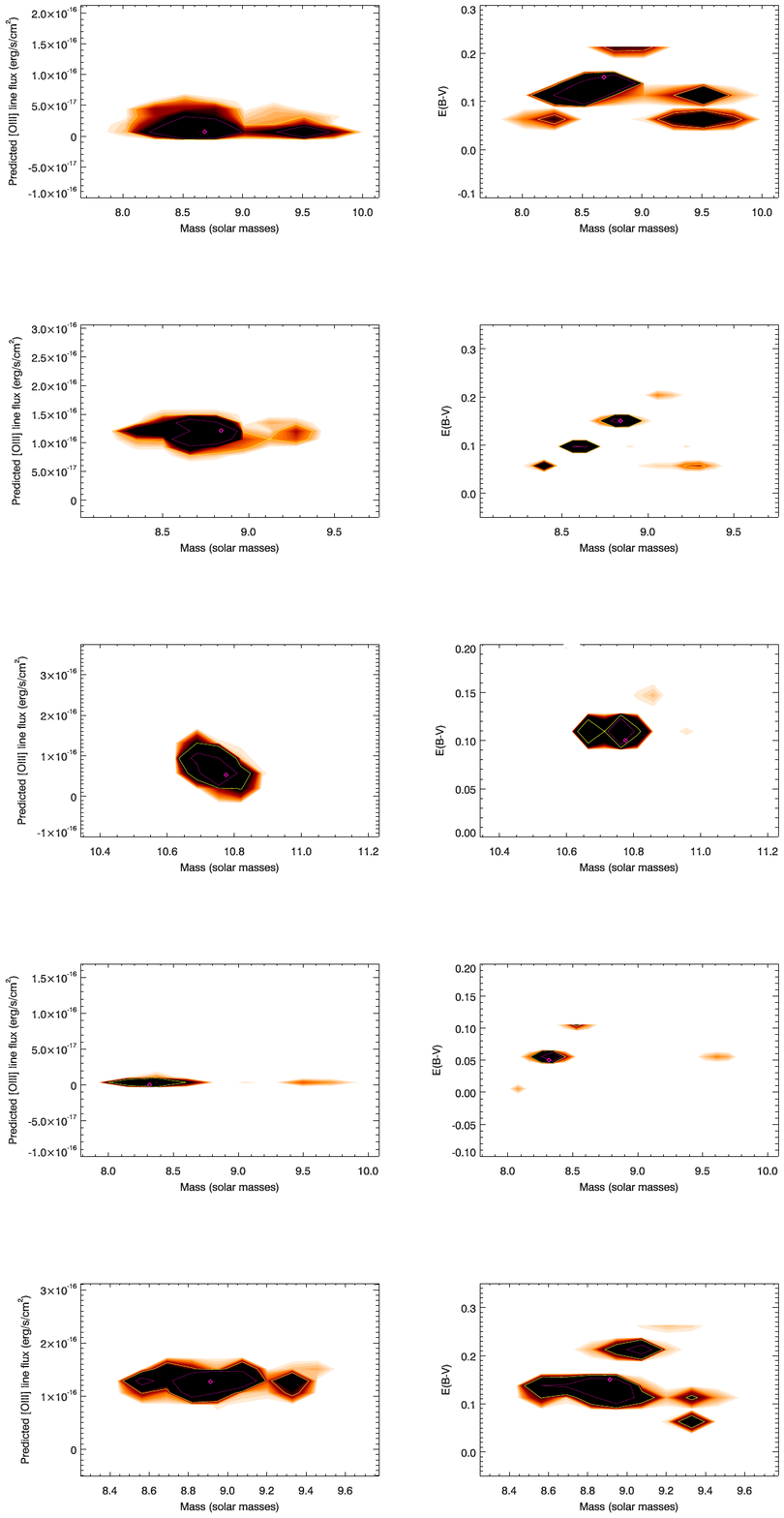}
\caption{Same as Figure \ref{fig:tbl17} for next 5 objects.}\label{fig:tbl20}
\end{figure*}

\begin{figure*}
\centering
\includegraphics[scale=0.8]{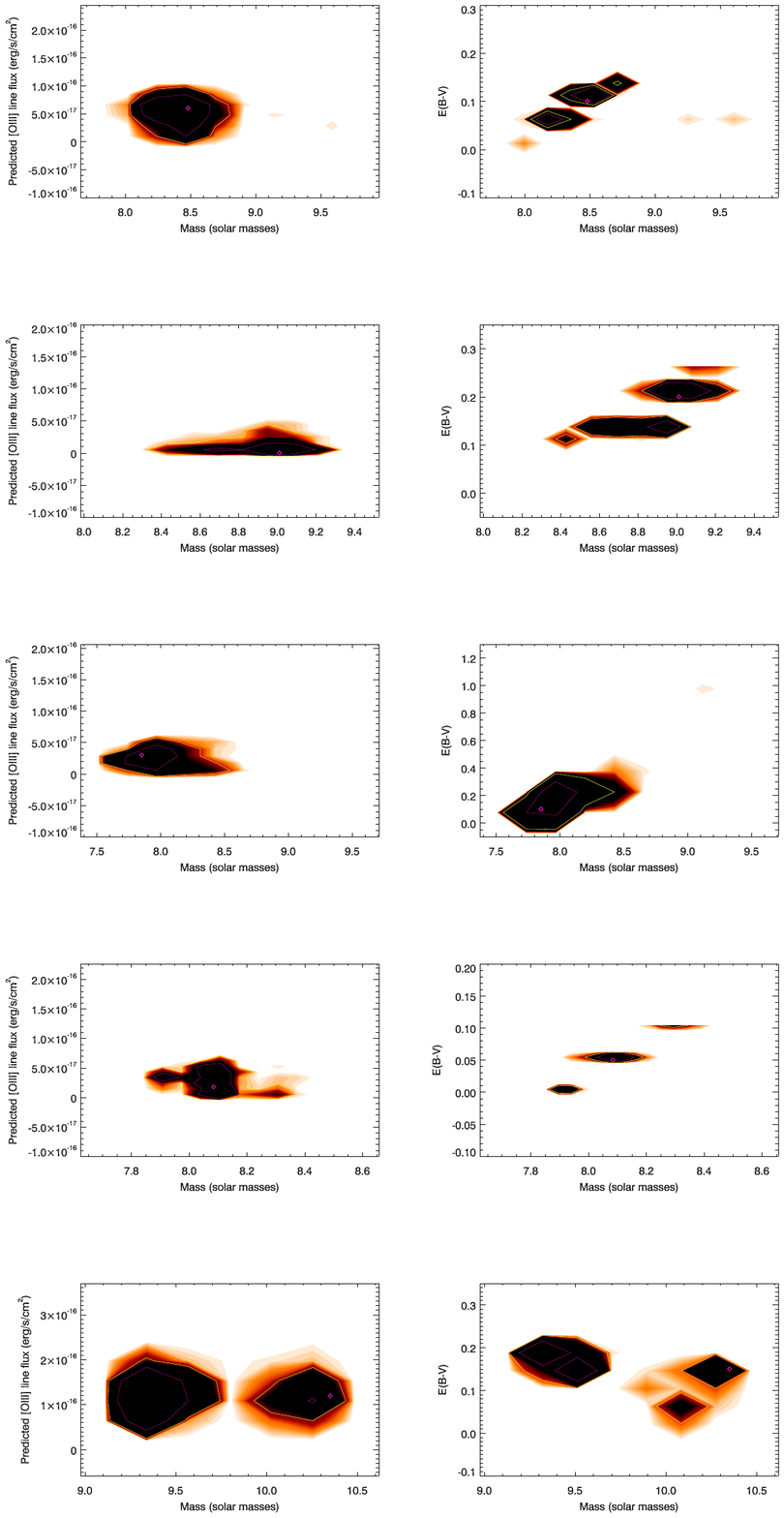}
\caption{Same as Figure \ref{fig:tbl17} for next 5 objects.}\label{fig:tbl21}
\end{figure*}

\begin{figure*}
\centering
\includegraphics[scale=0.8]{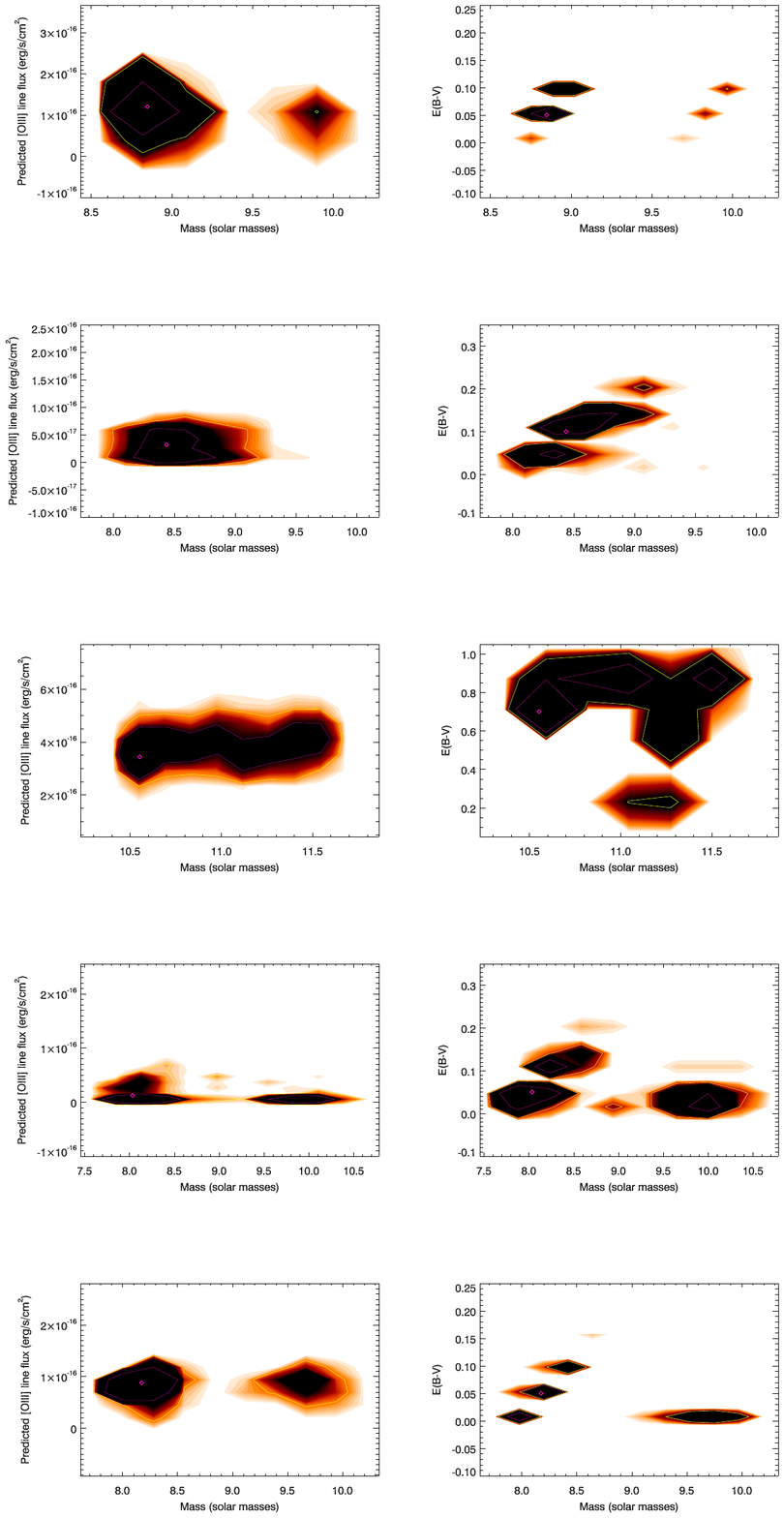}
\caption{Same as Figure \ref{fig:tbl17} for next 5 objects.}\label{fig:tbl22}
\end{figure*}

\begin{figure*}
\centering
\includegraphics[scale=0.8]{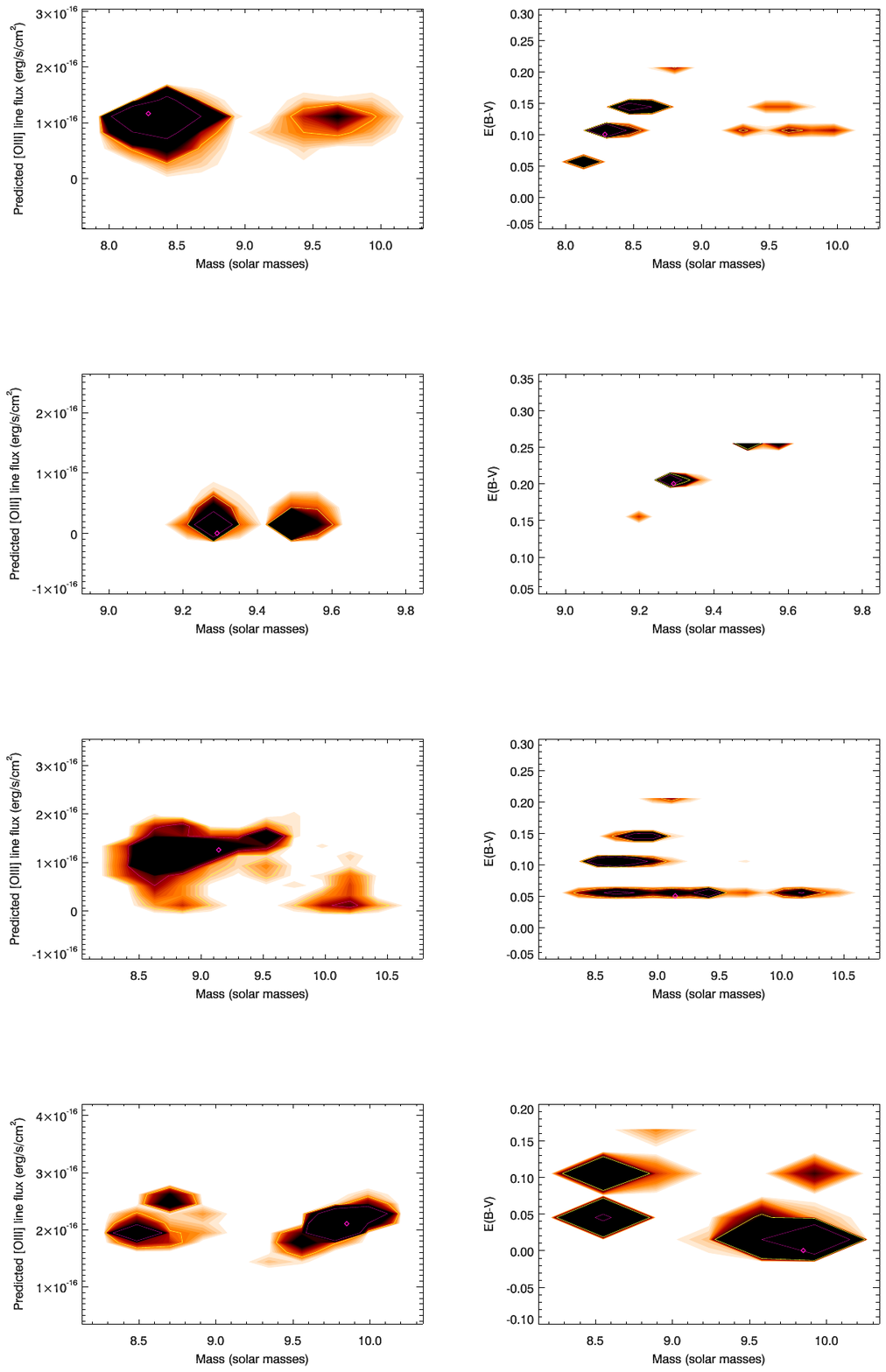}
\caption{Same as Figure \ref{fig:tbl17} for next 5 objects.}\label{fig:tbl23}
\end{figure*}
\clearpage

\bsp
\label{lastpage}

\end{document}